\documentclass{aa}
\usepackage{graphicx}
\usepackage{color}
\usepackage{lscape}
\usepackage{natbib}
\usepackage{txfonts}
\begin{document}

   \title{Morphology parameters: substructure identification in X-ray galaxy clusters}

   \author{Viral Parekh\thanks{E-mail:viral@rri.res.in}
          \inst{1,2}
          \and
          Kurt van der Heyden\inst{1}
          \and
          Chiara Ferrari\inst{3}
          \and
          Garry Angus\inst{1,4}
          \and
          Benne Holwerda\inst{5}
          }

   \institute{Astrophysics, Cosmology and Gravity Centre (ACGC), Astronomy Department, University of Cape Town,\\ 
              Private Bag X3, 7700, Rondebosch, Republic of South Africa
         \and
         Raman Research Institute, Sadashivanagar, Bangalore 560 080, India
         \and
             Laboratoire Lagrange, UMR7293, Universit\'{e} de Nice Sophia-Antipolis, CNRS, Observatoire de la C\^{o}te d' Azur, 06300 Nice, France
         \and 
             Department of Physics and Astrophysics, Vrije Universiteit Brussel, Pleinlaan 2, 1050 Brussels, Belgium    
         \and 
             European Space Agency (ESTEC), Keplerlaan 1, 2200 AG, Noordwijk, the Netherland
             }

  % \date{Received xxxx; accepted xxxx}
    \date {Accepted}
% \abstract{}{}{}{}{} 
% 5 {} token are mandatory
 
  \abstract
  % context heading (optional)
  % {} leave it empty if necessary  
   {In recent years multi-wavelength observations have shown the presence of substructures related to merging events in a high fraction of galaxy clusters. Clusters can be roughly grouped into two categories -- relaxed and non-relaxed -- and a proper characterisation of the dynamical state of these systems is of crucial importance both for astrophysical and cosmological studies.}
  % aims heading (mandatory)
   {In this paper we investigate the use of a number of morphological parameters (Gini, $M_{20}$, Concentration, Asymmetry, Smoothness, Ellipticity and Gini of the second order moment, $G_{M}$) introduced to automatically classify clusters as relaxed or dynamically disturbed systems.} {We apply our method to a sample of clusters at different redshifts extracted from the {\it Chandra} archive and we investigate possible correlations between morphological parameters and other X-ray gas properties. We conclude that a combination of the adopted parameters is a very useful tool to properly characterise the X-ray cluster morphology.}
  % results heading (mandatory)
   {According to our results three parameters -- Gini, $M_{20}$ and Concentration -- are very promising for identifying cluster mergers. The Gini coefficient is a particularly powerful tool, especially at high redshift, being independent from the choice of the position of the cluster centre. We find that high Gini ($>$ 0.65), high Concentration ($>$ 1.55) and low $M_{20}$ ($<$ -2.0) values are associated with relaxed clusters, while low Gini ($<$ 0.4), low Concentration ($<$ 1.0) and high $M_{20}$ ($>$ -1.4) characterise dynamically perturbed systems. We also estimate the X-ray cluster morphological parameters in the case of {\it radio loud} clusters. In excellent agreement with previous analyses we confirm that diffuse intracluster radio sources are associated with major mergers.}
  % conclusions heading (optional), leave it empty if necessary 
   {}

   \keywords{galaxies: X-ray clusters: substructure - merger: radio halo: radio relic
               }

   \maketitle
%
%________________________________________________________________

\section{Introduction}

\par  It is now well proven that massive galaxy clusters form and evolve at the intersection of cosmic web filaments through merging and accretion of smaller mass systems \cite[e.g.][and references therein]{2011A&A...525A..79M}. Huge gravitational energy is released during cluster collisions ($\approx 10^{64}$ ergs) and several billion years are required for the cluster to re-establish a situation of (quasi-)equilibrium after a major merger episode.

\par Both observations and numerical simulations have shown that merging events deeply affect the properties of the different cluster components \cite[e.g.][and references therein]{2008SSRv..134...93F}. Multiple merger episodes could for instance be responsible for disturbing the dynamically relaxed cores of the X-ray emitting hot intracluster medium \citep[ICM, e.g. ][]{2008ApJ...675.1125B}, as well as the ICM density, temperature and metallicity distribution \citep[e.g.][]{2006A&A...447..827K}. In addition, it is well known that a disturbed X-ray morphology is typical of dynamically perturbed galaxy clusters \citep[][and references therein]{2006A&A...447..827K}. The presence of substructures, a highly elliptical cluster X-ray morphology or an X-ray centroid variation are typical features { that suggests that a cluster is not virialized}. This has important implications both for using clusters as tools for cosmology, and for studying the complex gravitational and non-gravitational processes acting during large-scale structure formation and evolution, since in both cases we need to know if observed clusters are relaxed or not.

\par Joint X-ray and optical studies can provide detailed information about the dynamical state of a cluster \cite[e.g.][]{2006A&A...446..417F}, but they are extremely time demanding from the observational and analysis point of view. For statistical studies of large cluster samples we need to identify robust indicators that can somehow quantify the cluster dynamical state. Since the morphology of clusters is deeply related to their evolutionary history, different morphological estimators have been proposed. \cite{1992csg..conf...49J} classified galaxy clusters observed by the {\it Einstein} X-ray satellite into six morphological classes which include single, elliptical, offset center, primary with small secondary, bimodal and complex. Several studies have tried to quantify the fraction of dynamically disturbed clusters from morphological analyses. {\cite{1999ApJ...511...65J}} showed that around 30\% of clusters observed with the {\it Einstein} satellite contain substructure. More recently \cite{2005RvMA...18...76S} have noticed in ROSAT observations that $\sim$ 50$\%$ of clusters have substructure. With high resolution telescopes such as {\it Chandra} and {\it XMM}, it has become easier to identify subclusters, bimodality and X-ray centroid shifts in clusters. However some mergers are too complex to be identified only from X-ray morphological analysis and, particularly in the case of high redshift clusters, some special techniques and statistics are required, which can provide a more quantitative and robust measure of the degree of the cluster disturbance.

\par Various techniques have been suggested to provide a more quantitative and qualitative measure of the degree of the cluster disturbance: Power ratios \citep {1995ApJ...439...29B, 2005ApJ...624..606J, 2010A&A...514A..32B} and the emission centroid shift \citep {1993ApJ...413..492M, 2010A&A...514A..32B} are most commonly used to classify X-ray galaxy clusters from the morphological point of view. Recently, \cite {2012ApJ...746..139A} have used a {\it residual flux method}; in order  to calculate the substructure level in a given X-ray galaxy cluster, they take into account the ratio between number of counts on the residual (which they obtain by the subtraction of a surface brightness model from the original X-ray image) and on the original cluster images. \cite {2012arXiv1210.6445W} proposed to use the maximum of the third order power ratios calculated in annuli of fixed width and constantly increasing radius to measure the degree of substructure. \cite {2012arXiv1211.7040R} have used six different morphology parameters namely, asymmetry, fluctuation of the X-ray brightness (smoothness), hardness ratios, concentration, the centroid-shift method and third order power ratio to characterise simulated clusters. They took hydrodynamical simulation of 60 clusters and passed it through a {\it Chandra} telescope simulator with uniform exposure time (100 ks) for all clusters. Out of all of these parameters they found that only the asymmetry and concentration parameters could straightforwardly and clearly separate relaxed and non-relaxed systems. The smoothness parameter is affected by the choice of radii and smoothing kernel size. The centroid-shift parameter also works reasonably well and leaves only a few overlapping relaxed and non-relaxed clusters. The third order power ratio technique also depends on the choice of radius and is limited to only detecting substructure near to the cluster center. {More recently \cite{2013ApJ...779..112N} have used photon asymmetry and central Concentration parameters to quantify morphology of high-$z$ clusters which suffer from low photon counts.}

\par In this paper, we investigate seven morphology parameters, typically used for galaxy classification, to study X-ray galaxy cluster morphology and we motivate which parameters are the optimal for identifying substructure or characterising dynamical states. The combination of morphology parameters has been successfully used to classify different galaxy morphologies {\citep {2007ApJS..172..468Z, 2007ApJS..172..406S, 2011MNRAS.416.2426H}} and we want to investigate their usefulness in galaxy cluster classification. Our focus is not limited to separating galaxy clusters into relaxed and non-relaxed categories, but to study correlations between X-ray gas properties and morphology parameters, as well as the evolution of morphological properties of galaxy clusters from the high redshift universe to present.

\par We explore the usefulness of the non-parametric morphology parameters on a subset of the {\it ROSAT} 400 deg$^{2}$ cluster sample observed by the {\it Chandra} X-ray telescope (\citealt {2009ApJ...692.1033V}, hereafter V09). We choose this sample because it has good quality X-ray data and also a broad distribution of redshifts. There are in total 85 ({49 low-$z$ (0.02--0.3) and 36 high-$z$ (0.3--0.8)}) galaxy clusters in our analysis. In addition, V09 has measured global properties of the galaxy clusters ({such as luminosity, temperature, mass, etc.}) which we use to compare with our measured morphology parameters. 
     
\par  As a test case, in this paper, we study the dynamical activity of clusters hosting diffuse radio sources ({\it radio halos}), in particular we focus our attention on clusters taken from \cite {2009A&A...507.1257G}. Current results suggest a strong link between the presence of diffuse intracluster radio emission and cluster mergers \citep[][and references therein]{2008SSRv..134...93F,2012A&ARv..20...54F}. {Similar to what was done by \citet{2010ApJ...721L..82C}, but using our set of X-ray morphological parameters}, in this paper we analyze the X-ray morphology of relaxed clusters and non-relaxed clusters (which includes both {\it radio quiet} and {\it radio loud} mergers).

\par This paper is organized as follows. \S\ 2 gives a brief introduction to the morphology parameters. \S\ 3  gives the sample selection and X-ray data reduction. In \S\ 4, we present our results. \S\ 5 shows the systematic and possible bias on morphology parameters. In \S\ 6, we compare our parameter measures with available cluster global properties. Finally, \S\ 7 will give our discussions and conclusions. We assumed $H_{0}$=73 km s$^{-1}$ Mpc$^{-1}$ $\Omega_{M}$=0.3 and $\Omega_{\Lambda}$=0.7 throughout the paper, unless stated otherwise.  

\section{Introduction of morphology parameters}
\label{intro_sec}
\par The non-parametric morphology parameters (Gini, $M_{20}$, Concentration, Asymmetry, Smoothness, Ellipticity and Gini of the second order moment) are widely used to automatically separate galaxies of different Hubble types. As an example, they are used for galaxy morphology classification in the analysis of the HST and SDSS galaxy surveys {\citep {2003ApJ...588..218A, 2003ApJS..147....1C, 2004AJ....128..163L, 2007ApJS..172..468Z, 2011MNRAS.416.2426H, 2012ApJ...752..134W}}.
\par \cite {2003ApJ...588..218A}, \cite {2004AJ....128..163L} and \cite {2012ApJ...752..134W} also revealed the inter-relation between Gini, Concentration and $M_{20}$, as well as the possible inter-change between the Concentration and Gini parameter for high-$z$ galaxies. This encouraged us to investigate these parameters in more detail in order to characterise the dynamical state of galaxy clusters, particularly at high-$z$. In this paper we adopt the definition of Concentration, Asymmetry and Smoothness from \cite {2003ApJS..147....1C}, and of the Gini coefficient and $M_{20}$ from \cite {2004AJ....128..163L}. Gini of the second order moment was defined by \cite {2011MNRAS.416.2437H}. The required input parameters for computing the morphological indicators (except the Gini parameter) are the central position ($x_{c}, y_{c}$) of the galaxy clusters, as well as a fixed aperture size or area over which these morphology parameters are measured.

\par We calculate the center position by first assigning initial coordinates based on visual observation of each cluster image and then allow the flux weighted coordinates to iterate in a fixed aperture size of e.g. 500 kpc until they have converged. The centre coordinates are then the unique point at the center of the distribution of flux, essentially the light distribution equivalent to the ``centre of mass''.

\subsection{Gini coefficient and Gini of the second order moment ($G_{M}$)}
\par The Gini parameter is widely used in the field of economics, where it originated as the Lorenz curve \citep {1905PAmSA...9..209L}. It describes the inequality of wealth in a population. Here we use it as a calculation of flux distribution in a cluster image. If the total flux is equally distributed among the pixels, then the Gini value is equal to 0 (there is constant flux across the pixels regardless of whether those pixels are in the projected centre or not); but if the total flux is unevenly distributed and belongs to only a small number of pixels, then the Gini value is equal to 1. We adopt the following definition from \cite {2004AJ....128..163L}:
\begin{equation}
G=\frac{1}{\bar{K}n(n-1)} \sum_{i} (2i-n-1)K_{i} , 
\end{equation}
where $K_{i}$ is the pixel value in the $i^{th}$ pixel of a given image, $n$ is the total number of pixels in the image, and $\bar{K}$ is the mean pixel value of the image. 
\par We also apply a Gini value to the second order moment of each pixel, defining Gini of the second order moment as:
\begin{equation}
G_{M}=\frac{1}{\bar{F}n(n-1)} \sum_{i} (2i-n-1)F_{i} , 
\end{equation} 
where $F_{i}$ is the second order moment of each pixel:  \\
\begin{equation}
F_{i}=K_{i} \times [(x-x_{c})^2 + (y-y_{c})^2],
\end{equation}
where ($x$, $y$) is the pixel position with flux value $K_{i}$ in the cluster image; and $(x_{c}$,  $y_{c})$ is the coordinate of the cluster centre.

\subsection{Moment of light, $M_{20}$}
\par \cite {2004AJ....128..163L} define the total second order moment $F_{tot}$ as the flux in each pixel $K_{i}$ multiplied by the squared distance to the centre of the source, summed over all the selected pixels:
\begin{equation}
F_{tot}=\sum_{i} F_{i}=\sum_{i} K_{i}[(x_{i}-x_{c})^{2}+(y_{i}-y_{c})^{2}] ,
\end{equation}
where ($x_{c}$, $y_{c}$) is the centre of the cluster. \\
\par The second order moment can be used to trace various properties of galaxy clusters, such as the spatial distribution of multiple bright cores, substructure or mergers. $M_{20}$ is defined as the normalised second order moment of the relative contribution of the brightest 20\% of the pixels. To compute $M_{20}$, we rank-order the image pixels by flux, calculate $F_{i}$ over the brightest pixels until their sum equals 20\% of the total selected cluster flux, and then normalise by $F_{tot}$:
\begin{equation}
M_{20}=log\left(\frac{\sum_{i} F_{i}}{F_{tot}}\right), {\rm while} \sum_{i} K_{i}\leq 0.2K_{tot},
\label{m20eqn}
\end{equation}
where $K_{tot}$ is the total flux of the cluster image (image pixels are selected from the segmentation map\footnote{A map which defines the chosen circular aperture size with all pixels fixed to a value of 1.}), and $K_{i}$ is the flux value for each pixel {\it i} (where $K_{1}$ =  the brightest pixel, $K_{2}$ = the second brightest pixel, etc). 

\subsection{Concentration, Asymmetry and Smoothness (CAS)} 
\par Concentration, Asymmetry and Smoothness parameters are commonly known as CAS. 
\par Concentration is defined by \cite {2000AJ....119.2645B} and \cite {2003ApJS..147....1C} as:
\begin{equation}
C=5\times log\left(\frac{r_{80}}{r_{20}}\right) ,
\label{concequn}
\end{equation}
where $r_{80}$ and $r_{20}$ represent the radius within which 80\% and 20\% of the flux reside, respectively. Concentration is widely used in the classification of cool core, especially among distant clusters. \cite {2008A&A...483...35S} define the surface brightness concentration parameter for finding cool core clusters at high redshift as,
\begin{equation}
c_{sb} = \frac{C_{r} (r < 40 kpc)}{C_{r} (r < 400 kpc)} ,
\end{equation} 
where $C_{r}$ ($r$ $<$ 40 kpc) and $C_{r}$ ($r$ $<$ 400 kpc) are the integrated surface brightness within 40 kpc and 400 kpc, respectively. Instead of physical radii, we used the percentages of total flux within a given aperture size. This has an advantage { (atleast in low-$z$ clusters)} in that the flux is independent of angular bin size and galaxy cluster redshift. Our sample covers the redshift range 0.02 $<$ $z$ $<$ 0.9. This means that, by adopting a pixel size of \begin{math}2''\end{math}, 250 kpc correspond to a pixel range from $\sim$16 to $\sim$225. In order to avoid this large deviation in pixel spread, it is best to use various percentages of the total flux of the galaxy clusters. For the inner radii we use 20$\%$--50$\%$, and for the outer radii we used 80$\%$--90$\%$ of the total flux. For example, we use the $C_{5080}$ Concentration parameter which means 50\% of the flux within the inner radii and 80\% within the outer radii.

\par The Asymmetry value, which will give rotational symmetry around the cluster centre, is calculated when a cluster image is rotated by 180$^{\circ}$ around its centre ($x_{c}, y_{c}$) and is then subtracted from its original image.
\begin{equation}
A=\frac{\sum_{i,j}\mid K(i,j)-K_{180}(i,j)\mid}{\sum_{i,j}\mid K(i,j)\mid} ,
\end{equation}
where $K(i, j)$ is the value of the pixel at the image position $i, j$, and $K_{180}(i, j)$ is the value of the pixel in the cluster's image rotated by 180$^{\circ}$ around its centre. The Asymmetry value is sensitive to any region of the cluster that is responsible for asymmetric flux distribution. Hence, if the substructure affects the flux distribution at any scale, we can pick it up from the Asymmetry value for that galaxy cluster. 

\par The Smoothness parameter can be used to identify patchy flux distribution expected in non-relaxed clusters. By smoothing a cluster image with a filter of width $\sigma$,  high frequency structures can be removed from the image. At this point the original image is subtracted from this newly smoothed, lower resolution image. The effect is to produce a residual map that has only high-frequency components of the galaxy cluster's flux distribution. The flux of this residual image is then summed and divided by the total flux of the original cluster image in order to find its smoothness value,
\begin{equation}
S=\frac{\sum_{i,j}\mid K(i,j)-K_{s}(i,j)\mid}{\sum_{i,j}\mid K(i,j)\mid},
\end{equation}  
where $K_{s}(i,j)$ is the pixel in a smoothed image. Here we choose a Gaussian smoothing kernel of $\sigma$ = \begin{math}12''\end{math} as an arbitrary scale to smooth the cluster image. 
\subsection{Ellipticity}
Ellipticity is commonly defined by the ratio between a semi-major axis (A) and a semi-minor axis (B) as
\begin{equation}
E=1-\frac{B}{A},
\end{equation}
where A and B can be computed directly from the second order moments of the flux in the cluster image as
\begin{equation}
%\begin{small}
A^2=\frac { {\overline {x^2}} + {\overline {y^2}} } {2} + \sqrt {{\left (\frac { {\overline {x^2}} - {\overline {y^2}} } {2} \right) ^2} + \overline {xy}^2}
%\end{small}
\end{equation}
\begin{equation}
B^2=\frac { {\overline {x^2}} + {\overline {y^2}} } {2} - \sqrt {{\left (\frac { {\overline {x^2}} - {\overline {y^2}} } {2} \right) ^2} + \overline {xy}^2} ,
\end{equation}

where 

\begin{equation}
\begin{tiny}
\overline {x}=\frac{\displaystyle \sum_{i\in S} K_{i}x_{i}}{\displaystyle \sum_{i\in S} K_{i}},\hspace{0.5cm}
 \overline {y}=\frac{\displaystyle \sum_{i\in S} K_{i}y_{i}}{\displaystyle \sum_{i\in S} K_{i}}, 
\label{centroid_eqn}
\end{tiny}
\end{equation}

\begin{align}
\overline {x^2}=\frac{\displaystyle \sum_{i\in S} K_{i}x_{i}^2}{\displaystyle \sum_{i\in S} K_{i}} - \overline{x}^2, \nonumber  \\
\hspace{0.5cm}
\overline {y^2}=\frac{\displaystyle \sum_{i\in S} K_{i}y_{i}^2}{\displaystyle \sum_{i\in S} K_{i}} - \overline{y}^2, \nonumber  \\
\hspace{0.5cm} 
\overline{xy}=\frac{\displaystyle \sum_{i\in S} K_{i}x_{i}y_{i}}{\displaystyle \sum_{i\in S} K_{i}} - \overline{x} \hspace{0.1cm} \overline{y}.
\end{align}

Here ($x_{i}$, $y_{i}$) is the ($x$, $y$) coordinate of the image of a pixel $i$ of value $K_{i}$ inside an area $S$.
{
\subsection{Uncertainty estimation}
\label{para_uncer_sec}
\par There are three sources of uncertainty in the calculation of the morphology parameters. These are: (1) shot noise in the image pixel values, (2) uncertainties in the centre of the cluster, and (3) variation in the area over which morphology parameters are calculated. The first two uncertainties can be approximated using a number of iterations of the Monte Carlo method. For estimating the third uncertainty, we used a jackknifing technique. 

\par The shot noise effect can be approximated by replacing each pixel value with a Poisson random variable of the mean value of each pixel for the given image, and recalculating the parameters a number of times. After a set number of iterations (in our case 10), the rms of the spread in parameter values is an approximation of uncertainty in the parameters.

\par Resultant uncertainty from variation of the central position of a cluster is computed by deviating the input ($x_{c}$, $y_{c}$) coordinate within a fixed Gaussian width ($\sim$ \begin{math}30''\end{math} in our case). We then recalculate the parameters for several ($x_{c}$, $y_{c}$) value. After a number of iterations (10) \citep{2011MNRAS.416.2401H}, we compute the rms of the spread in parameter values as an approximate value of the uncertainty in the parameters.

\par The Gini coefficient is less sensitive to the shot noise than to changes in area, as the pixels are ordered first and do not depend on the adopted central position of the cluster in any way. We estimated its uncertainty from a shot noise, and the rms in Gini values from a series of subsets of the pixels in the image (by allowing the pixels to be varied using Poisson statistics for a given area). This is known as jackknife error estimation. \cite{1991JBES..9...235} suggested the use of the jackknife approach to estimate an uncertainty in the Gini coefficient. The jackknife and shot noise uncertainty estimates in the Gini coefficient are of similar order.  

\par We used an error propagation formula to combined the shot noise and central position uncertainties for all parameters except the Gini coefficient. For the Gini coefficient, we combined the uncertainties from the shot noise and the jackknife estimation. Hence, in this work, all reported uncertainties are a combination of those mentioned above. 

\section{Sample selection and data reduction}
\subsection{{\it Chandra} sample}
\label{chandra_sam_sec}
\par For our analysis, we required high quality data, particularly for high-$z$ clusters. \cite {2007ApJS..172..561B} compiled a 400 deg$^{2}$ galaxy cluster catalogue based on the ROSAT PSPC survey. They detected a large number of extended X-ray sources with $f>$ 1.4$\times$10$^{-13}$ erg s$^{-1}$ cm$^{-2}$ in the soft (0.5--2 keV) band. They compared all their detections with their optical counterparts, and confirmed 266 out of 283 as galaxy clusters. From this catalogue, a subsample of the low (0.02--0.3)-$z$ and high (0.3--0.9)-$z$ galaxy clusters has been observed with the $Chandra$ X-ray telescope (V09). Fig.~\ref{Fig.1}a shows the distribution of redshift, while Fig.~\ref{Fig.1}b shows the luminosity distribution. In this plot, median $z$ = 0.0853 and median luminosity = 1.8$\times$10$^{44}$ (erg/s) (0.5--2 keV). Tables \ref{tab:lowz} and \ref{tab:highz} list the low- and high-$z$ samples of galaxy clusters, respectively. {We normalised $H$ dependent quantities (e.g. $L_{X}$ and mass) with $H_{0}$=73 km s$^{-1}$ Mpc$^{-1}$.}
    
\begin{figure*}
\begin{center}
\includegraphics[scale=0.4]{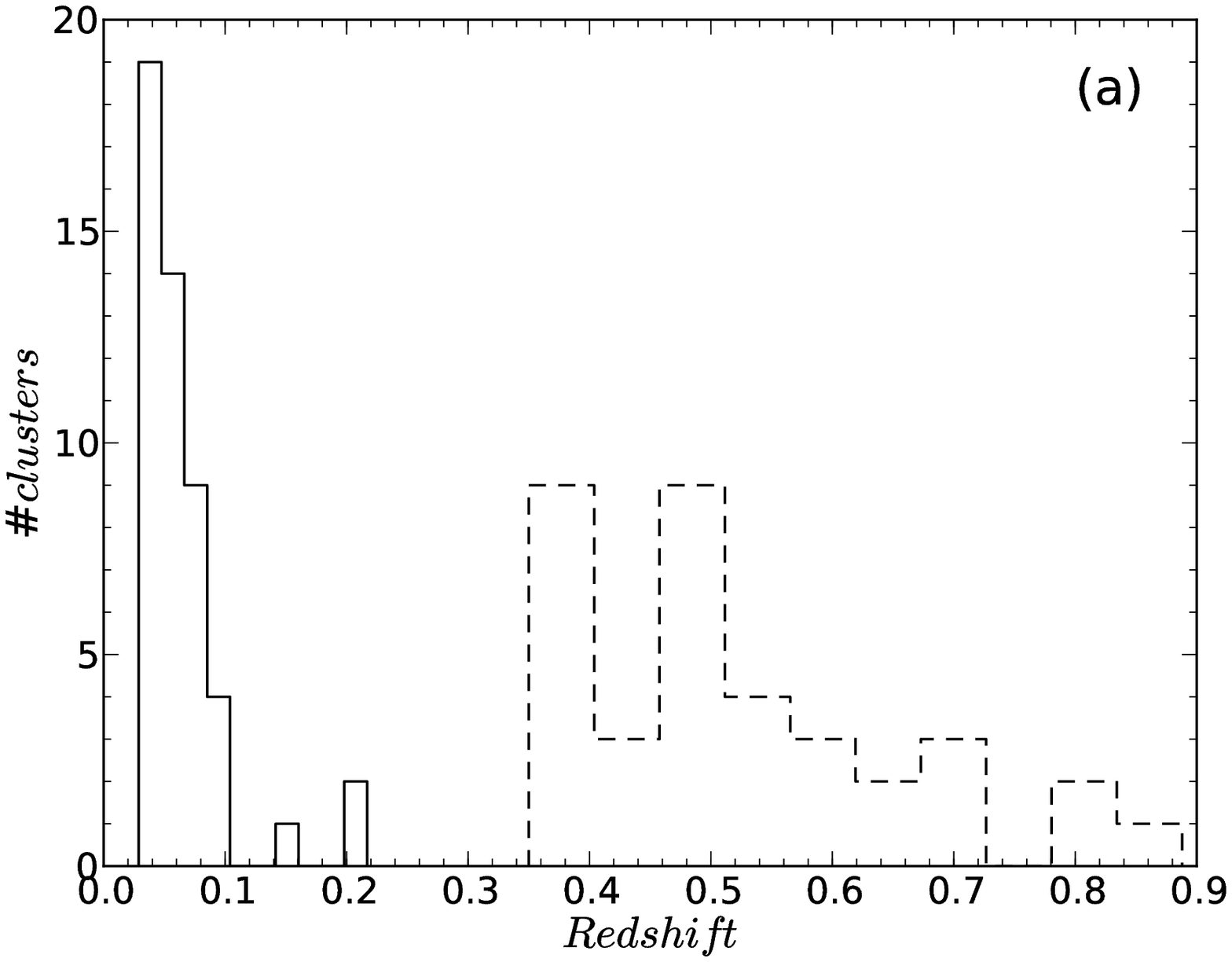}\includegraphics[scale=0.4]{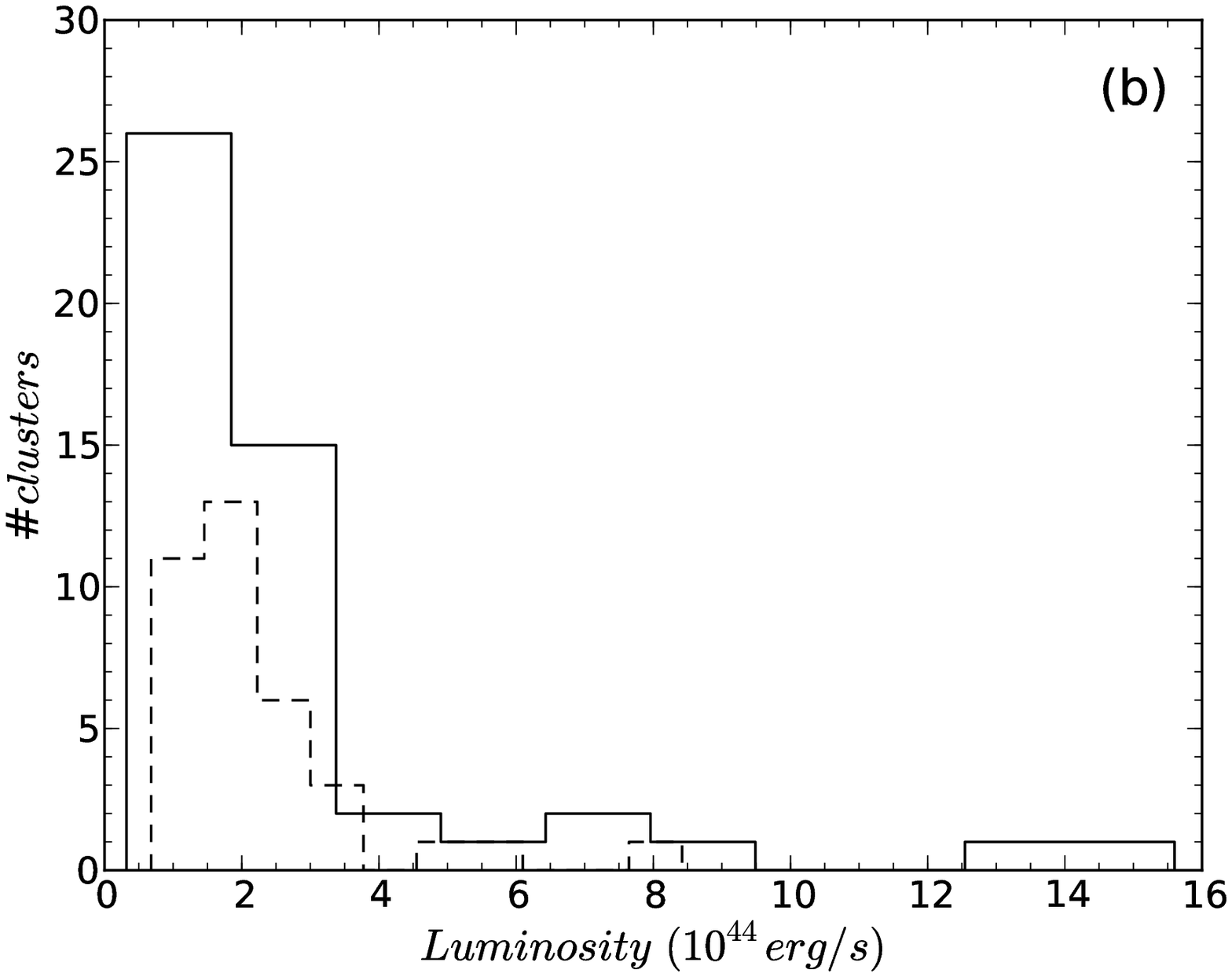}
\caption{\small (a) Redshift and (b) luminosity distribution for low- and high-$z$ clusters (from the V09). Solid line = low-$z$; and dashed line = high-$z$ clusters.}
\label{Fig.1}
\end{center}
\end{figure*}

\begin{table*}
\centering
\caption{Low-$z$ (0.02--0.3) cluster sample from the V09. (1) cluster name; (2) redshift; (3) total X-ray flux (0.5--2 keV); (4) total X-ray luminosity (0.5--2 keV); (5) average temperature in [0.15-1]$R_{500}$ annulus; (6) dynamical state (a) (according to the V09); (7) {dynamical state (b) (combination of morphology parameters (see \S\ \ref{combpara})) (SR = strong relaxed, R = relaxed, NR = non-relaxed, SNR = strong non-relaxed);} (8) exposure time.}
\begin{tiny}
%\hspace*{-1.5cm}   
\begin{center}
\begin{tabular}{ccccccccc}
\hline 
\hline
 Name & $z$ & flux & Luminosity & Temperature & Dynamical state&Dynamical state&Exposure time\\
       &    &(10 $^{-11}$cgs)&(erg s$^{-1}$)& (keV) & (a) & (b) &ks\\    
\hline  	
A3571 & 0.0386 &  7.42 & 2.30$\times$10$^{44}$ & 6.81 $\pm$ 0.10 & R & NR & 31.224 \\  
A2199 & 0.0304 &  6.43 & 1.23$\times$10$^{44}$ & 3.99 $\pm$ 0.10 & R & NR & 17.787 \\
2A 0335 & 0.0346 &  6.24 & 1.55$\times$10$^{44}$ & 3.43 $\pm$ 0.10& R& SR & 19.716 \\
A496 &  0.0328 &  5.33 & 1.19$\times$10$^{44}$ & 4.12 $\pm$ 0.07 &  R& NR & 88.878 \\
A3667 & 0.0557 &  4.64 & 3.05$\times$10$^{44}$ & 6.33 $\pm$ 0.06 &  M & NR &  60.403 \\  
A754 &  0.0542 &  4.35 & 2.70$\times$10$^{44}$ & 8.73 $\pm$ 0.00 &  M & SNR & 43.972 \\ 
A85 &   0.0557 &  4.30 & 2.83$\times$10$^{44}$ & 6.45 $\pm$ 0.10 &  R& NR & 38.201 \\ 
A2029 & 0.0779 &  4.23 & 5.56$\times$10$^{44}$ & 8.22 $\pm$ 0.16 &  R& SR & 10.739 \\ 
A478 &  0.0881 &  4.16 & 7.04$\times$10$^{44}$ & 7.96 $\pm$ 0.27 &  R& SR & 42.393 \\ 
A1795 & 0.0622 &  4.14 & 3.42$\times$10$^{44}$ & 6.14 $\pm$ 0.10 &  R & SR & 91.301 \\ 
A3558 & 0.0469 &  4.11 & 1.90$\times$10$^{44}$ & 4.88 $\pm$ 0.10 &  M& NR& 14.024  \\ 
A2142 & 0.0904 &  3.94 & 7.00$\times$10$^{44}$ & 10.04 $\pm$ 0.26 & R& NR &  44.367  \\ 
A2256 & 0.0581 &  3.61 & 2.58$\times$10$^{44}$ & 8.37 $\pm$ 0.24 &  M& SNR & 11.647  \\ 
A4038 & 0.0288 &  3.48 & 6.01$\times$10$^{43}$ & 2.61 $\pm$ 0.05 &  R& NR & 33.523  \\ 
A2147 & 0.0355 &  3.47 & 9.14$\times$10$^{43}$ & 3.83 $\pm$ 0.12 &  M& SNR& 17.879  \\ 
A3266 & 0.0602 &  3.39 & 2.61$\times$10$^{44}$ & 8.63 $\pm$ 0.18 &  M& SNR & 29.289  \\ 
A401 &  0.0743 &  3.19 & 3.79$\times$10$^{44}$ & 7.72 $\pm$ 0.30 &  R& NR & 18.005  \\ 
A2052 & 0.0345 &  2.93 & 7.26$\times$10$^{43}$ & 3.03 $\pm$ 0.07 &  R& NR & 36.754  \\ 
Hydra-A & 0.0549 &  2.91 & 1.87$\times$10$^{44}$ & 3.64 $\pm$ 0.06 & R& R & 89.809   \\ 
A119 &  0.0445 &  2.47 & 1.03$\times$10$^{44}$ & 5.72 $\pm$ 0.00 &  M& NR & 11.537   \\ 
A2063 & 0.0342 &  2.39 & 5.81$\times$10$^{43}$ & 3.57 $\pm$ 0.19 &  R& NR & 8.777 \\ 
A1644 & 0.0475 &  2.33 & 1.10$\times$10$^{44}$ & 4.61 $\pm$ 0.14 &  M& NR & 18.712   \\
A3158 & 0.0583 &  2.30 & 1.67$\times$10$^{44}$ & 4.67 $\pm$ 0.07 &  R& NR & 30.921  \\ 
MKW3s & 0.0453 &  2.08 & 9.02$\times$10$^{43}$ & 3.03 $\pm$ 0.05 &  R& NR & 57.123  \\ 
A1736 & 0.0449 &  2.04 & 8.69$\times$10$^{43}$ & 2.95 $\pm$ 0.09 &  M& SNR & 14.918  \\ 
EXO0422 & 0.0382 &  2.01 & 6.17$\times$10$^{43}$ & 2.84 $\pm$ 0.09 & R& R &10.001 \\
A4059 & 0.0491 &  2.00 & 1.02$\times$10$^{44}$ & 4.25 $\pm$ 0.08 &  R& NR & 13.236 \\ 
A3395 & 0.0506 &  1.95 & 1.06$\times$10$^{44}$ & 5.10 $\pm$ 0.17 &  M& SNR & 21.094  \\ 
A2589 & 0.0411 &  1.94 & 6.90$\times$10$^{43}$ & 3.17 $\pm$ 0.27 &  R& NR & 13.478  \\ 
A3112 & 0.0759 &  1.89 & 2.36$\times$10$^{44}$ & 5.19 $\pm$ 0.21 &  R& R & 15.466  \\ 
A3562 & 0.0489 &  1.84 & 9.32$\times$10$^{43}$ & 4.31 $\pm$ 0.12 &  R& NR & 19.283  \\ 
A1651 & 0.0853 &  1.80 & 2.85$\times$10$^{44}$ & 6.41 $\pm$ 0.25 &   R& NR & 9.643  \\ 
A399 &  0.0713 &  1.78 & 1.95$\times$10$^{44}$ & 6.49 $\pm$ 0.17 &  M& NR & 48.575  \\ 
A2204 & 0.1511 &  1.74 & 9.10$\times$10$^{44}$ & 8.55 $\pm$ 0.58 &   R& SR &9.609  \\ 
A576 &  0.0401 &  1.72 & 5.82$\times$10$^{43}$ & 3.68 $\pm$ 0.11 &   R& NR &29.078  \\ 
A2657 & 0.0402 &  1.62 & 5.50$\times$10$^{43}$ & 3.62 $\pm$ 0.15 &   R& NR & 16.148   \\ 
A2634 & 0.0305 &  1.61 & 3.11$\times$10$^{43}$ & 2.96 $\pm$ 0.09 &   R&- &49.528  \\ 
A3391 & 0.0551 &  1.58 & 1.02$\times$10$^{44}$ & 5.39 $\pm$ 0.19 &   R& NR &17.461  \\ 
A2065 & 0.0723 &  1.56 & 1.77$\times$10$^{44}$ & 5.44 $\pm$ 0.09 &   M& NR &49.416  \\ 
A1650 & 0.0823 &  1.53 & 2.26$\times$10$^{44}$ & 5.29 $\pm$ 0.17 &   R& R &27.258 \\ 
A3822 & 0.0760 &  1.48 & 1.85$\times$10$^{44}$ & 5.23 $\pm$ 0.30 &   M& NR & 8.067  \\ 
S 1101 &  0.0564 &  1.46 & 1.00$\times$10$^{44}$ & 2.44 $\pm$ 0.08&  R& R & 9.946\\ 
A2163 & 0.2030 &  1.38 & 1.33$\times$10$^{45}$ & 14.72 $\pm$ 0.31 &  M& NR &71.039   \\ 
ZwCl1215 & 0.0767 &  1.38 & 1.75$\times$10$^{44}$ & 6.54 $\pm$ 0.21& R& NR &11.999 \\ 
RXJ1504 & 0.2169 &  1.35 & 1.51 $\times$10$^{45}$ & 9.89 $\pm$ 0.53& R& SR &13.290 \\ 
A2597 & 0.0830 &  1.35 & 2.03$\times$10$^{44}$ & 3.87 $\pm$ 0.11 &   R& SR &26.414   \\
A133 &  0.0569 &  1.35 & 9.33$\times$10$^{43}$ & 4.01 $\pm$ 0.11 &   R& R & 34.471   \\
A2244 & 0.0989 &  1.34 & 2.89$\times$10$^{44}$ & 5.37 $\pm$ 0.12 &   R& NR &56.964   \\
A3376 & 0.0455 &  1.31 & 5.72$\times$10$^{43}$ & 4.37 $\pm$ 0.13 &   M & NR &44.267  \\
\hline
\end{tabular}
\end{center}
\label{tab:lowz}
\end{tiny}  
\end{table*}

\begin{table*}
\centering
\caption{As in Table \ref{tab:lowz}, but for high-$z$ (0.3--0.9) clusters.}
\begin{tiny}
%\hspace*{-1.5cm}  
\begin{center}
\begin{tabular}{cccccccccccccc}
        
\hline 
\hline
    Name & $z$ & flux & Luminosity & Temperature &Dynamical state &Dynamical state&Exposure time\\
         &    &(10 $^{-13}$cgs)&(erg s$^{-1}$)& (keV)& (a) & (b)  &ks   \\
\hline  	
0302-0423&0.3501& 15.34& 5.09$\times$10$^{44}$&4.78 $\pm$ 0.75 & R &SR &100.41\\
1212+2733&0.3533& 10.53& 3.51$\times$10$^{44}$&6.62 $\pm$ 0.89 &M& NR&14.581\\
0350-3801&0.3631&  1.68& 6.61$\times$10$^{43}$&2.45 $\pm$ 0.50 &M&NR&23.733\\
0318-0302&0.3700&  4.63& 1.77$\times$10$^{44}$&4.04 $\pm$ 0.63 &M&NR&14.578\\
0159+0030&0.3860&  3.30& 1.38$\times$10$^{44}$&4.25 $\pm$ 0.96 &R&R&19.880\\
0958+4702&0.3900&  2.22& 1.01$\times$10$^{44}$&3.57 $\pm$ 0.73 &R&R&25.231\\
0809+2811&0.3990&  5.40& 2.43$\times$10$^{44}$&4.17 $\pm$ 0.73 &M&SNR&19.338\\
1416+4446&0.4000&  4.01& 1.88$\times$10$^{44}$&3.26 $\pm$ 0.46 &R&NR&29.187\\
1312+3900&0.4037&  2.71& 1.33$\times$10$^{44}$&3.72 $\pm$ 1.06 &M&SNR&26.421\\
1003+3253&0.4161&  3.04& 1.48$\times$10$^{44}$&5.44 $\pm$ 1.40 &R&R&19.859\\
0141-3034&0.4423&  2.06& 1.28$\times$10$^{44}$&2.13 $\pm$ 0.38 &M&SNR&28.273\\
1701+6414&0.4530&  3.91& 2.32$\times$10$^{44}$&4.36 $\pm$ 0.46 &R&R&49.294\\
1641+4001&0.4640&  1.43& 9.20$\times$10$^{43}$&3.31 $\pm$ 0.62 &R&NR&45.345\\
0522-3624&0.4720&  1.47& 1.01$\times$10$^{44}$&3.46 $\pm$ 0.48 &M&NR&45.999\\
1222+2709&0.4720&  1.39& 9.61$\times$10$^{43}$&3.74 $\pm$ 0.61 &R&NR&49.117\\
0355-3741&0.4730&  2.48& 1.71$\times$10$^{44}$&4.61 $\pm$ 0.82 &R&R&27.190\\
0853+5759&0.4750&  1.22& 8.20$\times$10$^{43}$&3.42 $\pm$ 0.67 &M&NR&42.179\\
0333-2456&0.4751&  1.33& 9.52$\times$10$^{43}$&3.16 $\pm$ 0.58 &M&NR&34.160\\
0926+1242&0.4890&  2.04& 1.45$\times$10$^{44}$&4.74 $\pm$ 0.71 &M&NR&18.620\\
0030+2618&0.5000&  2.09& 1.52$\times$10$^{44}$&5.63 $\pm$ 1.13 &M&NR&57.362\\
1002+6858&0.5000&  2.19& 1.66$\times$10$^{44}$&4.04 $\pm$ 0.83 &M&NR&19.786\\ 
1524+0957&0.5160&  2.45& 2.01$\times$10$^{44}$&4.23 $\pm$ 0.51 &M&SNR&49.886\\
1357+6232&0.5250&  1.90& 1.58$\times$10$^{44}$&4.60 $\pm$ 0.69 &R&NR&43.862\\
1354-0221&0.5460&  1.45& 1.36$\times$10$^{44}$&3.77 $\pm$ 0.53 &M&NR&55.039\\
1120+2326&0.5620&  1.68& 1.74$\times$10$^{44}$&3.58 $\pm$ 0.44 &M&SNR&70.262\\
0956+4107&0.5870&  1.64& 1.79$\times$10$^{44}$&4.40 $\pm$ 0.50 &M&NR&40.165\\
0328-2140&0.5901&  2.09& 2.23$\times$10$^{44}$&5.14 $\pm$ 1.47 &R&NR&56.192\\
1120+4318&0.6000&  3.24& 3.64$\times$10$^{44}$&4.99 $\pm$ 0.30 &R&NR&19.837\\
1334+5031&0.6200&  1.76& 2.16$\times$10$^{44}$&4.31 $\pm$ 0.28 &M&NR&19.492\\
0542-4100&0.6420&  2.21& 2.83$\times$10$^{44}$&5.45 $\pm$ 0.77 &M&NR&50.008\\
1202+5751&0.6775&  1.34& 2.15$\times$10$^{44}$&4.08 $\pm$ 0.72 &M&SNR&57.210\\
0405-4100&0.6861&  1.33& 2.16$\times$10$^{44}$&3.98 $\pm$ 0.48 &M&NR&77.163\\
1221+4918&0.7000&  2.06& 3.25$\times$10$^{44}$&6.63 $\pm$ 0.75 &M&NR&78.887\\
0230+1836&0.7990&  1.09& 2.48$\times$10$^{44}$&5.50 $\pm$ 1.02 &M&SNR&67.178\\
0152-1358&0.8325&  2.24& 5.31$\times$10$^{44}$&5.40 $\pm$ 0.97 &M&NR&36.285\\
1226+3332&0.8880& 3.27& 8.19$\times$10$^{44}$& 11.08 $\pm$ 1.39 &M&NR&64.21\\
 
\hline
\end{tabular}
\end{center}
\label{tab:highz}
\end{tiny}
\end{table*}

\par {Properties of each cluster provided by the V09} are useful in comparison with their X-ray morphology. The total X-ray luminosity is calculated over the 0.5--2 keV band from accurate $Chandra$ flux. The average temperature is measured from the spectrum integrated over a [0.15-1]$R_{500}$\footnote{$R_{500}$ being the radius defined as that enclosing a region with an over-density $\bigtriangleup$ = 500 times the critical density at the cluster redshift.} annulus. 
\par {In agreement with V09, we used the `cuspiness' parameter defined by \cite{2007hvcg.conf...48V} to discriminate between cooling flow (relaxed) and non-cooling flow (unrelaxed) clusters}. Based on this classification, we used 49 (34 relaxed + 15 non-relaxed) low-$z$ clusters and 36 (12 relaxed + 24 non-relaxed) high-$z$ clusters.  
   
\par There are three high-$z$ and two low-$z$ clusters that have multiple components in their images:
\begin{itemize} 
\item There are two subclumps visible in the high redshift 1701+6414 cluster. We used the North clump in our calculation. The other clump (in the S-W) is a foreground low-$z$ X-ray cluster, confirmed by the NED database.
\item In the case of the high redshift system 1641+4001, there is a small clump (foreground galaxy) in the S-W, which we excluded from our analysis.
\item Double components are visible in the 0328-2140 system. One is in the East and the other in the West, the latter being a low redshift interloper. We used the high redshift East cluster. 
\item In A85, we focused on only the main North relaxed cluster, excluding the small southern clump \citep {2002ApJ...579..236K}.
\item In A1644 we excluded the small North component, and focused on the main larger  southern cluster only \citep {2010ApJ...710.1776J, 2004ApJ...608..179R}.
\end{itemize}      

\subsection{ {\it Chandra} data reduction and image preparation}   
\label{chan_data_red}
\par In our sample, each cluster had at least 1500--2000 photons (V09), which ensured good S/N. {We visually checked all these clusters, and verified that there was no cluster peak emission or centroid too close to the CCD edges. There are still a number of cases where cluster emission falls within CCD gaps and this can potential influence our results. We divided the counts image by the corresponding exposure map and applied light smoothing (as mentioned below) so as to minimise the effects of CCD gaps}. This inspection revealed that the relaxed cluster A2634 (Obs-Id 4816, ACIS-S) alone had low total counts. It was therefore excluded from our sample and now we have total of 84 clusters in sample.

\par We processed all {\it Chandra} archival data with CIAO version 4.3 and CALDB version 4.4.6. We first used the \verb"chandra_repro" task to reprocess all ACIS imaging data. This script creates a new second level event file as well as a bad pixel file by reading the data from the standard {\it Chandra} data distribution. After reprocessing, we removed any high background flares (3$\sigma$ clipping) with the task \verb"lc_sigma_clip" routine and then attached the good time interval (GTI) file to the events. All of our event files included the 0.3--7 keV broad energy band. We binned each event file with $\sim$ 2" pixels. {We detected point sources in observation using the \verb"wavdetect" task\footnote{http://cxc.harvard.edu/ciao4.3/threads/wavdetect/} with scale = 1, 2, 4, 8 and 16, which would be a reasonable default for {\it Chandra} data. This scale parameter refers the wavelet radii in image pixels. Then a wavelet transform is performed for each scale in the given radii list to estimate source properties. We also supplied information on the size of the PSF at each pixel in the image using energy value of 1.49.} We then selected elliptical background annuli (twice the point source detected region) around each detected point source. Then we removed the point sources using the \verb"dmfilth" task\footnote{http://cxc.harvard.edu/ciao4.3/threads/diffuse emission/}, which replaces pixel values in selected source regions of an image with values interpolated from the surrounding background region. We used the POISSON interpolation method, which randomly samples values in the selected source regions from the Poisson distribution, whose mean is that of the pixel values in the background region. We took care not to remove point source which was detected close to the cluster centre, {because in most cases, this point source was simply the peak of the surface brightness distribution rather than an actual X-ray point source}. We also aimed to study the effect of {cluster surrounded point sources} on parameters, hence we calculated morphology parameters for both images, i.e. the one with all point sources, the other without point sources (except central point source). For creating the exposure map we used the \verb"fluximage" task\footnote{http://cxc.harvard.edu/ciao/ahelp/fluximage.html} with broad band energy (0.3--7 keV), then divided the counts image by the exposure map to remove the CCD gaps,  vignetting, and telescope effects. This task generates the aspect histogram, instrument and exposure maps, automatically. We also used the \verb"dmimgthresh" task to perform a 5$\%$ total counts cut to enable uniform exposure everywhere in the image. Two or more of the available multiple observations were combined to make a single image. Finally we smoothed the cluster images with a $\sigma$ = 10" Gaussian width. Smoothing is very important in calculating the Gini coefficient, because there will be zero count pixels in the unsmoothed count image, which will not contribute to the flux distribution calculation and will change the Gini value greatly. For calculating the Smoothness parameter, we initially used unsmoothed cluster images. 
\par Because of limited {\it FOV} of {\it Chandra}, in the study of nearby clusters, we decided to use a fixed circular aperture size over which morphology parameters are calculated for individual clusters (200 kpc for $z$ $<$ 0.05 and 500 kpc for $z$ $>$ 0.05) with sufficient area, and to retain consistency with our morphology parameter calculations on the same relative scale. The possible effects of aperture size are discussed in \S\ \ref{aper_effe_sec}.

\section{Morphology parameter results}

\subsection{Distribution of morphology parameters}
\label{para_dis_sec}
\par Fig.~\ref{exp_cor_hist} shows the distribution of parameters for our sample of 84 clusters. Except for three parameters, Gini, $M_{20}$ and Concentration, most of the parameters show a similar distribution for relaxed and non-relaxed clusters based on the classification described in \S\ \ref{chandra_sam_sec}. The Smoothness parameter separates the two peaks of relaxed and non-relaxed clusters towards low Smoothness and high Smoothness, respectively, but there are large numbers of clusters in the overlap region. {To further investigate the Smoothness parameter, we (1) varied the $\sigma$ value between \begin{math}0.5'\end{math} to \begin{math}1'\end{math}, and (2) varied the smoothing size of input cluster images (initial smoothing) on the Smoothness parameter (see \S\ \ref{smth_asy}).} For the Gini and Concentration parameters, non-relaxed clusters are distributed towards low values of Gini and Concentration and, oppositely, relaxed clusters are characterised by high Gini and Concentration values. A similar trend is visible for the $M_{20}$, but the extreme of the relaxed clusters is on the left hand side (low $M_{20}$ values) and the extreme of the non-relaxed clusters is on the right hand side (high $M_{20}$ values). {Tables \ref{low_par_val_vo9} and \ref{high_par_val_vo9} list all parameter values together with 1$\sigma$ uncertainty (\S\ \ref{para_uncer_sec}) for the V09 low- and high-$z$ sample clusters.}
\begin{figure*}
\centering
%\hspace*{-1.0in}
\includegraphics[scale=0.60]{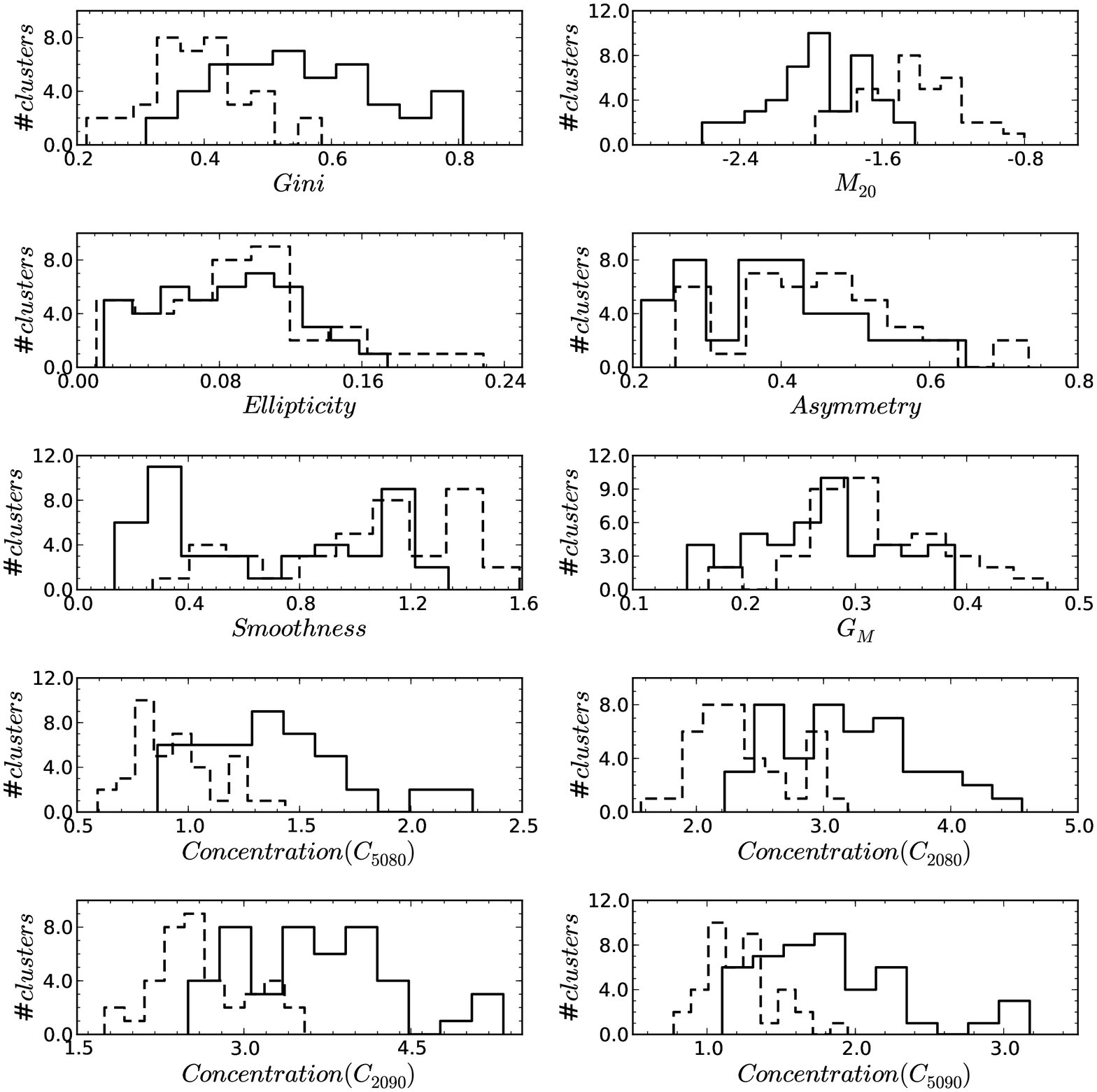} 
\caption{\small Seven morphology parameter distributions: solid line = relaxed cluster; dashed line = non-relaxed cluster. All four definitions of the Concentration parameter are displayed. Galaxy cluster separation (i.e. relaxed vs non-relaxed systems) is based on the V09.}
\label{exp_cor_hist}
\end{figure*}
%%%%%%%%

%%%%%%%%
\begin{table*}
\centering
\caption{Statistics for the subsample of relaxed (R) and non-relaxed (N-R) clusters, and the combined (C) cluster sample from the V09.}
\begin{small}
\begin{tabular}{cccccc}
\hline
\hline 
  & mean & median & K-S probability & R-S probability \\ 
        &    {\small R N-R C}          & {\small R N-R C} &  \\ 
\hline     
                                   
Gini & 0.55 0.39 0.47    &   0.54  0.39   0.44  & 1.42$\times$10$^{-07}$ & 8.73$\times$10$^{-09}$ \\ 
$M_{20}$ & -1.96 -1.44 -1.72  &  -1.96 -1.44  -1.73  & 8.65$\times$10$^{-09}$ & 8.63$\times$10$^{-11}$\\ 
Concentration & 1.38 0.95 1.18    &   1.33   0.92  1.11  & 8.87$\times$10$^{-08}$ & 2.26$\times$10$^{-9}$ \\ 
Asymmetry &           0.39 0.44 0.42    &   0.38   0.44   0.40 & 0.12 &  0.027  \\ 
Smoothness &          0.68 1.05 0.85    &   0.61   1.09   0.94 & 0.0020 & 6.7$\times$10$^{-5}$\\ 
$G_{M}$ &             0.27 0.31 0.29    &   0.28   0.30    0.28 & 0.012 & 0.003\\ 
Ellipticity &         0.08  0.09 0.09   &   0.08   0.09    0.08  & 0.95 & 0.66\\ 
\hline 
\end{tabular}
\end{small}
\label{kstest_dis_par}
\end{table*} 

\par We performed the Kolmogorov-Smirnov (K-S) test to investigate whether or not the relaxed and non relaxed cluster distributions are drawn from the same parent distribution. The results are given for each parameter in Table \ref{kstest_dis_par}, where a threshold of 1\% means that the probability is $<$ 0.01, implying a possible rejection of the null hypothesis. In the table we include the Wilcoxon rank sum (R-S) test which establishes the probability of whether the two samples with the null hypothesis have the same mean. A threshold of 0.1\% implies that we reject the null hypothesis (that they have the same mean) if the probability is $<$ 0.001. In addition, the mean and median values for all morphology parameters are supplied. From the statistical tests, we observed a significant separation between the two distributions (relaxed and non-relaxed clusters) for the Gini, $M_{20}$ and Concentration, which rejects the null hypotheses that the two samples (relaxed and non-relaxed clusters) are the same and they have the same mean values. {The K-S and R-S probabilities are $<$ 0.01 and 0.001, respectively, for the Smoothness parameter; but we observed a high incidence of overlapping between relaxed and non-relaxed clusters for the Smoothness distribution. We found that the right side peak of non-relaxed clusters (dotted line) in the Smoothness distribution corresponds mainly to the high-$z$ clusters which have low S/N compared with nearby clusters (which fall mainly on the left side).} Further details concerning the Smoothness parameter is given in \S\ \ref{smth_asy}. For the remaining parameters, we cannot reject the above null hypotheses, which means that Asymmetry, Gini of the second order moment and Ellipticity are not useful in separating relaxed and non-relaxed clusters. 
%This could suggests that the Smoothness parameter highly depends on S/N.

\subsection{Parameter vs Parameter planes}
\label{paravspara}
\par {Using combinations of morphology parameters we investigated relaxed vs non-relaxed clusters in the parameter-parameter plane to study the dynamical states of galaxy clusters and the correlation between each morphology parameter}. In Fig.~\ref{exp_cor} we plotted our results in parameter-parameter planes. Three parameters, Gini, Concentration and $M_{20}$, look particularly powerful after combining with other parameters to differentiate between non-relaxed and relaxed clusters. Clusters with different dynamical states (as classified by the V09) occupy distinct regions in the parameter-parameter planes; for e.g., in the Concentration vs Asymmetry plot, all relaxed clusters occupy the upper region, while the lower region is occupied by non-relaxed clusters. In our analysis we did not observe any correlation between cluster dynamical states and Ellipticity or Asymmetry. As seen in Fig.~\ref{exp_cor}, galaxy clusters within our sample show a range of different morphologies and are not concentrated in particular positions of the parameter-parameter space. This is probably because of the hierarchical cluster formation process, indicating that clusters pass through multiple (merger) phases in their evolution. Each phase is dynamically important, and traces the cluster properties. This could help in the understanding of large and complex structure formations in the standard cosmological model. 

\par We used the Spearman rank-order correlation coefficient, $\rho$, to quantify any correlation between parameter pairs. This resulted in a correlation coefficient between ranks of a group of individuals for a given pair of attributes. In order to calculate  $\rho$, it is necessary to assign ranks (low to high or high to low) to a given set of variables, individually. The next step is to measure the deviation between the ranks of variable pairs, square this rank difference, and sum up. The value of the rank order correlation coefficient varies between -1 and 1. If the variables are anti-correlated then $\rho$ falls between -1 and 0, while a value for $\rho$ between 0 and 1 implies a positive correlation between given variables. $\rho$ = 0 implies no correlation between variables.

\par Fig.~\ref{exp_cor} indicates that the Concentration is tightly correlated with the Gini, while anti-correlated with the $M_{20}$ (see also Table \ref{para-spearman}). The Concentration \citep {2008A&A...483...35S, 2010A&A...513A..37H, 2010ApJ...721L..82C} is a useful parameter in separating non-relaxed from relaxed clusters with almost all morphology parameters. Fig.~\ref{exp_cor} illustrates our plot of the $C_{5080}$ Concentration parameter. As seen in Fig.~\ref{exp_cor}, relaxed clusters occupy the upper left corner, while non-relaxed clusters occupy the bottom-right corner of the $M_{20}$ vs Gini and Concentration planes. In the Gini vs Concentration plane, relaxed clusters fall in the upper-right corner, while the bottom-left corner is occupied by non-relaxed clusters. In this study we found that the Gini coefficient was also useful in separating non-relaxed clusters from relaxed clusters when plotted against most other parameters. The Gini, Concentration and $M_{20}$ parameters are all inter-related as well (Table \ref{para-spearman}); and the Gini coefficient could be useful to use as a proxy of the Concentration for detecting substructure in high-$z$ clusters. The advantage of using the Gini coefficient is that it is independent of the precise location of a galaxy cluster's centre. The Asymmetry parameter is correlated with Gini of the second order moment and Smoothness parameters \citep {2010ApJ...721..875O, 2010ApJ...711.1033Z, 2012arXiv1211.7040R}. We did not observe any correlation between the other six parameters and Ellipticity. In general, three parameters - Gini, $M_{20}$ and Concentration - are very promising tools in the morphological classification of clusters.
\par Despite this reasonable division between relaxed and non-relaxed clusters, we observed an overlap between some clusters, such as A401, A3571, A1651, A3158, A3562, A576, A2063, ZwCl1215, A2657, A2589, A3391, 0355-3741, 1641+4001, 1120+4318, 1222+2709, 0328-2140 and 1357+6232. These are identified as relaxed clusters in the V09, but in our parameter space they fall into the non-relaxed region. A401 \citep {2010A&A...509A..86M} and A3562 \citep {2003A&A...402..913V} host {\it radio halos} (indicating possible signature of merger, see \S\ \ref{radio_halo_sec}) which are clearly identified as non-relaxed clusters by our parameters (although they are classified as relaxed in the V09). \cite {2007ApJ...668..781D} predicted a line of sight bullet-like cluster in A576, using a combination of $Chandra$ and $XMM$ observations. This gave us confidence that our parameter set provided a good indication of cluster dynamical states. The remaining clusters may be weak mergers (pre- or post-merger) or in intermediate state. \par In cluster A85 \citep {2002ApJ...579..236K}, in addition to the main cool core cluster, two subclusters are visible, in the far South and NW, respectively. We calculated parameters for the main relaxed cluster, and found that it falls into the relaxed category, although very close to the boundary of the non-relaxed side, which could indicate a weak interaction between the subcluster and the main cool core cluster. Based on our measurements, we categorised most relaxed clusters by Concentration $>$ 1.55 and $M_{20}$ $<$ -2.0, while intermediate clusters were categorised by 1.0 $<$ Concentration $<$ 1.55 and -2.0 $<$ $M_{20}$ $<$ -1.4, and distorted and non-relaxed clusters were categorised by Concentration $<$ 1.0 and $M_{20}$ $>$ -1.4. According to the Gini coefficient, most relaxed clusters have a Gini value $>$ 0.65 and non-relaxed have Gini value $<$ 0.40. We chose these boundaries based on visual identification of each cluster morphology (Fig.~\ref{exp_cor}). Mean and median values (Table \ref{kstest_dis_par}) produce a significant overlap due to the presence of a large number of intermediate clusters in our sample.

\begin{figure*}
\centering
\hspace*{-0.6in}
\includegraphics[scale=0.70]{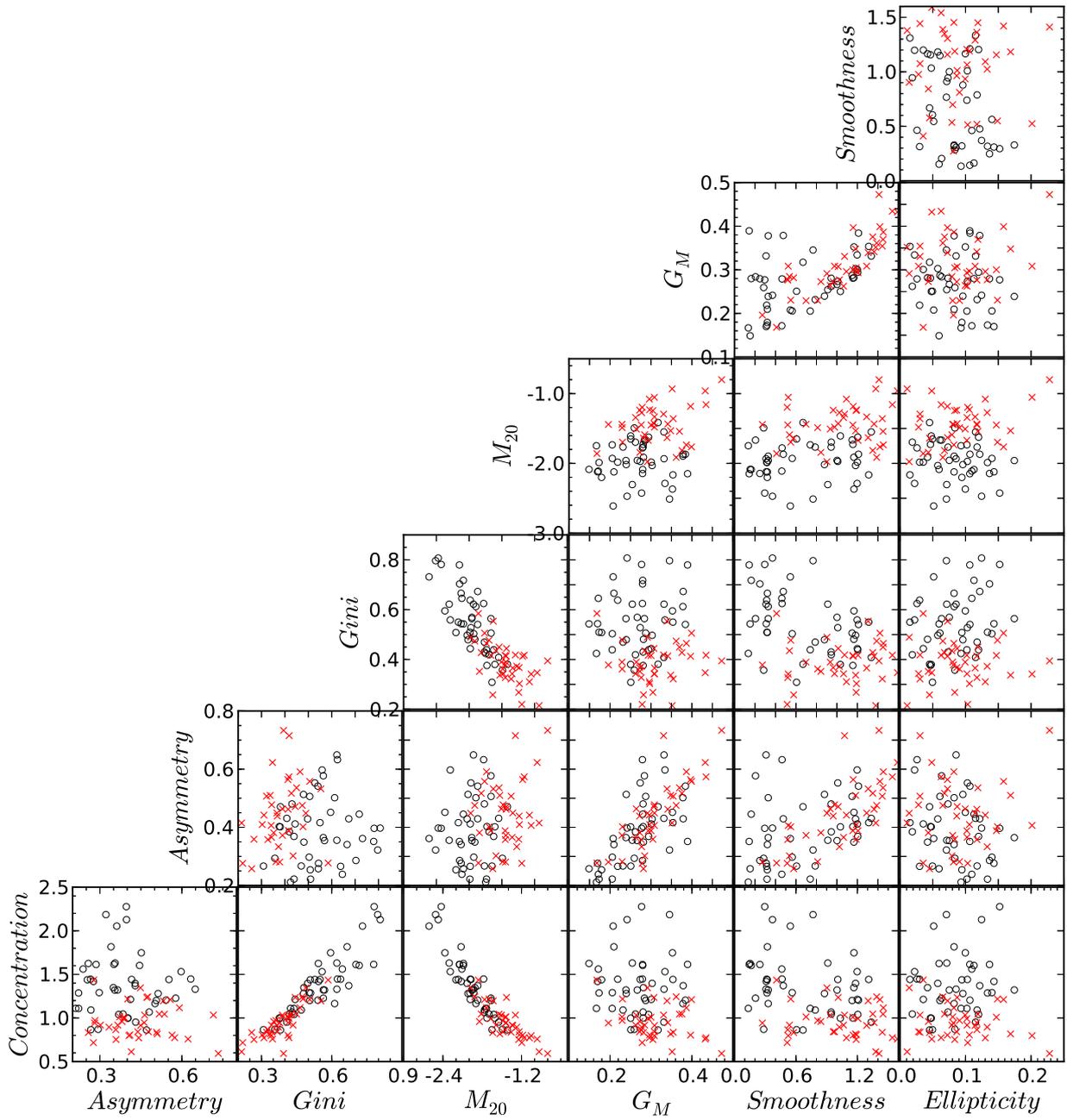}
\caption {\small Seven morphology parameters plotted in the parameter-parameter planes. $C_{5080}$ was plotted as the Concentration parameter. $\circ$ = relaxed cluster; {\small \color {red} $\times$} = non-relaxed cluster. Galaxy cluster separation is based on the V09.} 
\label{exp_cor}
\end{figure*}

\subsection{Combination of morphology parameters}
\label{combpara}
\par We decided to further classify dynamical states of each cluster based on the combination of Gini, $M_{20}$ and Concentration. In \S\ \ref{paravspara}, we defined three parameter boundaries. Based on the combination of these three parameters, we categorised the V09 sample of clusters into four stages - strong relaxed, relaxed, non-relaxed and strong non-relaxed clusters. We selected the following criteria for identifying the dynamical state of each cluster:
\begin{itemize}
\item If the Gini, $M_{20}$ and Concentration all indicate relaxed state, then the cluster will be ``strong relaxed''.

\item The intermediate state of clusters is further divided into two classes: if one or two parameters fall into the intermediate state and another is relaxed (or non-relaxed), then the cluster will be ``relaxed (or non-relaxed)''. 

\item If all three parameters indicate a non-relaxed state, then the cluster will be ``strong non-relaxed''.

\end{itemize}

\par Based on these categories, in the entire V09 cluster sample, 8 ($\sim$ 10\%) clusters were strong relaxed, 11 ($\sim$ 13\%) were relaxed, 52 ($\sim$ 62\%) were non-relaxed, and 13 ($\sim$ 15\%) were strong non-relaxed. In the low-$z$ sample, there were 7 ($\sim$ 15\%) strong relaxed, 6 ($\sim$ 12.5\%) relaxed, 29 ($\sim$ 60\%) non-relaxed, and 6 ($\sim$ 12.5\%) strong non-relaxed clusters. In the high-$z$ sample, there were 1 ($\sim$ 3\%) strong relaxed, 5 ($\sim$ 14\%) relaxed, 23 ($\sim$ 64\%) non-relaxed, and 7 ($\sim$ 19\%) strong non-relaxed clusters. This could suggest that the highest fraction of clusters (in the V09 sample) are evolving or showing some substructure activity (particularly in the high-$z$ sample) and that fewer clusters are fully evolved. Fig.~\ref{combipara} shows the 3-D plot of the Gini, $M_{20}$ and Concentration parameters. A combination of these three parameters classified galaxy clusters as strong relaxed ($\bullet$), relaxed ({\small $\diamond$}), non-relaxed ({\small \color {blue} $+$}) and strong non-relaxed clusters ({\small \color {red} $\times$}). 

\begin{figure*}
\centering
\hspace*{-0.6in}
\includegraphics[scale=0.70]{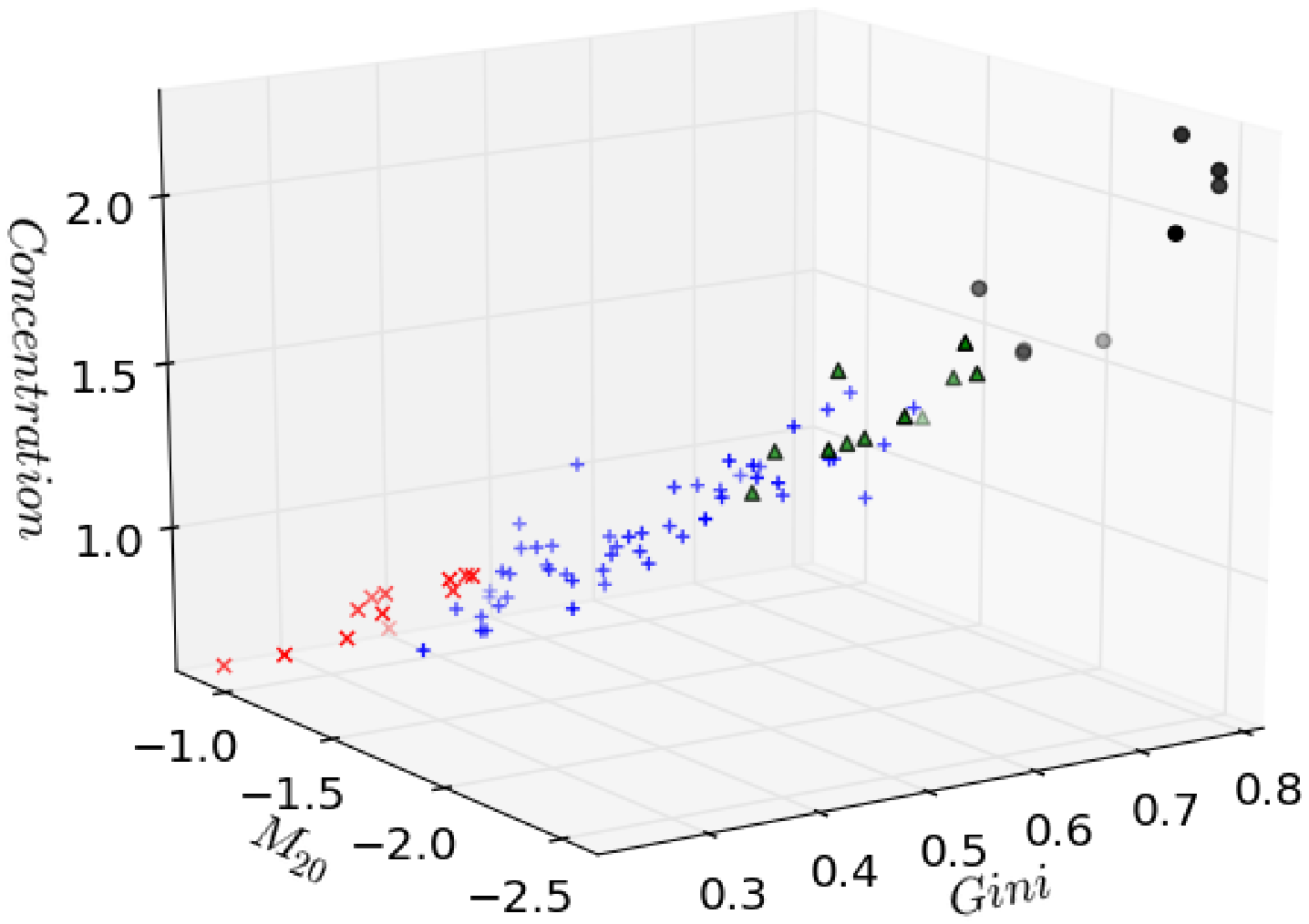}
\caption {\small Gini, $M_{20}$ and Concentration parameters plotted in the 3-D plot. We plotted $C_{5080}$ as the Concentration parameter. $\bullet$ = strong relaxed clusters, {\small $\vartriangle$} = relaxed clusters, {\small \color {blue} $+$} = non-relaxed clusters, and {\small \color {red} $\times$} = strong non-relaxed clusters.} 
\label{combipara}
\end{figure*}

{
\subsection{Smoothness and Asymmetry parameters}
\label{smth_asy}
\par In this section we will test different possible smoothing and angular size issues relating to the Smoothness and Asymmetry parameters. \cite{2010ApJ...721..875O} used a \begin{math}2'\end{math} smoothing scale for calculating the fluctuation parameter. They, however, used it for $R_{500}$ radius and the data of {\it XMM-Newton} which has a large FOV compared with {\it Chandra}. We used \begin{math}0.5'\end{math} (15 pix) and \begin{math}1'\end{math} (30 pix) smoothing scales, $\sigma$, with initial \begin{math}4''\end{math} (2 pix) and \begin{math}10''\end{math} (5 pix) smoothing of input cluster images (which is sufficient to not washed out any underlying substructure features). Results are shown in Fig.~\ref{smth_ang}. We find that the different smoothing scales ($\sigma$) had little affect on the Smoothness parameter. In this analysis, we used a fixed angular size of \begin{math}2''\end{math}/pix to bin each cluster image and the same angular smoothing scale ($\sigma$) for low- and high-$z$ clusters. This angular scaling could affect the Smoothness parameter. To overcome this, and instead of using a fixed angular size, we scaled each cluster image (of low- and high-$z$) in terms of fixed physical size of 10 kpc/pix. We then used 50 and 150 kpc smoothing scales ($\sigma$) to smooth cluster images in order to calculate the Smoothness parameter. Both of these $\sigma$ values were chosen based on the available aperture size (segmentation map) of cluster for which we calculated the Smoothness parameter. We also smoothed cluster input images with (1) 15 (1.5 pix) and (2) 30 (3 pix) kpc Gaussian kernel sizes (which is sufficient to not washed out any underlying substructure features). Fig.~\ref{smth_phy} gives the calculated smoothness parameter for fixed physical smoothing kernel size. The Smoothness parameter had barely changed from the previous results, and had still not separated the two classes of cluster, i.e. relaxed and non-relaxed. Furthermore, most clusters (mainly high-$z$) do not have $\gg$ 1 count in each (binned) pixel. So, in general, the Smoothness is not a good parameter for a large number of clusters in which each has a different exposure time. In \S\ \ref{exp_time_eff_sec}, further we show that the Smoothness parameter depends highly on S/N.

\par We also investigated the Asymmetry parameter using various smoothing Gaussian kernel sizes to smooth cluster input images. We rotated this smoothed image by 180$^{\circ}$, subtracted it from the input image, and normalised it. Our results for the fixed angular and physical scale cluster images are given in Fig.~\ref{asy_ang_phy}. None of the results indicates the separation of relaxed and non-relaxed clusters. As for the Smoothness parameter, Asymmetry parameter is not useful because the S/N is inadequate in a given large sample of clusters (see \S\ \ref{exp_time_eff_sec}).

\begin{figure*}
\centering
\hspace*{-0.6in}
\includegraphics[scale=0.70]{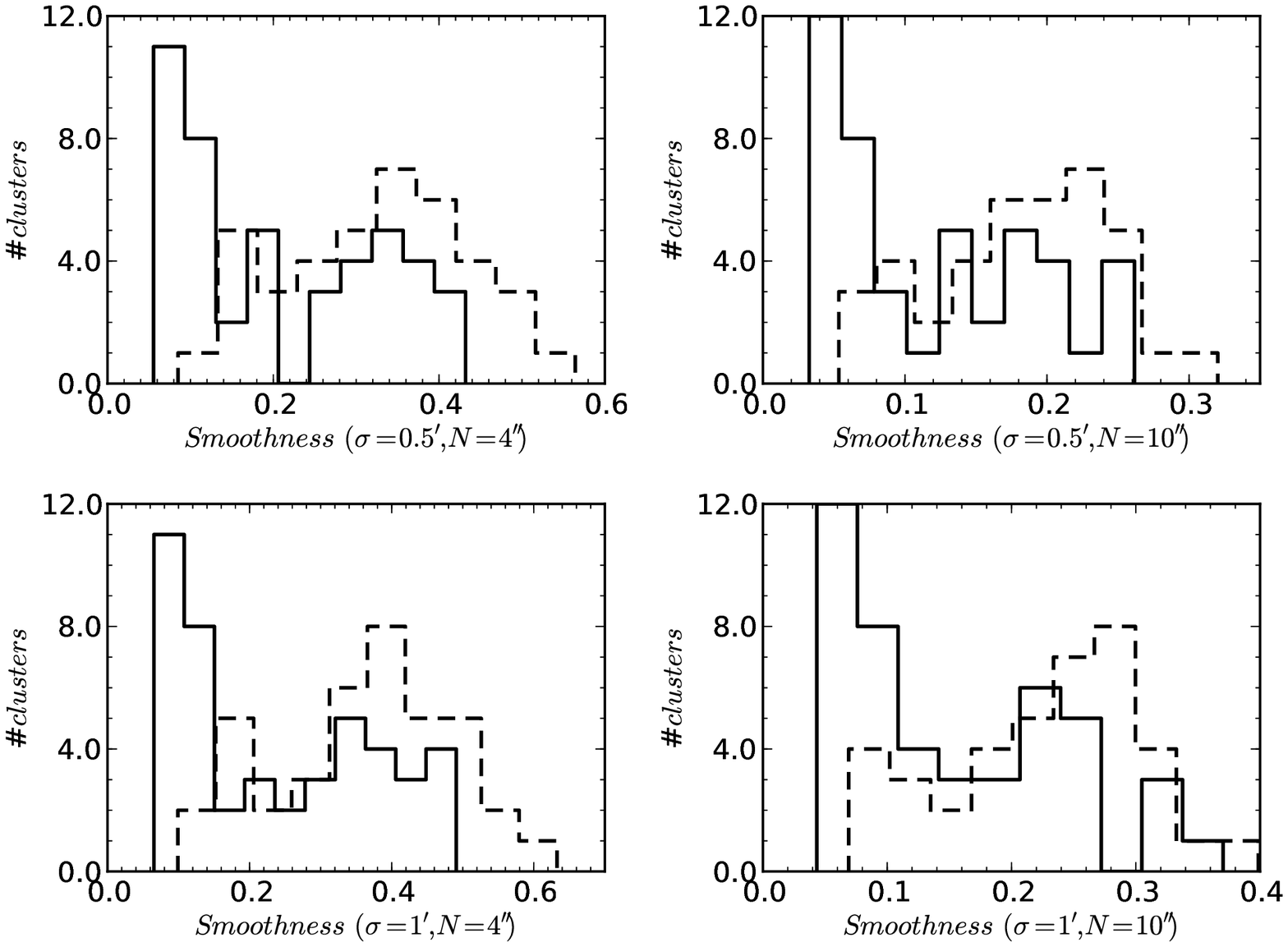}
\caption {\small Calculation of the Smoothness parameter for two different Gaussian kernel sizes of fixed angular scale: {solid line = relaxed cluster; dashed line = non-relaxed cluster}. N shows smoothing size for input cluster image. Galaxy cluster separation is based on the V09.} 
\label{smth_ang}
\end{figure*}

\begin{figure*}
\centering
\hspace*{-0.6in}
\includegraphics[scale=0.70]{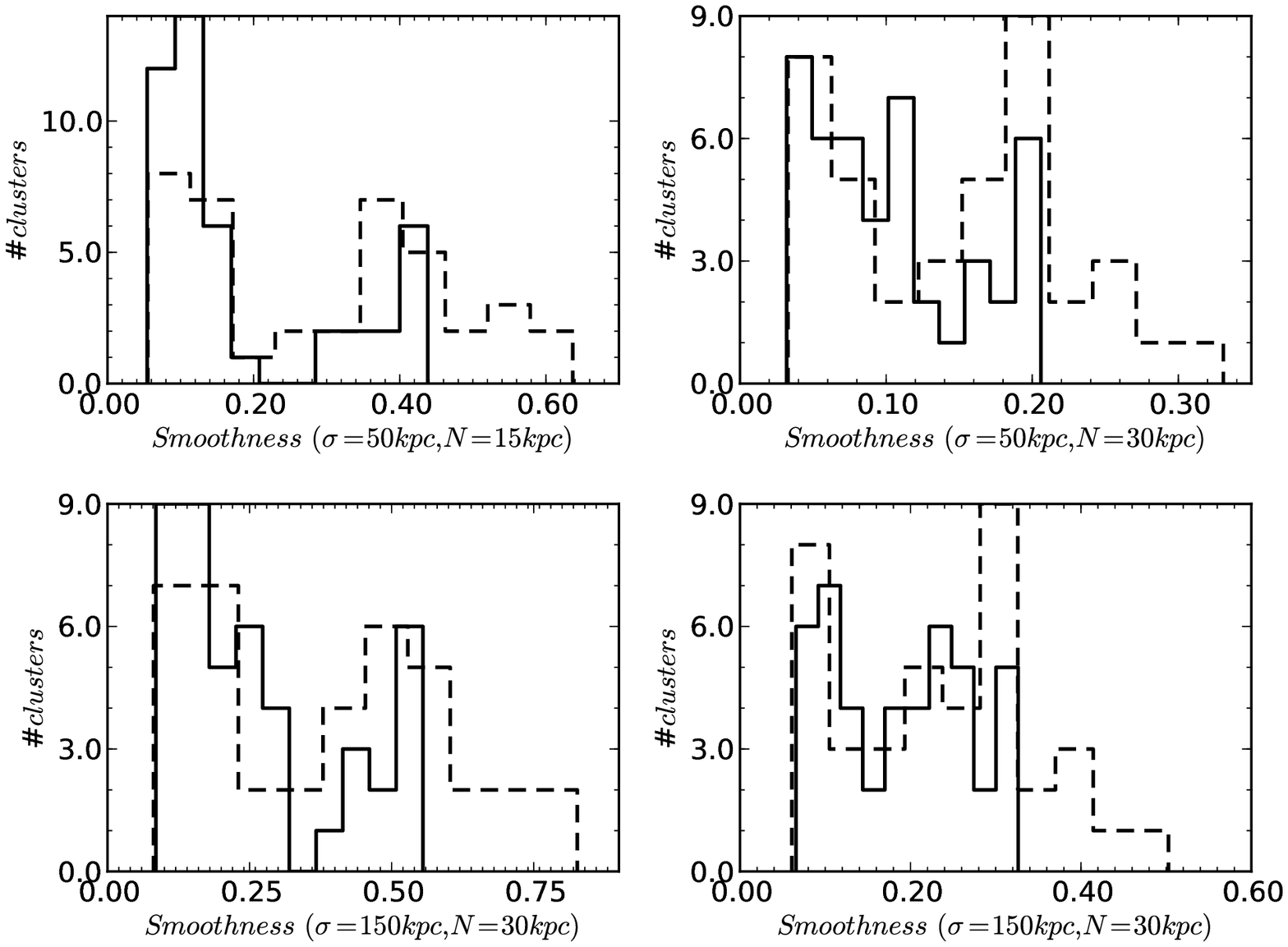}
\caption {\small Calculation of the Smoothness parameter for two different Gaussian kernel sizes of fixed physical scale: {solid line = relaxed cluster; dashed line = non-relaxed cluster}. N shows smoothing size for input cluster image. Galaxy cluster separation is based on the V09.} 
\label{smth_phy}
\end{figure*}

\begin{figure*}
\centering
\hspace*{-0.6in}
\includegraphics[scale=0.70]{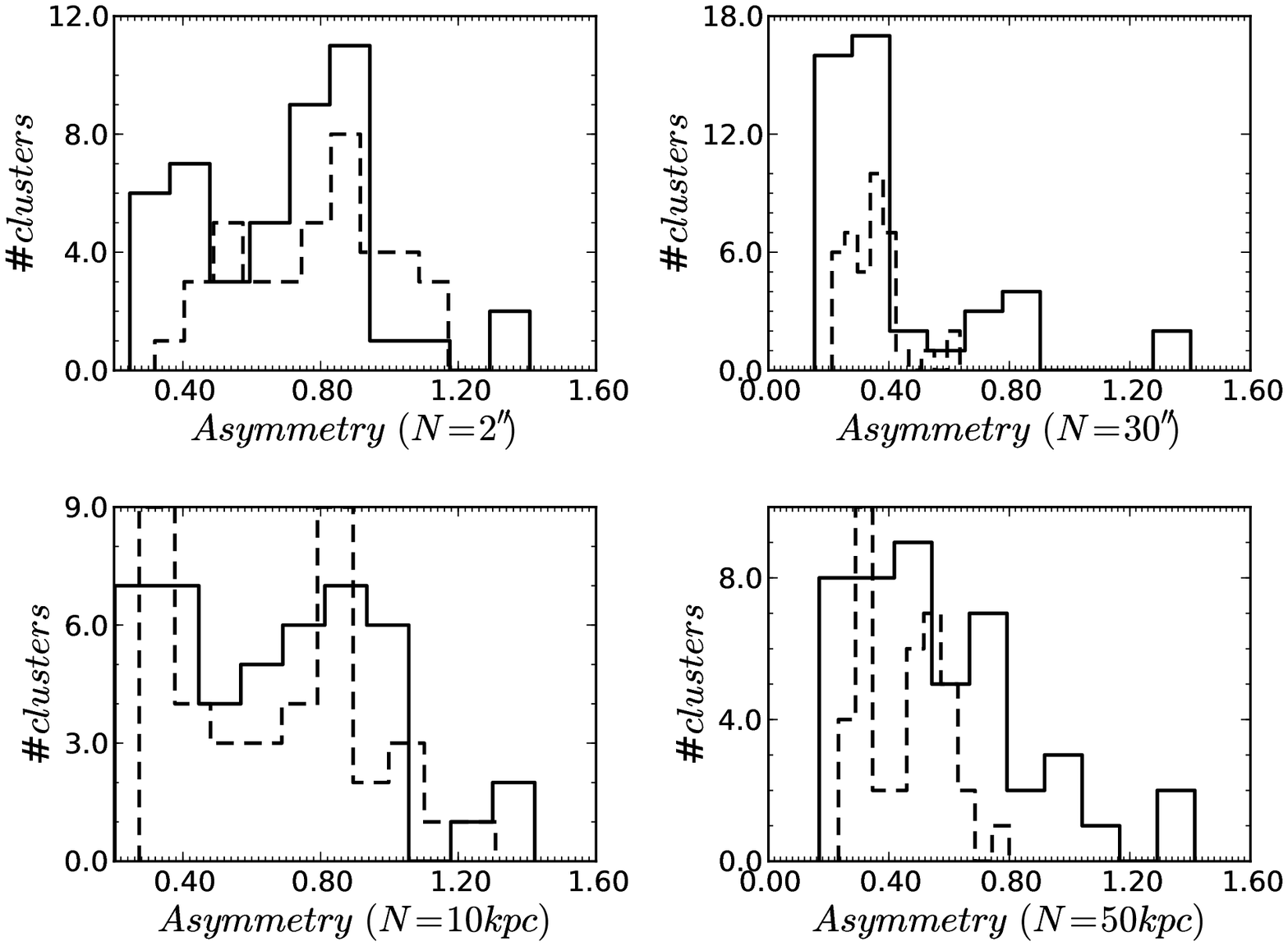}
\caption {\small Calculation of the Asymmetry parameter for two different Gaussian kernel sizes of fixed angular and physical scale: {solid line = relaxed cluster; dashed line = non-relaxed cluster}. N shows smoothing size for input cluster image. Galaxy cluster separation is based on the V09.} 
\label{asy_ang_phy}
\end{figure*}

\begin{landscape}
\begin{table}
\vspace*{6.5cm}
\centering
\caption{Spearman coefficient, $\rho$, for the subsample of relaxed, non-relaxed, and combined clusters, between each morphology parameter.}
\begin{center}
\begin{tabular}{@{}cccccccc@{}}
\hline
\hline 
 & Gini & $M_{20}$ & Concentration & Asymmetry & Smoothness & $G_{M}$ & Ellipticity \\ 
 & R  N-R  C & R  N-R  C &R  N-R  C &R  N-R  C &R  N-R  C &R  N-R  C &R  N-R  C \\
\hline 
Gini & - & -0.75 -0.58 -0.83 & 0.87 0.81 0.93 & 0.05 0.34 -0.02 & -0.40 0.14 -0.38 & 0.11 0.14 -0.12 & 0.25 -0.04 0.05 \\ 
 
$M_{20}$ & -0.75 -0.58 -0.83 & - & -0.91 -0.80 -0.92 & 0.13 0.12 0.24 & 0.18 0.15 0.40 & 0.07 0.21 0.29 & -0.15 0.18 0.04 \\ 
 
Concentration & 0.87 0.81 0.93 & -0.91 -0.80 -0.92 & - & -0.06 0.04 -0.16  & -0.29 -0.10 -0.41  & -0.05 -0.20 -0.28 & 0.13 -0.10 -0.01 \\ 
 
Asymmetry & 0.05 0.34 -0.02 & 0.13 0.12 0.24 & -0.06 0.04 -0.16  & - & 0.38 0.76 0.56 & 0.68 0.83 0.72 & -0.22 -0.05 -0.12 \\ 
 
Smoothness & -0.40 0.14 -0.38 & 0.18 0.15 0.40 & -0.29 -0.10 -0.41  & 0.38 0.76 0.56 & - & 0.42 0.85 0.65 & -0.33 -0.01 -0.13 \\ 
 
$G_{M}$ & 0.11 0.14 -0.12 & 0.07 0.21 0.29 & -0.05 -0.20 -0.28 &0.68 0.83 0.72  &0.42 0.85 0.65  & - & -0.20 -0.03 -0.06 \\ 
 
Ellipticity & 0.25 -0.04 0.05 & -0.15 0.18 0.04 &0.13 -0.10 -0.017  & -0.22 -0.05 -0.12 &-0.33 -0.019 -0.13 & -0.20 -0.03 -0.06 & - \\ 
\hline 
\end{tabular}
\end{center}
\label{para-spearman} 
\end{table}
\end{landscape}

}

\section{Systematics}
\par We investigated a number of possible systematic effects, discussed below, to study how robust these parameters are in various conditions.  
\subsection{Effect of point sources}
\par In order to test the effect of {point sources detected around cluster} on the calculation of parameters, we calculated morphology parameters on (exposure corrected and smoothed) cluster images without removing any of the detected point source. We noticed that parameters are fairly robust against the inclusion (or exclusion) of point sources into the parameter calculation. Fig.~\ref{eff_point} indicates how we plotted the offset for four parameters (Gini, $M_{20}$, Concentration and Smoothness) calculated with and without point sources. As we observed, less offset between parameters (with or without point sources) suggests that they are quite robust. The Gini and $M_{20}$, however, each show a small extended tail on the right hand side (Fig.~\ref{eff_point}). This result is obvious, because the Gini coefficient includes bright pixels from point sources in the calculation, and thus increases by a small value. In the case of the $M_{20}$, if the point sources are very close to the centre, its value will decrease by a small factor.
\begin{figure*}
\centering
%\hspace*{-0.6in}
\includegraphics[scale=0.50]{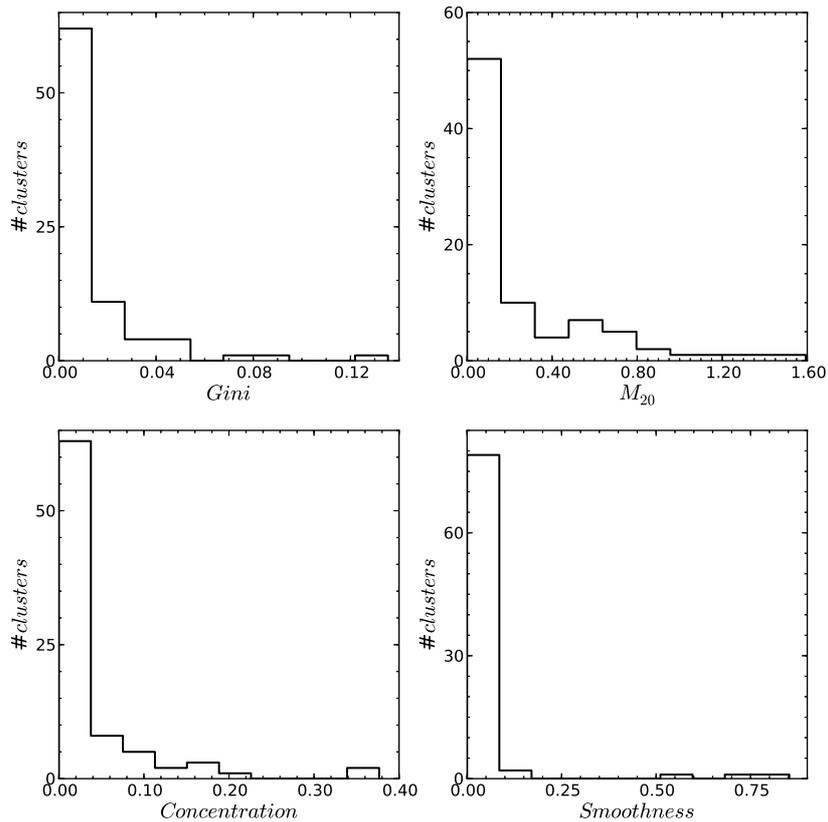}
\caption{\small Four parameters calculated with and without point sources, with plotted offset between them. Here we plotted $C_{5080}$ as the Concentration parameter. Less offset suggests that parameters are quite robust.} 
\label{eff_point}
\end{figure*}

\subsection{Aperture size effect}
\label{aper_effe_sec}
\par Aperture radii have an important effect when measuring morphology parameters. We chose to use a fixed physical size radius rather than a fixed overdensity ($R_{500}$) radius, because it is difficult to measure $R_{500}$ accurately for non-relaxed clusters. {Although a study of the possible bias effects of this choice goes beyond the purpose of this paper, we note that possible bias effects could be introduced when comparing clusters at the same redshift, but characterised by different luminosities (masses).} The large variation in aperture size does affect the parameters because they are related to surface brightness. {To check the effect of various aperture radii on the parameters, we chose a subsample that included both distant and nearby clusters and calculated the morphology parameters in the radius sequence of 100 kpc to 1 Mpc for distant clusters ($z$ $>$ 0.05), while 100 kpc to 500 kpc for nearby clusters ($z$ $<$ 0.05). Fig.~\ref{radii_para} illustrates our result}. In Fig.~\ref{radii_para}, the distant clusters are plotted with a continuous line, while the nearby clusters are plotted with a dotted line. As per Fig.~\ref{radii_para}, in the case of distant clusters, a few parameters (Asymmetry, Smoothness, $G_{M}$, and Ellipticity) remain constant in spite of aperture size, while the Gini, $M_{20}$ and Concentration parameters are sensitive to the aperture radius within which they are calculated, while tending to be stable for aperture radii greater than 400 kpc. The Gini parameter increases with radius as more (faint) sky pixels are included in the extraction aperture. Parameter values for nearby clusters (in particular $z$ $<$ 0.05) are very sensitive to the chosen aperture size.

\begin{figure*}
\centering
%\hspace*{-0.6in}
\includegraphics[scale=0.50]{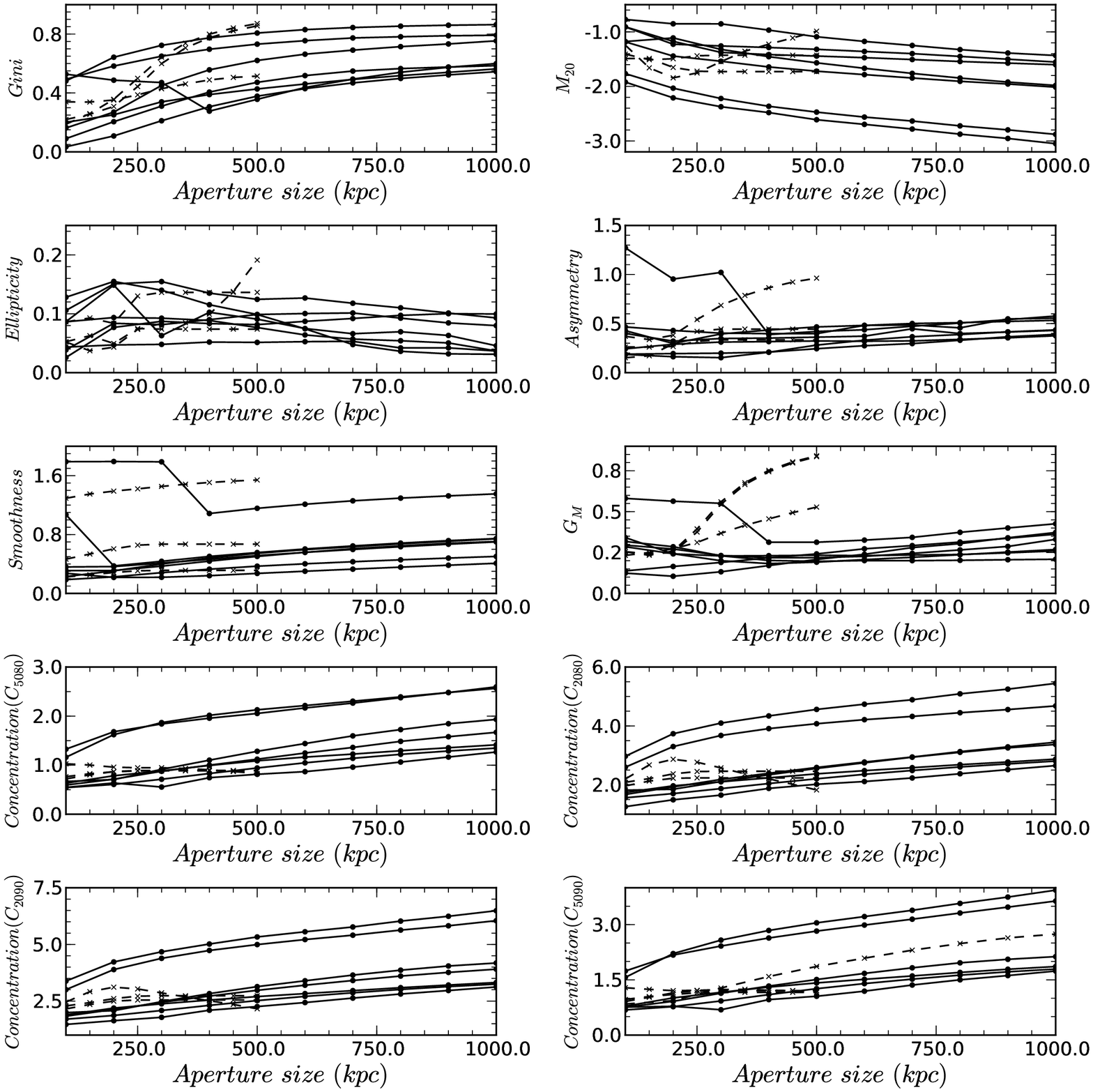}
\caption{\small Morphology parameters calculated for a range of aperture sizes from 100 kpc to 1 Mpc for distant clusters. Nearby clusters are calculated in a range of radii from 100 kpc to 500 kpc only. Solid lines with {\tiny $\bullet$} symbol are plotted for distant clusters, and dashed lines with {\tiny $\times$} symbol are plotted for nearby clusters. } 
\label{radii_para}
\end{figure*}

\subsection{Exposure time effect}
\label{exp_time_eff_sec}
\par It is important to check the consistency of the parameters over different exposure times. Observations with a shorter exposure time are likely to have lower S/N. Consequently, we were able to check the robustness of the parameters for cluster images with lower S/N. However, to rescale the real data by exposure time and then add Poisson noise gives an image which has an excessive amount of Poisson noise, in addition to the intrinsic noise present in the real observation. The simplest solution is to simulate a cluster image with no intrinsic noise, rather than using real data. To achieve this, we needed to estimate the different complex characteristics of a model to simulate galaxy clusters. This task is considered difficult in the cases of non-relaxed and disturbed clusters.
To overcome this problem, \cite {2007A&A...467..485H} suggested a novel technique for galaxy cluster simulations called ``adaptive scalings'', using real data and adding noise to the rescaled image. We refer the reader to \cite {2007A&A...467..485H}, for more details about this technique.

\par An example of this adaptive method is given in Fig.~\ref{exp_change_img}; an example of the low exposure time effect on a cluster. Subsequently we calculated our morphology parameters for each short to long exposure cluster image.

\begin{figure*}
\centering
\includegraphics[scale=0.5]{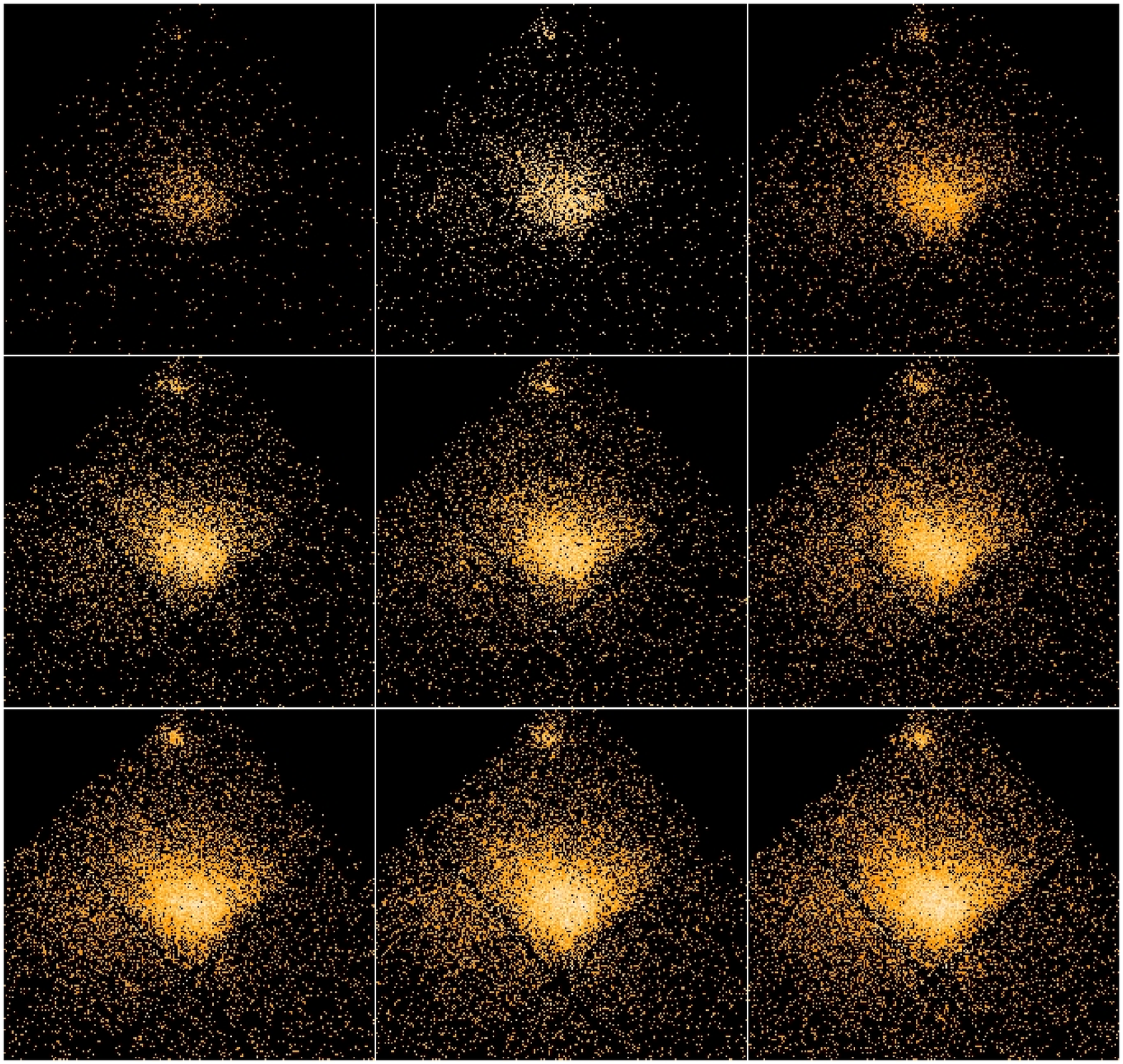} 
\caption{Simulated clusters for different exposure times. Horizontally, from top left to bottom right, the images were arranged from 2 ks to 18 ks, respectively, in order of 2 ks exposure time.}
\label{exp_change_img}
\end{figure*} 
\begin{figure*}
\centering
\includegraphics[scale=0.5]{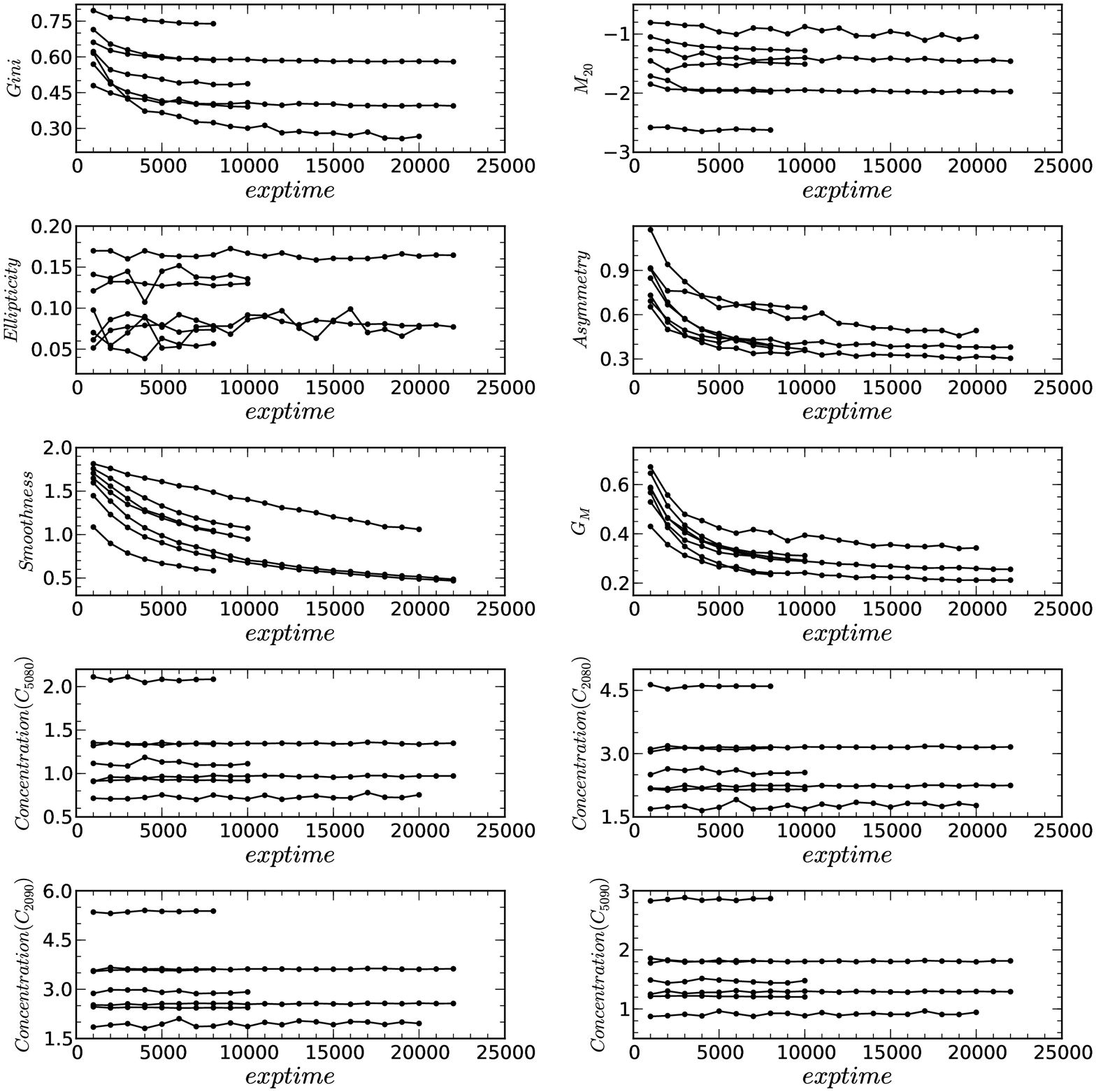} 
\caption{Robustness of morphology parameters calculated for several clusters against various simulated exposure lengths.}
\label{system_exp}
\end{figure*} 
\par Fig.~\ref{system_exp} shows how morphology parameters behave with different exposure lengths. We simulated a few original clusters with various exposure times, as described above, and re-calculated the morphology parameters for each simulated cluster. As depicted in Fig.~\ref{system_exp}, all parameters are robust against different exposure times, with the exception of the Smoothness parameter. We noted, however, that the Gini, Gini of the second order moment and Asymmetry were not reliable when exposure time was $\lesssim$ 5 ks. In the cases of the Gini and Gini of the second order moment, low exposure means low S/N, indicating that the Gini and Gini of the second order moment values are high for low exposure observations, although fairly consistent for exposure times $\gtrsim$ 5 ks. For the short exposure times, there are few bright pixels within the given aperture radius (the rest are scattered to very low $\sim$ 0). In other words, low S/N causes broader flux distribution in the faintest pixels, resulting in strong variation in the flux distribution, for which we calculated the Gini coefficient; and hence the Gini coefficient gives a high value compared with the long exposure time. The value of the Smoothness decreases continuously with increasing exposure time. This is to be expected because, for high S/N and exposure time, the flux distribution in any given cluster image becomes smooth and less patchy. Care must therefore be taken when calculating the Smoothness parameter. We found that the Smoothness is weakly correlated with the Gini of the second order moment and Asymmetry parameters, yet it is not established whether we can use these latter parameters as a substitute for the Smoothness. The remainder of the parameters were relatively constant over different exposure lengths.

\subsection{Redshift effect}
\par It is hoped that these morphology parameters could be used to trace the evolution of galaxy clusters with redshifts. It is therefore important to understand the robustness as a function of redshift and, to this end, it is necessary to check the following criteria: 

\begin{enumerate}
\item{The effect of various angular bin sizes on morphology parameters.}
\item{The surface brightness dimming effect on morphology parameters.}
\end{enumerate}  
We simulated several observations of galaxy clusters at a higher redshift ($z_{1}$) than its real redshift ($z_{0}$) using real data from our sample clusters. Again, we adopted the procedure describe by \cite {2007A&A...467..485H}.

\par In Fig.~\ref{z_change_img} we give an example of clusters simulated via the method described above that illustrates how a cluster appears at high redshift. Subsequent to the simulation, we calculated morphology parameters for each cluster redshift and plotted them (Fig.~\ref{system_z}). 

\begin{figure*}
\centering
%\hspace*{-0.6in}
\includegraphics[scale=0.3]{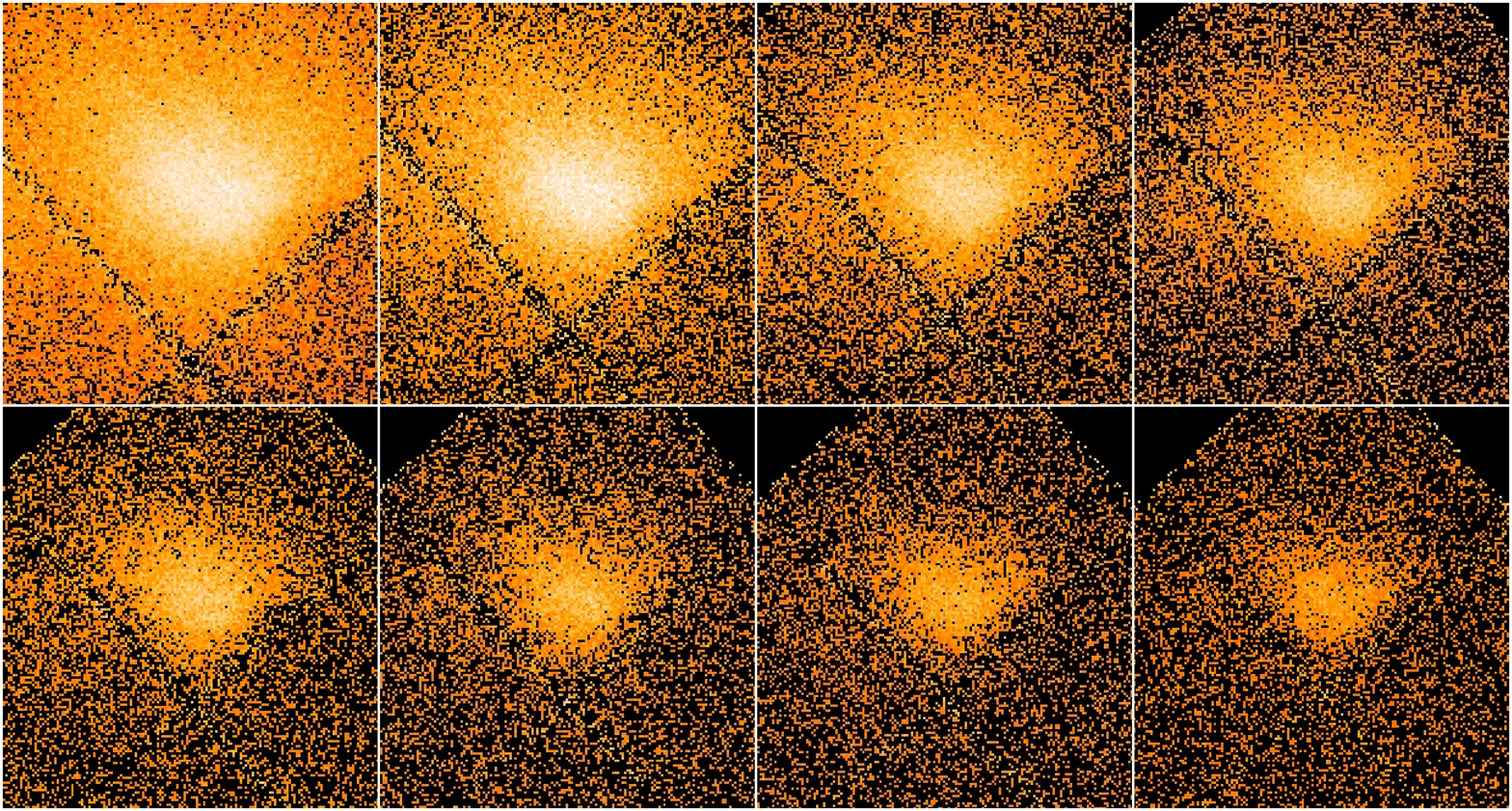} 
\caption{Simulated clusters for different redshifts. Figures are read from top left to bottom right in order of low- to high-$z$, respectively.}
\label{z_change_img}
\end{figure*} 

\begin{figure*}
\centering
\includegraphics[scale=0.5]{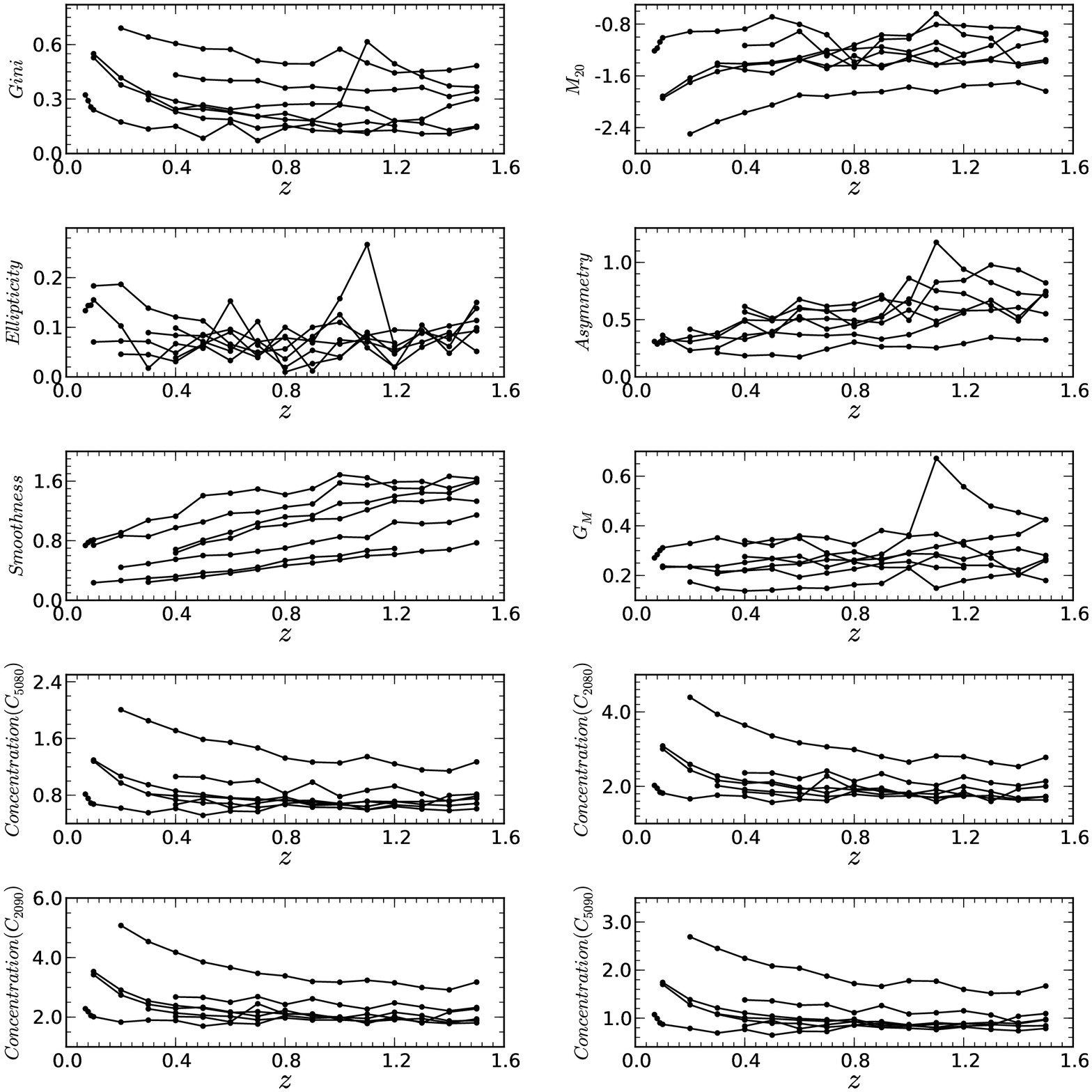} 
\caption{Robustness of morphology parameters which are calculated for several clusters against simulated clusters at different redshifts.}
\label{system_z}
\end{figure*} 

\par We simulated a few original clusters with various redshifts, using the method described above, and re-calculated morphology parameters for each simulated cluster. Fig.~\ref{system_z} illustrates that the Smoothness is systematically increased with redshift. The reason could be that high-$z$ clusters are noisy and appear patchy. The Ellipticity parameter appears particularly noisy. Except for a few clusters, and after $z$ $\gtrsim$ 0.5, the Gini, $M_{20}$, Concentration, Asymmetry and Gini of the second order moment parameters are fairly constant, and the systematics caused by redshifts are small.  

{

\section{Comparison of parameters with X-ray gas properties}
\par We expect that the dynamical states of clusters to be influence by factors like merger histories, which in turn could influence a number of properties like luminosity, temperature structure, cooling flows etc. In order to test the potential of our parameters as probes of the physical conditions in clusters, we investigated the parameter correlations with source redshift, global X-ray properties, cooling times and the presence of diffuse radio continuum emission.

\subsection{Redshift evolution}
\par According to the concordance model, massive galaxy clusters start to form around $z$ $\sim$ 1 and have continue to evolve up to the present epoch. To look into the evolutionary effect on the distribution of the parameters, we divided our entire sample into low-$z$ (0.02--0.3) and high-$z$ (0.3--0.9) clusters and classified the samples based on our morphology parameters. We also performed the K-S and R-S tests on each parameter distribution to observe the difference between the two redshift bins. Table~\ref{kstest_redshift} lists the mean, median and statistical test results for the two redshift bins.

\par Below are more descriptions for the results of our three most promising parameters (viz. Gini, $M_{20}$ \& Concentration). The combination of the three parameters are plotted in Fig.~\ref{diff_red}, with boundaries between relaxed and non-relaxed clusters. \begin{enumerate}

\item {Based on the Gini, we found that 9 clusters are relaxed, 48 are intermediate and 27 are non-relaxed. In the low-$z$ subsample there are 8 ($\sim$ 17\%) relaxed, 25 intermediate ($\sim$ 52\%) and 15 ($\sim$ 31\%) non-relaxed clusters; in the high-$z$ subsample there are 1 ($\sim$ 3\%) relaxed, 23 ($\sim$ 64\%) intermediate and 12 ($\sim$ 33\%) non-relaxed clusters.}

\item {Based on the $M_{20}$, we found that 18 clusters are relaxed, 50 are intermediate and 16 are non-relaxed. In the low-$z$ subsample there are 12 ($\sim$ 25\%) relaxed, 28 intermediate ($\sim$ 58\%) and 8 ($\sim$ 17\%) non-relaxed clusters; in the high-$z$ subsample there are 6 ($\sim$ 17\%) relaxed, 22 ($\sim$ 61\%) intermediate and 8 ($\sim$ 22\%) non-relaxed clusters.}

\item {Based on the Concentration, we found 12 clusters are relaxed, 41 clusters are intermediate and 31 clusters are non-relaxed. In the low-$z$ subsample there are 10 ($\sim$ 21\%) relaxed, 20 intermediate ($\sim$ 42\%) and 18 ($\sim$ 37\%) non-relaxed clusters; in the high-$z$ subsample there are 2 ($\sim$ 6\%) relaxed, 21 ($\sim$ 58\%) intermediate and 13 ($\sim$ 36\%) non-relaxed clusters.}

\end{enumerate}

\par As seen in Table~\ref{kstest_redshift}, the K-S and R-S probabilities are $<$ 1\% and $<$ 0.1\%, respectively, for the Asymmetry and Smoothness parameters implying that we can reject both null hypotheses mentioned in \S\ \ref{para_dis_sec}. Clearly most of the clusters in our sample are in the intermediate stage ($\sim$ 57$\%$ based on the Gini coefficient, $\sim$ 60$\%$ based on the $M_{20}$, and $\sim$ 49$\%$ based on the Concentration parameter). Weak evolution is visible in the Gini and Gini of the second order moment, which indicates the possibility that high redshift clusters are more extended (which could mean that they do not have a density peak at the cluster centre) compared with those of low redshift. The $M_{20}$, Concentration and Ellipticity do not show significant evolution. From our results, there is indication that relaxed clusters are more dominant within the low-$z$ sample, which could indicate that, in the current epoch, clusters show less substructure and are (fully) evolved as compared to distant clusters, but these results are marginal. }
\begin{table*}
\begin{small}
\centering
\caption{Statistics for two redshift bins: low-$z$ (0.02--0.3) and high-$z$ (0.3--0.9).}
\begin{tabular}{cccccccc}
\hline
\hline
Parameters & mean of low-$z$ & mean of high-$z$ &median of low-$z$ & median of high-$z$& K-S probability & R-S probability \\ 
\hline 
Gini & 0.50 & 0.44 &0.47 &0.43 &0.08 & 0.09 \\ 
$M_{20}$ & -1.75 & -1.68 &-1.76 & -1.67 & 0.74 & 0.35 \\ 
Concentration & 1.21 & 1.14 &1.14&1.06& 0.57 & 0.47 \\ 
Asymmetry & 0.38 & 0.46 &0.36&0.46 & 0.0002 & 0.0008 \\ 
Smoothness & 0.65 & 1.11 &0.53&1.16& 1.57 $\times$ 10$^{-7}$ & 2 $\times$ 10$^{-6}$ \\ 
$G_{M}$ & 0.28 & 0.31 &0.28&0.30& 0.0016 & 0.004 \\ 
Ellipticity & 0.10 & 0.08 &0.09&0.07& 0.08 & 0.024 \\ 
\hline 
\label{kstest_redshift} 
\end{tabular} 
\end{small}
\end{table*}

\begin{figure*}
\centering
%\hspace*{-0.6in}
\includegraphics[scale=0.40]{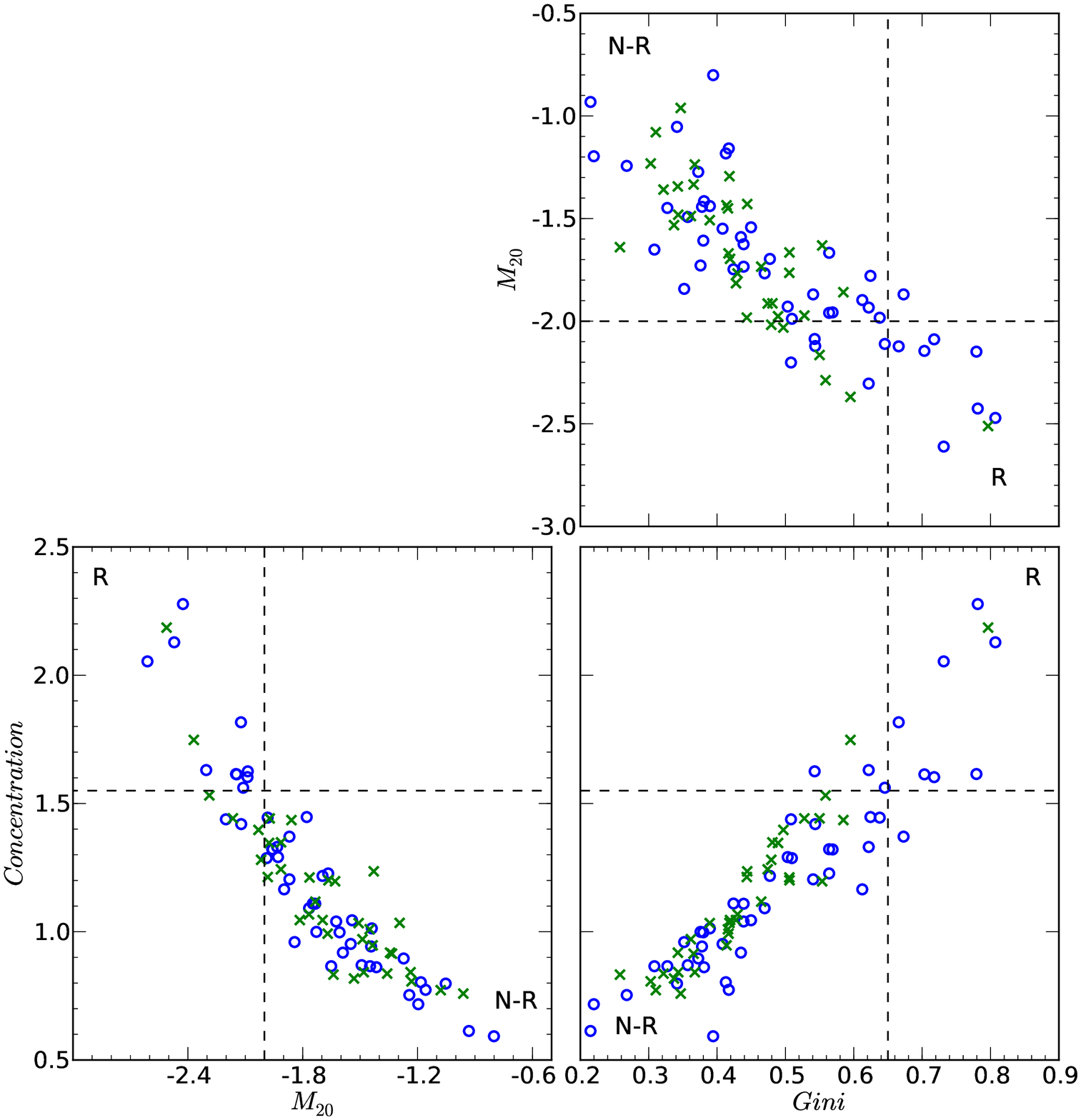}
\caption{\small Three parameters plotted in the parameter-parameter planes to show cluster evolution with redshift. Here we plotted $C_{5080}$ as the Concentration parameter. {\small \color {blue} $\circ$} = low-$z$ cluster (0.02--0.3); {\small \color {green} $\times$} = high-$z$ cluster (0.3--0.9). The dashed lines represent the boundaries between relaxed and non-relaxed clusters. Boundary values for Gini = 0.65, $M_{20}$ = -2.0 and Concentration = 1.55. R indicates relaxed clusters and N-R indicates non-relaxed clusters.} 
\label{diff_red}
\end{figure*}

\subsection{X-ray luminosity, temperature and mass}
\label{para_com_pro_mor}
\par We compared seven morphology parameters with three global cluster properties (luminosity, temperature and mass) taken from the V09 to search for any possible correlation of these global properties with cluster morphology. Fig.~\ref{paraprop} shows a comparison, while Table \ref{spearprop} lists the Spearman coefficient values calculated for the clusters' global properties and galaxy cluster morphologies. Fig.~\ref{paraprop} defines each clusters' dynamical state according to the combination of the Gini, $M_{20}$ and Concentration morphology parameters (\S\ \ref{combpara}).

\par No obvious correlation between cluster morphology and X-ray global properties was found during our analysis. This may be due to the fact that quantities such as X-ray luminosity and temperature are not solely related to the dynamical state of clusters \citep {2007A&A...467..485H}. These properties also depends strongly on cluster mass and on non-gravitational processes such as supernovae feedback, central AGN heating, etc. \citep{1999ApJ...513..690D, 2001ApJ...546...63T, 2003A&A...400..811N}. \cite {1996ApJ...458...27B} also compared their power ratio measurements for X-ray clusters with the ICM temperature and luminosity, without finding any strong correlation between subclustering and global X-ray properties. In agreement with our conclusion, they pointed out that this lack of correlation is reasonable, since power ratios are a measure of cluster evolution that do not take into account the cluster mass, to which all the other X-ray quantities (e.g. luminosity and temperature) are sensitive. We observed that all relaxed clusters occupy high Gini, low $M_{20}$ and high Concentration values in all three plots. We therefore concluded that, to identify the dynamical state of a galaxy cluster, X-ray global properties are not as useful as the surface brightness distribution or cluster morphology, quantifiable using morphology parameters.

\begin{figure*}
\begin{center}
%\hspace*{-1cm}
{\label{fig:parlum}\includegraphics[scale=0.5]{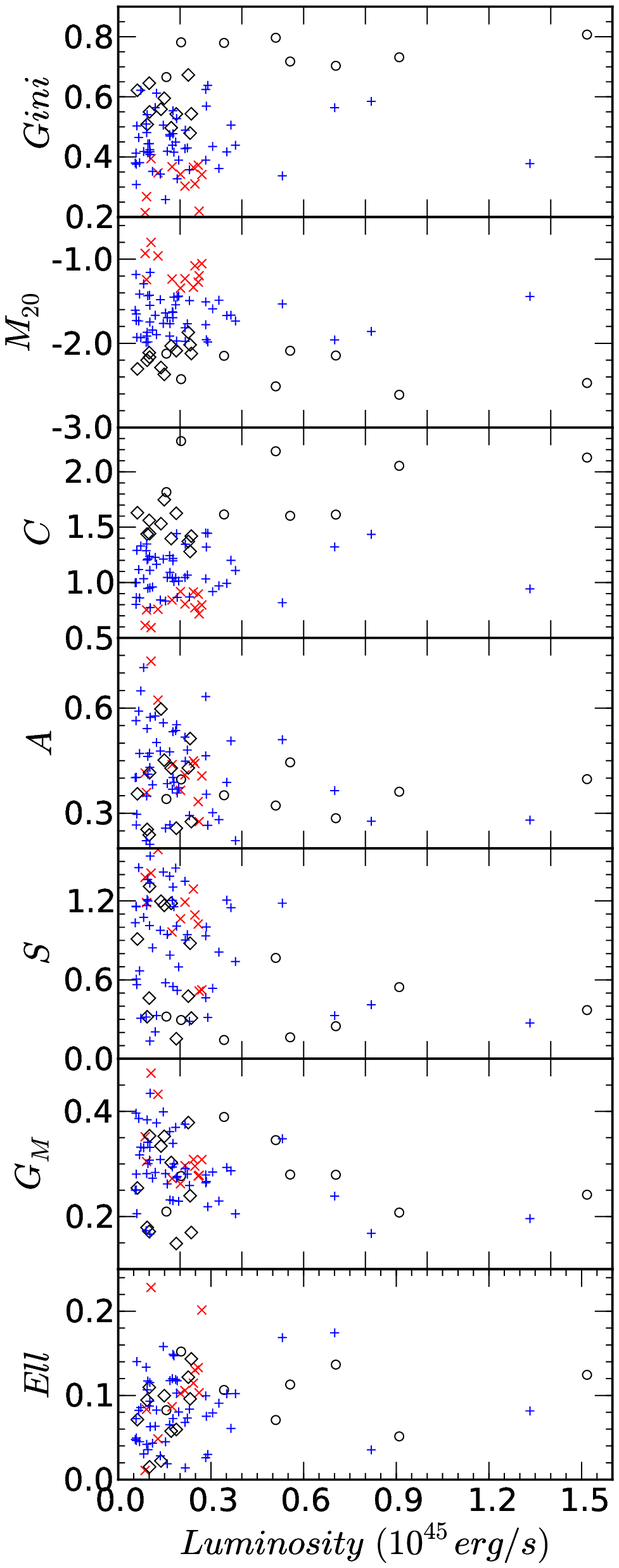}}{\label{fig:partemp}\includegraphics[scale=0.5]{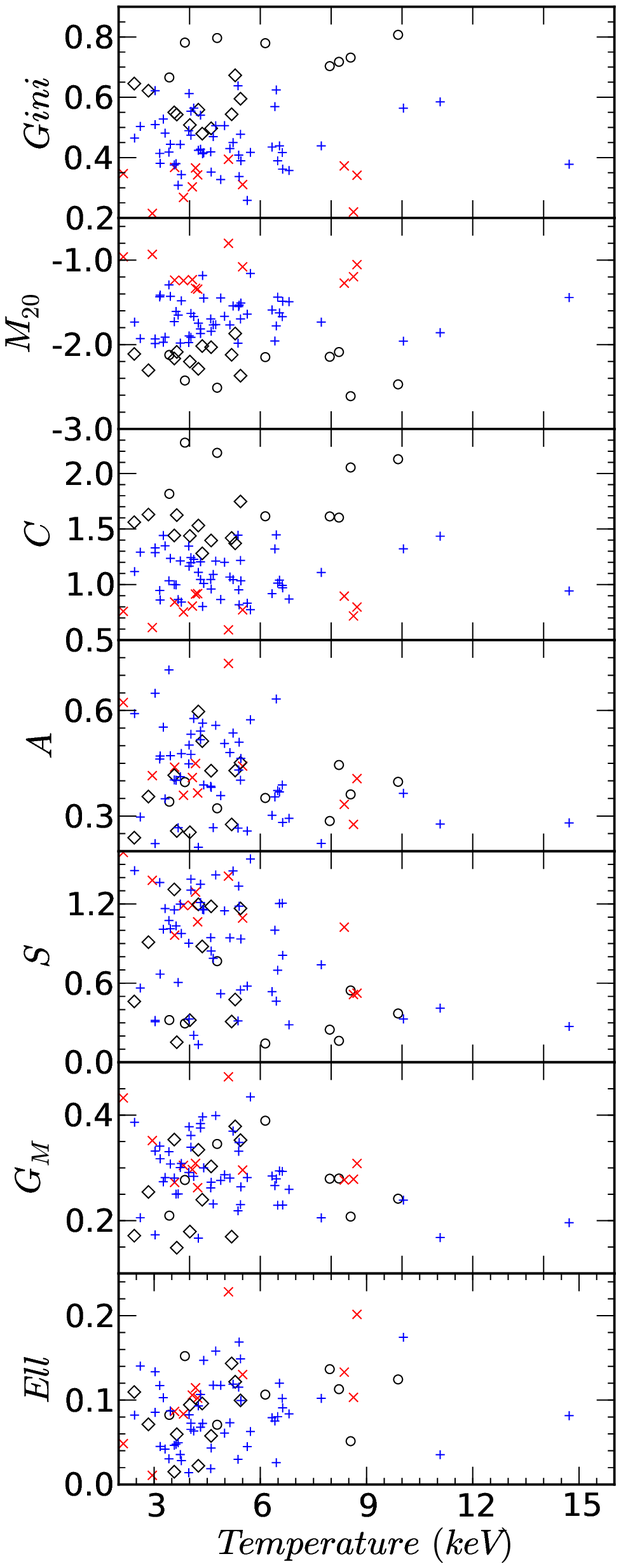}}{\label{fig:parmass}\includegraphics[scale=0.5]{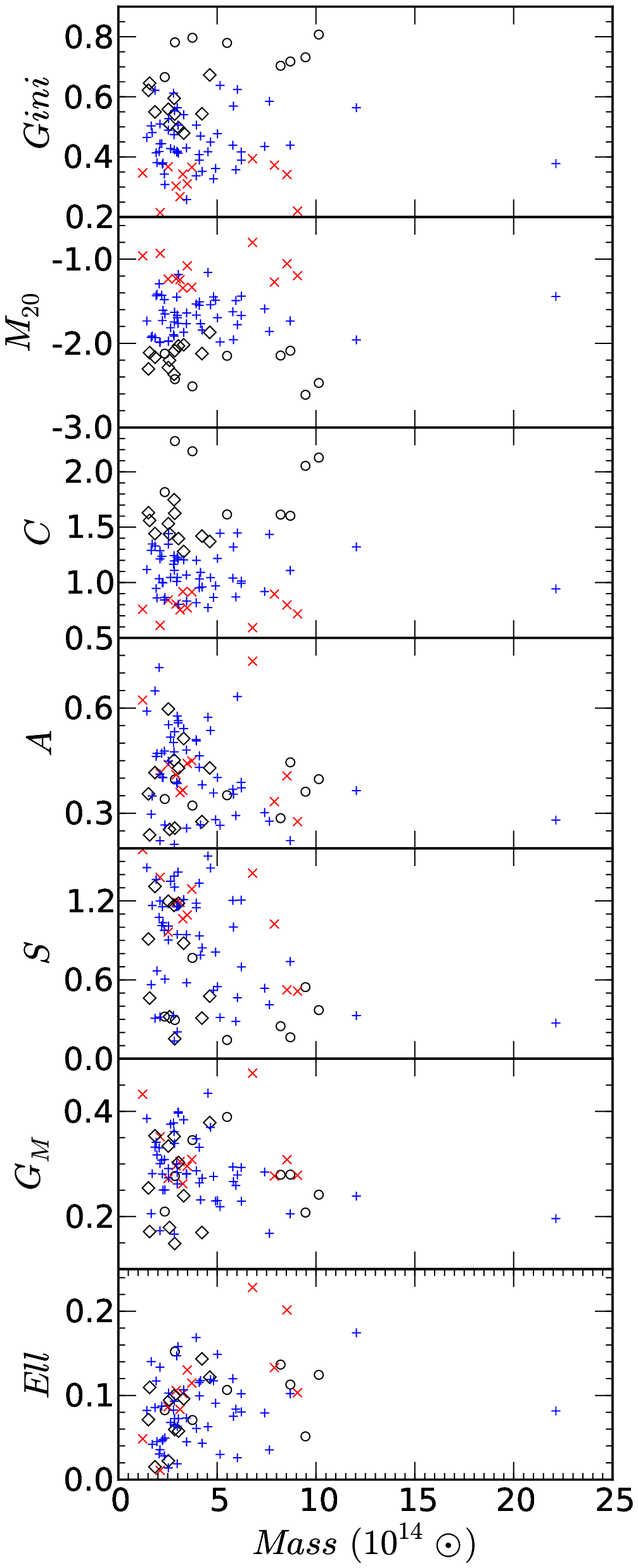}}
\caption{\small Comparison of {\bf Left} Luminosity, {\bf Middle} Temperature and {\bf Right} Mass (estimated from the $Y_{X}$ parameter) with morphology parameters. $\circ$ = strong relaxed clusters; {\small $\lozenge$} = relaxed clusters; {\small \color {blue} $+$} = non-relaxed clusters; and {\small \color {red} $\times$} = strong non-relaxed clusters. We plotted $C_{5080}$ as the Concentration parameter. We take all these global properties (luminosity, temperature and mass) values from the V09. }
\label{paraprop}
\end{center}
\end{figure*}

\begin{table*}
\caption{Spearman coefficient, $\rho$, for morphology parameters and X-ray global properties (luminosity, temperature and mass). }
\centering
\label{spearprop}
\begin{tabular}{ccccc}
\hline 
\hline
 & Luminosity & Temperature & Mass \\ 
% & R N-R C & R N-R C & R N-R C \\
\hline
Gini & 0.19 & 0.03 & 0.03  \\ 
$M_{20}$ & -0.13 & 0.03 & 0.07  \\ 
Concentration & 0.18 & -0.03 & -0.05 \\ 
Asymmetry & -0.22 & -0.23 & -0.25 \\ 
Smoothness & -0.33 & -0.26 & -0.30 \\ 
$G_{M}$ & -0.25 & -0.17 & -0.21 \\ 
Ellipticity & 0.26 & 0.30 & 0.35 \\ 
\hline 
\end{tabular} 
\end{table*}

\subsection{X-ray cluster cooling time}
\label{cooling_eff}
\par X-ray emission is considered to be the primary cooling process for the ICM. The cooling time is much shorter than the Hubble time at the centre of a cluster, allowing gas to cool from a high ICM temperature $\sim$ 10$^{8}$ K down to $\sim$ 10$^{7}$ K.
Numerical simulations suggest that cluster mergers may disturb gas cooling  \citep{2002MNRAS.329..675R, 2008ApJ...675.1125B, 2010A&A...513A..37H}. We aimed to  investigate the relationship between the degree of substructure and cooling times or rates of a cluster. We used the cooling time information supplied by \cite {2010A&A...513A..37H}. Table \ref{cool_list} lists the cooling time values for low-$z$ clusters.      

\begin{table*}
\centering
\caption{Available cooling time, $t_{cool}$, values from \citep {2010A&A...513A..37H} for the low-$z$ clusters.}
\begin{tabular}{cccc}
\hline 
\hline
Cluster name & $t_{cool}$ (Gyr) & Cluster name & $t_{cool}$ (Gyr)\\
\hline
A3571 & 2.13$_{0.71}^{-0.43}$   &      A2597 & 0.42$_{0.04}^{-0.03}$ \\ 
A2199 & 0.6$_{0.07}^{-0.06}$    &      A133 & 0.47$_{0.03}^{-0.03}$ \\ 
2A 0335 & 0.31$_{0.01}^{-0.01}$   &      A2244 & 1.53$_{0.27}^{-0.2}$ \\ 
A496 & 0.47$_{0.02}^{-0.02}$    &      RXJ1504 & 0.59$_{0.02}^{-0.02}$ \\
A85 & 0.51$_{0.04}^{-0.04}$     &      A2204 & 0.25$_{0.01}^{-0.01}$ \\ 
A478 & 0.43$_{0.1}^{-0.07}$     &      A2029 & 0.53$_{0.04}^{-0.04}$ \\ 
A1795 & 0.61$_{0.02}^{-0.02}$   &      A2142 & 1.94$_{0.16}^{-0.14}$\\
A4038 & 1.68$_{0.12}^{-0.11}$   &      A3562 & 5.15$_{0.72}^{-0.57}$ \\ 
A2052 & 0.51$_{0.02}^{-0.02}$   &      A401 & 8.81$_{1.41}^{-1.08}$ \\   
Hydra-A & 0.41$_{0.02}^{-0.02}$ &      A3558 & 1.69$_{1.15}^{-0.5}$ \\
A2063 & 2.36$_{0.14}^{-0.13}$   &      A2147 & 17.04$_{3.64}^{-2.72}$ \\
A3158 & 8.22$_{0.54}^{-0.47}$   &      A3266 & 7.62$_{2.63}^{-1.6}$ \\
MKW3s & 0.86$_{0.18}^{-0.13}$   &      A119 & 14.03$_{5.95}^{-3.43}$ \\
EXO0422 & 0.47$_{0.07}^{-0.05}$ &      A1644 & 0.84$_{0.32}^{-0.18}$ \\
A4059 & 0.7$_{0.07}^{-0.06}$    &      A1736 & 16.59$_{13.17}^{-5.52}$ \\
A2589 & 1.18$_{0.33}^{-0.22}$   &      A3395 & 12.66$_{3.04}^{-2.18}$ \\
A3112 & 0.37$_{0.08}^{-0.05}$   &      A2065 & 1.34$_{0.37}^{-0.24}$ \\
A1651 & 3.63$_{0.43}^{-0.37}$   &      A3667 & 6.14$_{0.52}^{-0.45}$ \\
A576 & 3.62$_{0.84}^{-0.59}$    &      A754 & 9.53$_{2.37}^{-1.64}$ \\
A2657 & 2.68$_{1.2}^{-0.66}$    &      A2256 & 11.56$_{2.43}^{-1.81}$ \\
A3391 & 12.46$_{2.49}^{-1.89}$  &      A399 & 12.13$_{1.44}^{-1.22}$ \\
A1650 & 1.25$_{0.29}^{-0.2}$    &      A2163 & 9.65$_{0.73}^{-0.78}$ \\
S 1101 & 0.88$_{0.2}^{-0.14}$    &      A3376 & 16.47$_{3.1}^{-2.35}$ \\  
ZwCl1215 & 10.99$_{2.09}^{-1.61}$ &                                  \\    
                                                                           
\hline
\label{cool_list}
\end{tabular}
\end{table*}

\par Of the relaxed systems (as identified by the V09), A3158, A3391, ZwCl1215, A3562 and A401 have relatively high cooling time values. These systems were classified as non-relaxed clusters by our combination of morphology parameters. In the non-relaxed sample (again from the V09), only three clusters - A3558, A1644, and A2065 have relatively short cooling times. The short cooling times may suggest that the cores of these clusters might not yet be disturbed by cluster merger. \cite {2007A&A...463..839R} analysed the {\it Chandra} and {\it XMM} observations of A3558 and found that its cool core had survived a merger. The {\it Chandra} observations show bright cluster nuclei, which also supports this idea. \cite {2006ApJ...643..751C} showed that A2065 is an unequal mass merger, which could be a reason behind the survival of one of its cool cores. According to the classical definition of a cooling flow, $t_{cool}$ $<$ $t_{age}$, where $t_{age}$ (age of galaxy cluster) $\sim$ 10 Gyr. We found the mean $t_{cool}$ value for strong relaxed clusters to be 0.44 Gyr, for relaxed clusters 0.64 Gyr, for non-relaxed clusters 4.72 Gyr, and for strong non-relaxed clusters 12.5 Gyr. This implies that the cooling mechanism is completely disturbed in strong non-relaxed clusters, while only $\sim$ 17\% are completely disturbed among non-relaxed clusters (five non-relaxed clusters have $t_{cool}$ $>$ $t_{age}$). Unfortunately, we did not have cooling time information for the high-$z$ clusters in our sample.

\par The correlation of morphology parameters with cooling time is plotted in Fig.~\ref{parvscool} for the Concentration, Gini and $M_{20}$, respectively, on a log-log scale. Figs.~\ref{parvscool}a, \ref{parvscool}b and \ref{parvscool}c show that two of our parameters, the Concentration and Gini, are anti-correlated, while the $M_{20}$ is correlated with the cooling time of clusters. This indicates the possibility that surface brightness imaging data could be useful in deriving the cooling time information of the central intracluster gas using simple morphology parameters. Table \ref{cool_spear} lists the Spearman coefficient, $\rho$, between the Concentration, Gini and $M_{20}$ parameters and cluster cooling time.

\begin{table*}
\centering
\caption{Spearman coefficient, $\rho$, between morphology parameters and cluster cooling time. }
\label{cool_spear}
\begin{tabular}{cc}
\hline 
\hline
 Morphology parameters &  Cooling time (Gyr) \\ 
\hline
Concentration  & -0.82 \\ 
Gini &  -0.71 \\ 
$M_{20}$ & 0.83 \\
\hline 
\end{tabular}        

\end{table*}

\begin{figure*}
%\hspace*{-1.2in}
\centering
{\label{concvscool}\includegraphics[scale=0.4]{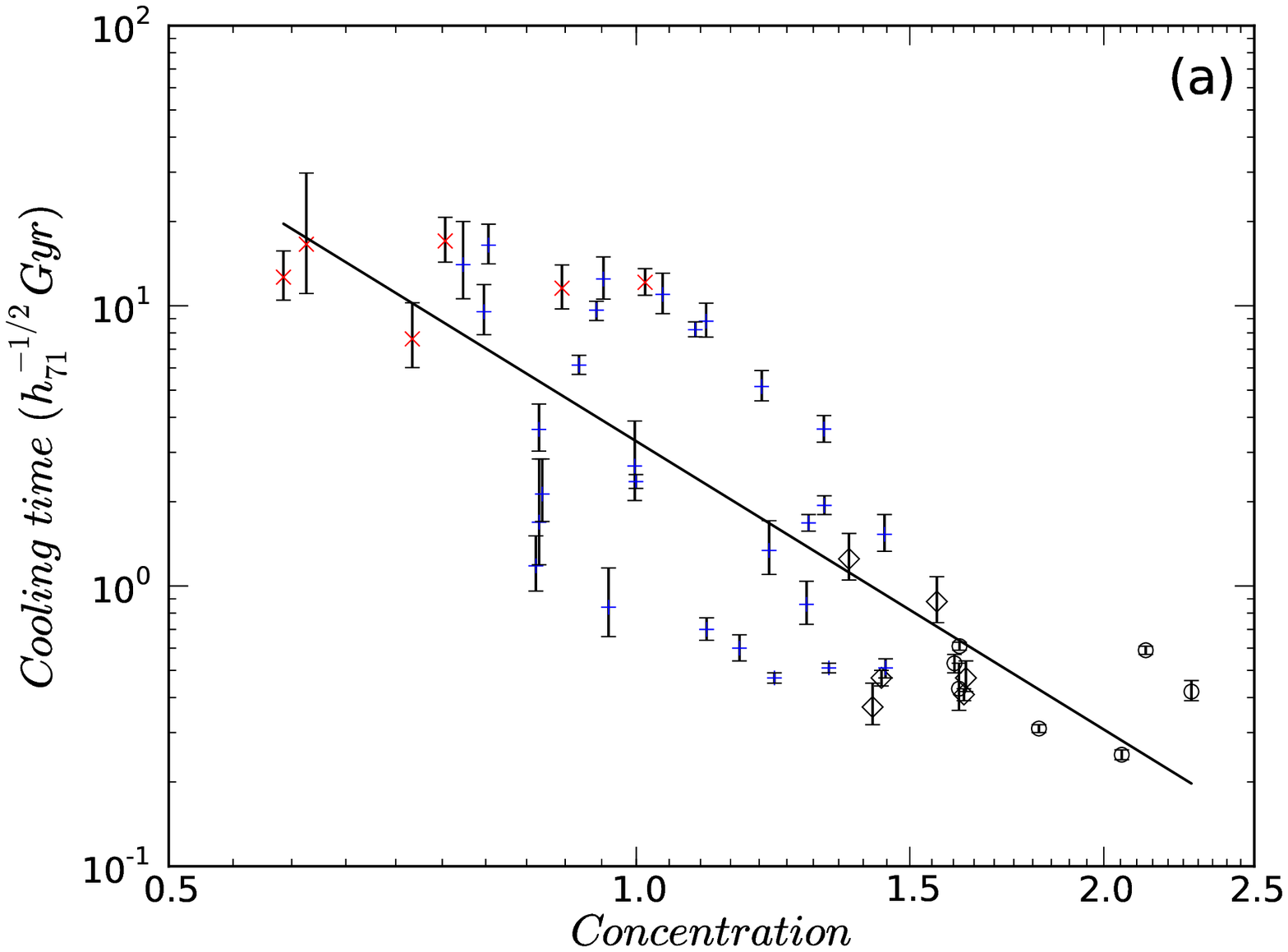}}
{\label{ginivscool}\includegraphics[scale=0.4]{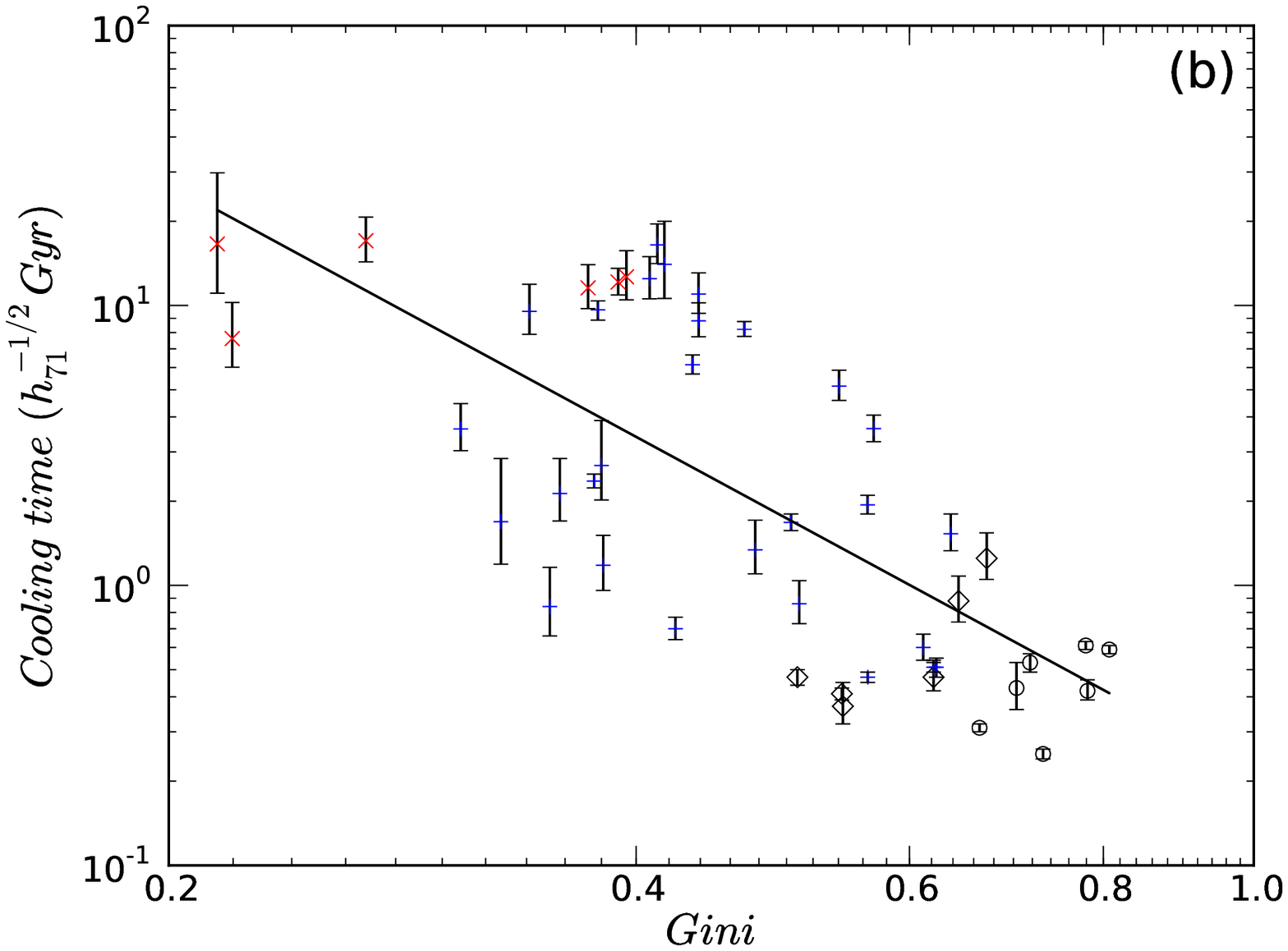}}
{\label{M20vscool}\includegraphics[scale=0.4]{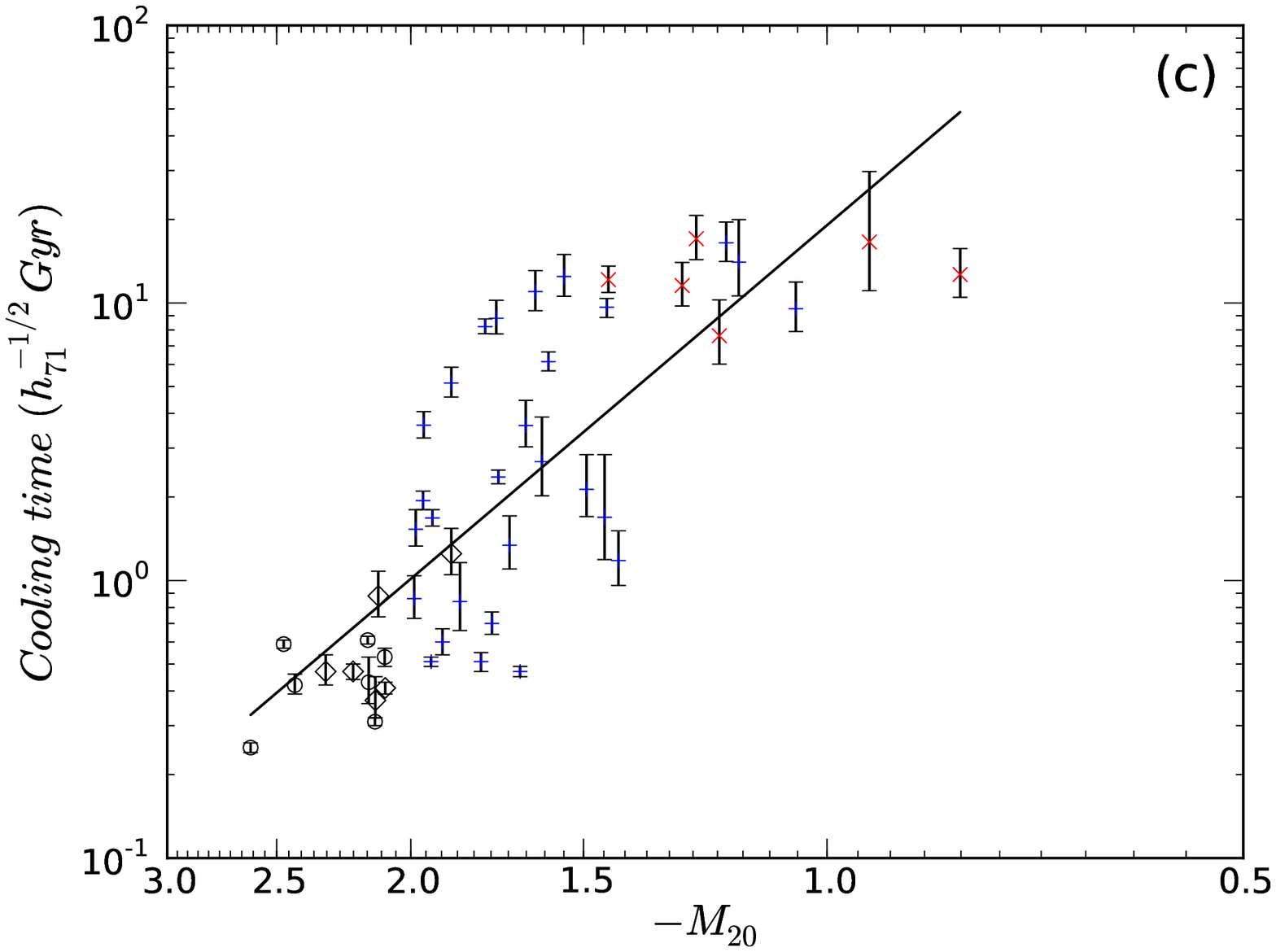}}

\caption{(a) Concentration ($C_{5080}$) parameter negatively correlated with cooling time for fixed radii. (b) Gini coefficient negatively correlated with cooling time. (c) $M_{20}$ correlated with cooling time. $\circ$ = strong relaxed clusters; {\small $\lozenge$} = relaxed clusters; {\small \color {blue} $+$} = non-relaxed clusters; and {\small \color {red} $\times$} = strong non-relaxed clusters. We used the power-law fitting to show the linear correlation.}
\label{parvscool}
\end{figure*}

\par In order to investigate linear fitting, we simply used the power-law model to establish a relationship between the morphology parameters and $t_{cool}$ (Fig.~\ref{parvscool}). For all three parameters, compositely, we constituted $t_{cool}$ $\propto$ Conc$^{-3.42\pm0.55}$, $t_{cool}$ $\propto$ Gini$^{-3.0\pm0.35}$ and $t_{cool}$ $\propto$ $M_{20}^{4.23\pm0.32}$.

\subsection{{\it Radio halo} cluster dynamical states}
\label{radio_halo_sec}
\par {Previous studies have shown the importance of joint X-ray and radio data to study the origin of non-thermal radio emission from galaxy clusters, in the form of a cluster wide {\it radio halo} which is generally situated at the cluster center (\citealt {2012A&ARv..20...54F}, and references therein). Current results indicate the presence of diffuse intracluster radio sources only in dynamically disturbed clusters and it is expected that future radio-surveys will reveal diffuse radio emission from a large fraction of major and minor mergers \citep{2010A&A...509A..68C}. }

\par {The relations between cluster properties derived from X-ray observations (luminosity ($L_{x}$), temperature (T) and mass (M)) and {\it radio halo} luminosity have been widely studied in the last few years \citep{2007MNRAS.378.1565C, 2009A&A...507.1257G,2009A&A...507..661B, 2011MmSAI..82..499V,2011JApA...32..519C}. They all show a strong correlation between {\it radio halo} and X-ray emission in galaxy clusters. \cite {2001ApJ...553L..15B}, \cite {2001A&A...378..408S} and \cite {2010ApJ...721L..82C} showed a relation between non-thermal radio sources and X-ray cluster morphology. \cite {2001ApJ...553L..15B} noted the linear relation between radio power ($P_{1.4GHz}$) and power ratio ($P_{1}$/$P_{0}$) \citep {1995ApJ...452..522B, 1996ApJ...458...27B} for {\it ROSAT} observed X-ray clusters. He concluded that approximately $P_{1.4GHz}$ $\propto$ $P_{1}$/$P_{0}$, which means the clusters that host the powerful {\it radio halos} are experiencing the largest departures from a virialized state. Recently, \cite {2010ApJ...721L..82C} used three parameters, namely centroid shift \citep {1993ApJ...413..492M, 2006MNRAS.373..881P, 2008ApJS..174..117M, 2010A&A...514A..32B}, third order power ratio ($P_{3}$/$P_{0}$) and Concentration \citep {2008A&A...483...35S} to demonstrate a relation between cluster mergers and the presence of a {\it radio halo}. }
  
\begin{table*}
\caption{{\it Radio halo} sample clusters. \small (1) cluster name; (2) redshift; (3) conversion factor (angular size to linear size); (4) radio flux density (1.4 GHz); (5) error in estimated radio flux density; (6) total radio power (1.4 GHz); (7) radio largest linear size; (8) total X-ray luminosity (0.1--2.4 keV); (9) Temperature (keV); (9) Exposure time; (10) References. {We normalised $H$ dependent quantities (LLS and $L_{X}$) with $H_{0}$=73 km s$^{-1}$ Mpc$^{-1}$.}}
\begin{center}
\begin{small}
%\hspace*{-1.0in}
\begin{tabular}{cccccccccccccc}

\hline 
\hline
Name&$z$&kpc/\begin{math}''\end{math}&S (1.4)& $\Delta$S & log P(1.4)&LLS&Luminosity & Temperature &Exposure time & Ref.\\
& & & mJy &mJy&W/Hz&Mpc&(10$^{44}$ erg/s)& (keV)& ks &      \\
    \hline  	

A1914&              0.1712 & 2.80 & 64.0 &  3.0 & 24.04 &   1.01 &  9.86 &  10.50 & 19  & 1,2\\
A2218&              0.1756 & 2.86 & 4.7  &  0.1 & 22.96 &   0.37 &  5.46  &   6.70 &  59 & 1,2 \\
A665&               0.1819 & 2.95 & 43.1 &  2.2 & 23.92 &   1.77 &  9.13  &   8.30 &  39 & 1,2 \\
A520&               0.1990 & 3.16 & 34.4 &  1.5 & 23.91 &   1.08 &  7.85  &   7.40 &  9 & 1,2 \\
A773&               0.2170 & 3.38 & 12.7 &  1.3 & 23.57 &   1.21 &  7.52  &   8.53 &   20  & 1,2\\
IE0657-56&         0.2960 & 4.26 & 78.0 &  5.0 & 24.64 &   2.0  &  21.37 &   11.64& 84  & 1,2\\
A2255&              0.0806 & 1.46 & 56.0 &  3.0 & 23.28 &   0.8 &  2.50  &   6.42 &  39  & 1,2  \\
A2319&              0.0557 & 1.04 & 153.0&  8.0 & 23.38 &   1.0 &  8.00  &   9.49 &  14  & 1,2 \\
A754&               0.0542 & 1.01 & 86.0 &  4.0 & 23.12 &   0.96 &  2.10  &   9.94 &  44   & 1,2\\
A2256&              0.0581 & 1.08 & 103.4&  1.1 & 23.26 &   0.79 &  3.55  &   6.90 &  12   & 1,2\\
A401&               0.0737 & 1.34 & 17.0 &  1.0 & 22.70 &   0.50 &  6.17  &   8.07 &  18  & 1,2\\
A3562&              0.0490 & 0.92 & 20.0 &  2.0 & 22.41 &   0.27 &  1.48  &   3.80 &  19   & 1,2\\
A399&               0.0718 & 1.31 & 16.0 &  0   & 22.67 &   0.55 &  3.60  &   5.80 &  48 & 3,2 \\
A2163&              0.2030 & 3.22 & 155.0&  2.0 & 24.57 &   2.21 &  21.50 &   12.12&  71  & 1,2\\
A1300&              0.3072 & 4.37 & 20.0 &  2.0 & 24.10 &   1.26  & 13.0  &   9.2  &   14 & 1,4 \\
A1758&              0.2790 & 4.08 & 16.7 &  0.8 & 23.93 &   1.47 & 6.70   &   7.95 &  7   & 1,2\\
A1995&              0.3186 & 4.48 & 4.1  &  0.7 & 23.47 &   0.80 & 8.35   &   8.60 &  56  & 1,2 \\
A2034&              0.1130 & 1.97 & 13.6 &  1.0 & 23.00 &   0.59 & 3.60   &   7.15 &  195  & 1,2  \\
A209&               0.2060 & 3.25 & 16.9 &  1.0 & 23.65 &   1.36 & 5.84   &   8.28 &  20   & 1,2  \\
A2219&              0.2256 & 3.48 & 81.0 &  4.0 & 24.40 &   1.67 & 11.53  &   9.81 &  42  & 1,2 \\
A2294&              0.1780 & 2.90 & 5.8  &  0.5 & 23.06 &   0.52 & 3.70   &   7.10 &  10  & 1,2 \\
A2744&              0.3080 & 4.38 & 57.1 &  2.9 & 24.55 &   1.84 & 12.16  &   9.61 &  24  & 1,2 \\
A521&               0.2533 & 3.80 & 5.9  &  0.5 & 23.40 &   1.14 & 8.01   &   6.74 &  37  & 1,2 \\
A697&               0.2820 & 4.11 & 7.8  &  1.0 & 23.62 &   0.63 & 9.84  &   9.06 &  27  & 1,2 \\
RXCJ2003.5-2323&    0.3171 & 4.46 & 35.0 &  2.0 & 24.40 &   1.36 & 8.63   &   9.1  & 50   & 1,5\\

\hline

\end{tabular}
	\label{tab:halo/relic}
\end{small}
\end{center}
{\small References: 1 = \cite[and references therein]{2009A&A...507.1257G}; 2 =  \cite{2009ApJS..182...12C}; 3 = \cite[and references therein]{2012A&ARv..20...54F}; 4 = \cite{2012MNRAS.420.2480Z}; 5 = \cite{2009A&A...505...45G}. }
\end{table*}  

\begin{figure*}
\centering
\includegraphics[scale=0.6]{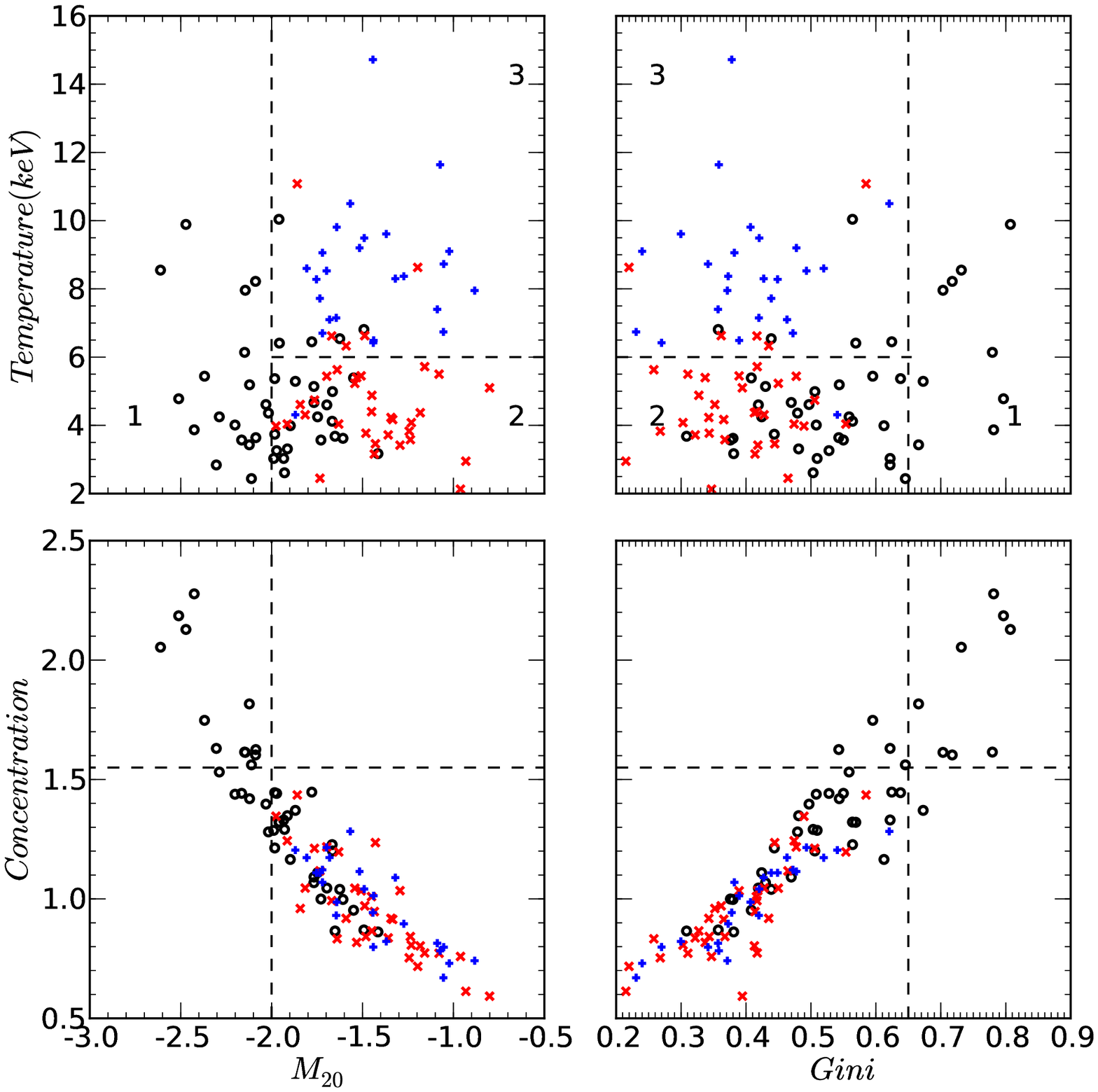}
\caption{Bottom left and right plots show the $M_{20}$ and Gini vs Concentration, respectively. Top left and right plots show the $M_{20}$ and Gini vs temperature, respectively. {$\circ$ = relaxed cluster; and {\small \color {red} $\times$} = non-relaxed cluster}. {\it Radio halo} clusters are identified with a {\small \color{blue} ``+''} symbol. Galaxy cluster separation is based on the V09.}
\label{halopara}
\end{figure*}

\par Since we have shown that {our morphology parameters are useful} to characterise the dynamical state of galaxy clusters, we investigate any possible correlation between our set of parameters and the presence of diffuse intracluster radio emission. We take 25 halo clusters from \cite {2009A&A...507.1257G} (as listed in Table \ref{tab:halo/relic}) for which some of them are already present in the V09 cluster sample (A754, A2256, A401, A3562, A399, and A2163). We reduced the X-ray data of the sample of \cite {2009A&A...507.1257G} in a similar way to that described in \S\ \ref{chan_data_red}. We subsequently calculated the morphology parameters (Gini, $M_{20}$ and Concentration) for each cluster (see Table \ref{halo_par_val_vo9}).

\par In Fig.~\ref{halopara} we show that some parameters are useful for studying the dynamical state of {\it radio halo} clusters. In the bottom left and right plots, we can see that the {\it radio halo} clusters are separated from the relaxed clusters and overlap with non-relaxed clusters in $M_{20}$ and Gini vs. Concentration parameter planes, respectively (similar results were observed by \cite {2010ApJ...721L..82C}, \cite{2001ApJ...553L..15B}). In these plots, we can roughly separate relaxed and non-relaxed clusters based on $M_{20}$ and Concentration, where the region contained by $M_{20}$ $<$ -2.0 and Concentration $>$ 1.55 gives exclusively relaxed clusters. Based on the Gini and Concentration plot, all non-relaxed clusters have Gini $<$ 0.65 and Concentration $<$ 1.55. 
The upper left and right plots in Fig.~\ref{halopara} show $M_{20}$ vs. Temperature and Gini vs. Temperature, respectively. We subdivided the $M_{20}$ vs. Temperature plot into three regions: (1) the $M_{20}$ $<$ -2.0 region has all dynamically relaxed clusters, (2) the $M_{20}$ $>$ -2.0 and Temperature $<$ 6 keV region has {\it radio quiet} merger clusters, and (3) the $M_{20}$ $>$ -2.0 and Temperature $>$ 6 keV region has {\it radio loud} merger clusters. Similarly, the Gini vs. Temperature plot has three regions defined as (1) the Gini $>$ 0.65 region has only dynamically relaxed clusters, (2) the Gini $<$ 0.65 and Temperature $<$ 6 keV region has {\it radio quiet} merger clusters, and (3) the Gini $<$ 0.65 and Temperature $>$ 6 keV region has {\it radio loud} merger clusters.

\section{Discussion and Conclusion}
\par {In this work we used a set of morphological parameters, among which some have to date been applied only to quantifying galaxy morphologies, to obtain constraints on the dynamical state of clusters. These parameters do not depend on any model parameter fits, such as the Beta model or power-law, and can therefore be applied to both disturbed and regular clusters. In principle, the Gini, Concentration and $M_{20}$ are promising for the detection of disturbances in clusters at any scale. In order to measure the Concentration, it was necessary to fix the cluster centre position (either optical position of the BCG, X-ray peak or barycentre) and then define the inner and outer regions around the chosen centre. This is not always possible in many of the high-$z$ ($\sim$ 1) clusters due to low photon counts and distorted cluster morphology. In this work, we showed a tight correlation between the Concentration and the Gini coefficient for both the low- and high-$z$ clusters. This suggested that the Gini coefficient could be used as a replacement for the Concentration mainly for high-$z$ clusters. The chief benefit of using the Gini coefficient is that we do not need a well-defined cluster centre. It works well for low photon counts as well as accurately quantifying substructure. It is completely independent of cluster shape, it does not depend on any symmetry underlying the cluster, and is independent of the location of projected (bright) pixels (whether they be at the centre or the edge) in the given aperture. }

\par The Gini coefficient was also correlated with the $M_{20}$, which could suggest that, in a relaxed system, 20$\%$ of the X-ray flux is centrally located where the electron density is high, while this is not possible for distorted and non-relaxed clusters where mergers may change the cluster density distribution. Most relaxed and cool-core clusters have Gini $>$ 0.65, which means that most of the flux comes from only a small number of bright pixels, mainly found at the cluster centre. Weak cool-core clusters were found between 0.4 $<$ Gini $<$ 0.65, and most disturbed clusters had Gini $<$ 0.4. This is obvious because, in non-relaxed clusters, the X-ray emission (or bright pixels) is not compact but equally distributed in the given aperture radius which gives a low Gini value.

{
\par We found that, unlike the Gini, $M_{20}$ and Concentration parameters, the Smoothness and Asymmetry parameters did not show any reliable signs of being able to classify galaxy clusters based on their dynamical state. This was unexpected as these parameters trace the 2-D structure and should be sensitive to a range of substructure types. In addition, \cite {2012arXiv1211.7040R} found that the Smoothness $\&$ Asymmetry were promising as indicators to probe the presence of substructure. \cite {2012arXiv1211.7040R}, however, tested their parameters on simulated cluster data and all the simulations had a uniform exposure time. We found that the value of these two parameters largely depends on cluster exposure time or S/N ratio. We have also shown that, in clusters sample that have heterogeneous exposure times, as in our situation, the Asymmetry and Smoothness are not reliable parameters. Our sample does not have uniform exposure time for all clusters, which largely affects S/N and photon counts between clusters.

}
\par We did not find any relationship between the morphology parameters and X-ray global properties of galaxy clusters. This is most likely due to the fact that morphology parameters are estimated without regard for the mass of clusters, to which other quantities, such as X-ray luminosity and temperature, are instead related. The Gini, Concentration and $M_{20}$ appear to be correlated with the cooling time of clusters. X-ray imaging and surface brightness maps are therefore useful in investigating the cooling time of relaxed clusters.

\par The morphology parameters studied in this paper were robust in various physical or observational conditions (high redshift, low exposure, etc). They helped us to investigate, with sufficient accuracy, cluster morphology and dynamical state. We noticed, however, that the Gini, Gini of the second order moment and Asymmetry are fairly weak parameters at $<$ 5 ks exposure time. The Smoothness parameter systematically decreases with increasing exposure time and is therefore not suitable when comparing objects with a range of redshifts and/or exposure times. The Concentration and $M_{20}$ are reasonably robust at low exposure times. Systematics associated with redshift effects are low, but some caution is required in the interpretation of the results based on simulated clusters. We concluded that the Concentration and $M_{20}$ are more robust than the other parameters. {We also tested (Gini, $M_{20}$ and Concentration) parameters against different background level and found that these parameters are robust.}

\par We have taken 25 {\it radio loud} clusters, which are well-known non-relaxed clusters which host {\it radio halos}. Our morphology parameters are quite useful to study their dynamical activities. Based on Gini, $M_{20}$ and Concentration parameters and in agreement with previous results \citep{2010ApJ...721L..82C}, we find that only -- but not all -- merging systems host {\it radio halos}. We can approximately separate {\it radio halo} clusters from relaxed clusters in parameter space with Concentration $<$ 1.5, $M_{20}$ $>$ -2.0 and Gini $<$ 0.65. 

\par In conclusion, we have shown that the combination of Gini, Concentration and $M_{20}$ show good potential for identifying substructure and perturbed dynamical states. In future we will also investigate scaling relations between cluster mass and luminosity or temperature by dividing clusters based on parameter boundary values.

\begin{acknowledgements}
V. Parekh acknowledges financial support from the South African Square Kilometer Array Project. C. Ferrari acknowledges financial support by the ``{\it Agence Nationale de la Recherche}'' through grant ANR-09-JCJC-0001-01. We are grateful to Prof. Alexei Vikhlinin (CFA) for providing the {\it Chandra} archival data. G. Angus acknowledges support from the Claude-Leon foundation.
\end{acknowledgements} 

%\bibliography{references}
%\input{morphology_paper.bbl}

%\label{lastpage}

\newpage

\begin{appendix}
\section{}
\begin{landscape}
\begin{table}
\caption{Morphology parameters value of relaxed clusters (V09). Values are listed with appropriate uncertainties of 1$\sigma$ for each parameter.}
\begin{center}
\begin{small}
\begin{tabular}{ccccccccc}
\hline
\hline
Cluster name & Gini & $M_{20}$ & Concentration & Asymmetry & Smoothness & $G_{M}$ & Ellipticity \\
\hline
A3571  & 0.357 $\pm$ 0.002 & -1.492 $\pm$ 0.217 & 0.870 $\pm$ 0.091 & 0.293 $\pm$ 0.232 & 0.284 $\pm$ 0.004 & 0.259 $\pm$ 0.001 & 0.084 $\pm$ 0.043 \\
A2199  & 0.612 $\pm$ 0.001 & -1.897 $\pm$ 0.188 & 1.165 $\pm$ 0.097 & 0.502 $\pm$ 0.078 & 0.329 $\pm$ 0.005 & 0.378 $\pm$ 0.002 & 0.083 $\pm$ 0.037 \\
2A 0335  & 0.666 $\pm$ 0.001 & -2.123 $\pm$ 0.258 & 1.817 $\pm$ 0.237 & 0.341 $\pm$ 0.128 & 0.321 $\pm$ 0.008 & 0.209 $\pm$ 0.001 & 0.082 $\pm$ 0.041 \\
A496  & 0.564 $\pm$ 0.002 & -1.667 $\pm$ 0.320 & 1.227 $\pm$ 0.086 & 0.577 $\pm$ 0.081 & 0.205 $\pm$ 0.005 & 0.284 $\pm$ 0.001 & 0.063 $\pm$ 0.036 \\
A3667  & 0.435 $\pm$ 0.001 & -1.590 $\pm$ 0.109 & 0.918 $\pm$ 0.087 & 0.302 $\pm$ 0.132 & 0.536 $\pm$ 0.002 & 0.285 $\pm$ 0.001 & 0.079 $\pm$ 0.053 \\
A754  & 0.341 $\pm$ 0.001 & -1.053 $\pm$ 0.076 & 0.798 $\pm$ 0.052 & 0.406 $\pm$ 0.058 & 0.524 $\pm$ 0.003 & 0.308 $\pm$ 0.001 & 0.202 $\pm$ 0.006 \\
A85  & 0.624 $\pm$ 0.001 & -1.779 $\pm$ 0.260 & 1.447 $\pm$ 0.067 & 0.633 $\pm$ 0.117 & 0.464 $\pm$ 0.003 & 0.279 $\pm$ 0.001 & 0.026 $\pm$ 0.006 \\
A2029  & 0.718 $\pm$ 0.002 & -2.088 $\pm$ 0.315 & 1.603 $\pm$ 0.192 & 0.445 $\pm$ 0.165 & 0.163 $\pm$ 0.009 & 0.280 $\pm$ 0.001 & 0.113 $\pm$ 0.023 \\
A478  & 0.703 $\pm$ 0.003 & -2.145 $\pm$ 0.255 & 1.613 $\pm$ 0.162 & 0.286 $\pm$ 0.165 & 0.248 $\pm$ 0.008 & 0.279 $\pm$ 0.001 & 0.137 $\pm$ 0.062 \\
A1795  & 0.779 $\pm$ 0.002 & -2.148 $\pm$ 0.296 & 1.615 $\pm$ 0.097 & 0.351 $\pm$ 0.053 & 0.143 $\pm$ 0.012 & 0.389 $\pm$ 0.001 & 0.107 $\pm$ 0.029 \\
A3558  & 0.327 $\pm$ 0.001 & -1.448 $\pm$ 0.226 & 0.866 $\pm$ 0.087 & 0.358 $\pm$ 0.153 & 0.520 $\pm$ 0.002 & 0.276 $\pm$ 0.003 & 0.117 $\pm$ 0.068 \\
A2142  & 0.564 $\pm$ 0.002 & -1.959 $\pm$ 0.218 & 1.322 $\pm$ 0.199 & 0.365 $\pm$ 0.263 & 0.329 $\pm$ 0.003 & 0.239 $\pm$ 0.000 & 0.174 $\pm$ 0.044 \\
A2256  & 0.373 $\pm$ 0.000 & -1.273 $\pm$ 0.064 & 0.896 $\pm$ 0.047 & 0.333 $\pm$ 0.119 & 1.024 $\pm$ 0.002 & 0.277 $\pm$ 0.000 & 0.133 $\pm$ 0.041 \\
A4038  & 0.503 $\pm$ 0.002 & -1.929 $\pm$ 0.193 & 1.291 $\pm$ 0.181 & 0.297 $\pm$ 0.191 & 0.563 $\pm$ 0.001 & 0.206 $\pm$ 0.000 & 0.140 $\pm$ 0.038 \\
A2147  & 0.268 $\pm$ 0.001 & -1.243 $\pm$ 0.102 & 0.753 $\pm$ 0.178 & 0.359 $\pm$ 0.150 & 1.188 $\pm$ 0.002 & 0.305 $\pm$ 0.000 & 0.083 $\pm$ 0.057 \\
A3266  & 0.220 $\pm$ 0.001 & -1.196 $\pm$ 0.232 & 0.717 $\pm$ 0.172 & 0.276 $\pm$ 0.256 & 0.514 $\pm$ 0.005 & 0.279 $\pm$ 0.002 & 0.103 $\pm$ 0.069 \\
A401  & 0.439 $\pm$ 0.000 & -1.734 $\pm$ 0.121 & 1.109 $\pm$ 0.122 & 0.222 $\pm$ 0.207 & 0.739 $\pm$ 0.002 & 0.205 $\pm$ 0.000 & 0.102 $\pm$ 0.040 \\
A2052  & 0.622 $\pm$ 0.004 & -1.934 $\pm$ 0.237 & 1.330 $\pm$ 0.195 & 0.649 $\pm$ 0.035 & 0.308 $\pm$ 0.003 & 0.332 $\pm$ 0.003 & 0.086 $\pm$ 0.046 \\
Hydra-A  & 0.543 $\pm$ 0.004 & -2.087 $\pm$ 0.383 & 1.625 $\pm$ 0.248 & 0.258 $\pm$ 0.293 & 0.152 $\pm$ 0.013 & 0.149 $\pm$ 0.002 & 0.060 $\pm$ 0.065 \\
A119  & 0.417 $\pm$ 0.000 & -1.158 $\pm$ 0.066 & 0.774 $\pm$ 0.085 & 0.574 $\pm$ 0.022 & 1.542 $\pm$ 0.001 & 0.434 $\pm$ 0.000 & 0.063 $\pm$ 0.057 \\
A2063  & 0.376 $\pm$ 0.001 & -1.729 $\pm$ 0.169 & 0.999 $\pm$ 0.156 & 0.403 $\pm$ 0.132 & 1.156 $\pm$ 0.001 & 0.281 $\pm$ 0.001 & 0.046 $\pm$ 0.057 \\
A1644  & 0.352 $\pm$ 0.003 & -1.843 $\pm$ 0.183 & 0.960 $\pm$ 0.286 & 0.381 $\pm$ 0.282 & 0.843 $\pm$ 0.003 & 0.273 $\pm$ 0.001 & 0.043 $\pm$ 0.067 \\
A3158  & 0.470 $\pm$ 0.001 & -1.767 $\pm$ 0.104 & 1.091 $\pm$ 0.138 & 0.267 $\pm$ 0.221 & 0.788 $\pm$ 0.001 & 0.232 $\pm$ 0.000 & 0.118 $\pm$ 0.043 \\
MKW3s  & 0.509 $\pm$ 0.003 & -1.988 $\pm$ 0.273 & 1.287 $\pm$ 0.193 & 0.222 $\pm$ 0.338 & 0.317 $\pm$ 0.006 & 0.173 $\pm$ 0.001 & 0.134 $\pm$ 0.056 \\
A1736  & 0.215 $\pm$ 0.001 & -0.932 $\pm$ 0.101 & 0.613 $\pm$ 0.208 & 0.415 $\pm$ 0.109 & 1.379 $\pm$ 0.001 & 0.352 $\pm$ 0.001 & 0.011 $\pm$ 0.080 \\
EXO0422  & 0.622 $\pm$ 0.002 & -2.304 $\pm$ 0.228 & 1.630 $\pm$ 0.268 & 0.355 $\pm$ 0.183 & 0.911 $\pm$ 0.005 & 0.254 $\pm$ 0.000 & 0.071 $\pm$ 0.071 \\
A4059  & 0.424 $\pm$ 0.002 & -1.747 $\pm$ 0.267 & 1.110 $\pm$ 0.107 & 0.211 $\pm$ 0.170 & 0.135 $\pm$ 0.005 & 0.167 $\pm$ 0.001 & 0.093 $\pm$ 0.044 \\
A3395  & 0.394 $\pm$ 0.001 & -0.801 $\pm$ 0.098 & 0.593 $\pm$ 0.093 & 0.734 $\pm$ 0.071 & 1.411 $\pm$ 0.001 & 0.473 $\pm$ 0.001 & 0.228 $\pm$ 0.090 \\
A2589  & 0.381 $\pm$ 0.001 & -1.415 $\pm$ 0.230 & 0.862 $\pm$ 0.166 & 0.471 $\pm$ 0.055 & 0.669 $\pm$ 0.003 & 0.317 $\pm$ 0.001 & 0.045 $\pm$ 0.037 \\
A3112  & 0.544 $\pm$ 0.004 & -2.121 $\pm$ 0.406 & 1.420 $\pm$ 0.204 & 0.276 $\pm$ 0.340 & 0.309 $\pm$ 0.013 & 0.170 $\pm$ 0.004 & 0.143 $\pm$ 0.072 \\
A3562  & 0.541 $\pm$ 0.001 & -1.870 $\pm$ 0.099 & 1.204 $\pm$ 0.175 & 0.542 $\pm$ 0.019 & 1.211 $\pm$ 0.001 & 0.384 $\pm$ 0.000 & 0.107 $\pm$ 0.051 \\
A1651  & 0.569 $\pm$ 0.001 & -1.957 $\pm$ 0.207 & 1.321 $\pm$ 0.180 & 0.354 $\pm$ 0.228 & 1.001 $\pm$ 0.002 & 0.266 $\pm$ 0.000 & 0.075 $\pm$ 0.049 \\
A399  & 0.389 $\pm$ 0.001 & -1.439 $\pm$ 0.178 & 1.013 $\pm$ 0.169 & 0.372 $\pm$ 0.137 & 0.698 $\pm$ 0.001 & 0.229 $\pm$ 0.001 & 0.080 $\pm$ 0.079 \\
A2204  & 0.732 $\pm$ 0.005 & -2.611 $\pm$ 0.580 & 2.054 $\pm$ 0.485 & 0.361 $\pm$ 0.159 & 0.545 $\pm$ 0.033 & 0.208 $\pm$ 0.002 & 0.051 $\pm$ 0.062 \\
A576  & 0.308 $\pm$ 0.001 & -1.651 $\pm$ 0.162 & 0.866 $\pm$ 0.147 & 0.267 $\pm$ 0.064 & 0.606 $\pm$ 0.002 & 0.251 $\pm$ 0.001 & 0.050 $\pm$ 0.068 \\
A2657  & 0.380 $\pm$ 0.001 & -1.607 $\pm$ 0.194 & 0.998 $\pm$ 0.222 & 0.402 $\pm$ 0.157 & 1.034 $\pm$ 0.003 & 0.250 $\pm$ 0.001 & 0.048 $\pm$ 0.051 \\
A3391  & 0.408 $\pm$ 0.001 & -1.549 $\pm$ 0.079 & 0.952 $\pm$ 0.123 & 0.431 $\pm$ 0.096 & 1.335 $\pm$ 0.001 & 0.332 $\pm$ 0.000 & 0.115 $\pm$ 0.051 \\
A2065  & 0.477 $\pm$ 0.001 & -1.697 $\pm$ 0.166 & 1.218 $\pm$ 0.091 & 0.402 $\pm$ 0.155 & 0.549 $\pm$ 0.003 & 0.230 $\pm$ 0.001 & 0.149 $\pm$ 0.049 \\
A1650  & 0.673 $\pm$ 0.001 & -1.869 $\pm$ 0.318 & 1.371 $\pm$ 0.165 & 0.429 $\pm$ 0.087 & 0.476 $\pm$ 0.004 & 0.379 $\pm$ 0.002 & 0.122 $\pm$ 0.093 \\
A3822  & 0.450 $\pm$ 0.001 & -1.542 $\pm$ 0.146 & 1.045 $\pm$ 0.197 & 0.536 $\pm$ 0.071 & 1.450 $\pm$ 0.002 & 0.370 $\pm$ 0.000 & 0.119 $\pm$ 0.066 \\
S 1101  & 0.645 $\pm$ 0.002 & -2.111 $\pm$ 0.214 & 1.561 $\pm$ 0.241 & 0.238 $\pm$ 0.199 & 0.461 $\pm$ 0.011 & 0.172 $\pm$ 0.002 & 0.110 $\pm$ 0.062 \\
A2163  & 0.378 $\pm$ 0.002 & -1.443 $\pm$ 0.179 & 0.943 $\pm$ 0.168 & 0.281 $\pm$ 0.158 & 0.272 $\pm$ 0.004 & 0.196 $\pm$ 0.002 & 0.082 $\pm$ 0.061 \\
ZwCl1215  & 0.439 $\pm$ 0.001 & -1.625 $\pm$ 0.109 & 1.040 $\pm$ 0.103 & 0.369 $\pm$ 0.117 & 1.204 $\pm$ 0.002 & 0.294 $\pm$ 0.000 & 0.120 $\pm$ 0.086 \\
RXJ1504  & 0.807 $\pm$ 0.004 & -2.471 $\pm$ 0.400 & 2.128 $\pm$ 0.322 & 0.397 $\pm$ 0.188 & 0.370 $\pm$ 0.038 & 0.242 $\pm$ 0.003 & 0.124 $\pm$ 0.067 \\
A2597  & 0.781 $\pm$ 0.003 & -2.426 $\pm$ 0.558 & 2.277 $\pm$ 0.389 & 0.397 $\pm$ 0.266 & 0.295 $\pm$ 0.010 & 0.277 $\pm$ 0.001 & 0.152 $\pm$ 0.077 \\
A133  & 0.508 $\pm$ 0.003 & -2.201 $\pm$ 0.396 & 1.438 $\pm$ 0.301 & 0.254 $\pm$ 0.292 & 0.320 $\pm$ 0.008 & 0.179 $\pm$ 0.002 & 0.094 $\pm$ 0.079 \\
A2244  & 0.638 $\pm$ 0.001 & -1.983 $\pm$ 0.287 & 1.445 $\pm$ 0.232 & 0.266 $\pm$ 0.151 & 0.314 $\pm$ 0.003 & 0.219 $\pm$ 0.005 & 0.030 $\pm$ 0.059 \\
A3376  & 0.413 $\pm$ 0.000 & -1.183 $\pm$ 0.063 & 0.803 $\pm$ 0.153 & 0.564 $\pm$ 0.045 & 1.159 $\pm$ 0.001 & 0.397 $\pm$ 0.001 & 0.072 $\pm$ 0.0869 \\
\hline
\end{tabular}
\label{low_par_val_vo9}
\end{small}
\end{center}
\end{table}
\end{landscape}

\begin{landscape}
\begin{table}
\caption{Morphology parameters value of (V09) non-relaxed clusters. Values are listed with appropriate uncertainties of 1$\sigma$ for each parameter.}
\begin{center}
\begin{small}
\begin{tabular}{ccccccccc}
\hline
\hline
Cluster name & Gini & $M_{20}$ & Concentration & Asymmetry & Smoothness & $G_{M}$ & Ellipticity \\
\hline
0302-0423  & 0.796 $\pm$ 0.010 & -2.511 $\pm$ 0.750 & 2.186 $\pm$ 0.769 & 0.322 $\pm$ 0.281 & 0.767 $\pm$ 0.027 & 0.345 $\pm$ 0.006 & 0.071 $\pm$ 0.231 \\
1212+2733  & 0.417 $\pm$ 0.004 & -1.670 $\pm$ 0.349 & 0.993 $\pm$ 0.553 & 0.388 $\pm$ 0.377 & 1.206 $\pm$ 0.003 & 0.293 $\pm$ 0.003 & 0.102 $\pm$ 0.212 \\
0350-3801  & 0.465 $\pm$ 0.003 & -1.734 $\pm$ 0.538 & 1.117 $\pm$ 1.046 & 0.591 $\pm$ 0.288 & 1.453 $\pm$ 0.000 & 0.386 $\pm$ 0.001 & 0.082 $\pm$ 0.000 \\
0318-0302  & 0.554 $\pm$ 0.003 & -1.631 $\pm$ 0.675 & 1.197 $\pm$ 0.900 & 0.532 $\pm$ 0.474 & 1.306 $\pm$ 0.002 & 0.339 $\pm$ 0.003 & 0.073 $\pm$ 0.000 \\
0159+0030  & 0.559 $\pm$ 0.008 & -2.288 $\pm$ 1.139 & 1.531 $\pm$ 0.628 & 0.597 $\pm$ 0.400 & 1.198 $\pm$ 0.002 & 0.334 $\pm$ 0.002 & 0.022 $\pm$ 0.000 \\
0958+4702  & 0.550 $\pm$ 0.007 & -2.165 $\pm$ 0.557 & 1.442 $\pm$ 1.034 & 0.416 $\pm$ 0.354 & 1.310 $\pm$ 0.000 & 0.354 $\pm$ 0.001 & 0.015 $\pm$ 0.000 \\
0809+2811  & 0.366 $\pm$ 0.003 & -1.334 $\pm$ 0.738 & 0.914 $\pm$ 1.160 & 0.449 $\pm$ 0.370 & 1.291 $\pm$ 0.002 & 0.309 $\pm$ 0.003 & 0.115 $\pm$ 0.307 \\
1416+4446  & 0.528 $\pm$ 0.010 & -1.972 $\pm$ 0.466 & 1.442 $\pm$ 1.005 & 0.553 $\pm$ 0.472 & 1.009 $\pm$ 0.013 & 0.274 $\pm$ 0.003 & 0.103 $\pm$ 0.000 \\
1312+3900  & 0.322 $\pm$ 0.002 & -1.359 $\pm$ 0.279 & 0.837 $\pm$ 0.612 & 0.507 $\pm$ 0.453 & 1.443 $\pm$ 0.000 & 0.354 $\pm$ 0.001 & 0.030 $\pm$ 0.000 \\
1003+3253  & 0.595 $\pm$ 0.008 & -2.369 $\pm$ 0.779 & 1.748 $\pm$ 1.035 & 0.451 $\pm$ 0.259 & 1.166 $\pm$ 0.000 & 0.353 $\pm$ 0.002 & 0.100 $\pm$ 0.242 \\
0141-3034  & 0.347 $\pm$ 0.003 & -0.961 $\pm$ 0.170 & 0.759 $\pm$ 0.293 & 0.623 $\pm$ 0.303 & 1.591 $\pm$ 0.000 & 0.432 $\pm$ 0.002 & 0.048 $\pm$ 0.000 \\
1701+6414  & 0.479 $\pm$ 0.008 & -2.017 $\pm$ 0.817 & 1.281 $\pm$ 0.855 & 0.513 $\pm$ 0.384 & 0.878 $\pm$ 0.005 & 0.240 $\pm$ 0.002 & 0.096 $\pm$ 0.000 \\
1641+4001  & 0.481 $\pm$ 0.005 & -1.913 $\pm$ 0.483 & 1.348 $\pm$ 0.203 & 0.349 $\pm$ 0.384 & 1.166 $\pm$ 0.000 & 0.282 $\pm$ 0.002 & 0.042 $\pm$ 0.000 \\
0522-3624  & 0.444 $\pm$ 0.008 & -1.429 $\pm$ 0.356 & 1.236 $\pm$ 1.353 & 0.471 $\pm$ 0.415 & 1.013 $\pm$ 0.003 & 0.307 $\pm$ 0.002 & 0.087 $\pm$ 0.351 \\
1222+2709  & 0.444 $\pm$ 0.005 & -1.983 $\pm$ 0.241 & 1.213 $\pm$ 0.135 & 0.412 $\pm$ 0.299 & 1.200 $\pm$ 0.000 & 0.301 $\pm$ 0.001 & 0.036 $\pm$ 0.000 \\
0355-3741  & 0.497 $\pm$ 0.007 & -2.031 $\pm$ 0.239 & 1.397 $\pm$ 1.198 & 0.429 $\pm$ 0.357 & 1.182 $\pm$ 0.000 & 0.303 $\pm$ 0.002 & 0.058 $\pm$ 0.000 \\
0853+5759  & 0.418 $\pm$ 0.004 & -1.294 $\pm$ 0.432 & 1.034 $\pm$ 0.746 & 0.715 $\pm$ 0.361 & 1.075 $\pm$ 0.000 & 0.331 $\pm$ 0.004 & 0.030 $\pm$ 0.000 \\
0333-2456  & 0.414 $\pm$ 0.002 & -1.435 $\pm$ 0.218 & 0.947 $\pm$ 0.199 & 0.462 $\pm$ 0.315 & 1.363 $\pm$ 0.000 & 0.341 $\pm$ 0.001 & 0.117 $\pm$ 0.000 \\
0926+1242  & 0.506 $\pm$ 0.003 & -1.764 $\pm$ 0.584 & 1.212 $\pm$ 1.019 & 0.558 $\pm$ 0.336 & 1.420 $\pm$ 0.000 & 0.399 $\pm$ 0.002 & 0.158 $\pm$ 0.000 \\
0030+2618  & 0.258 $\pm$ 0.004 & -1.639 $\pm$ 0.336 & 0.833 $\pm$ 0.392 & 0.258 $\pm$ 0.502 & 0.578 $\pm$ 0.006 & 0.282 $\pm$ 0.002 & 0.045 $\pm$ 0.130 \\
1002+6858  & 0.474 $\pm$ 0.005 & -1.914 $\pm$ 0.625 & 1.243 $\pm$ 0.861 & 0.475 $\pm$ 0.217 & 1.388 $\pm$ 0.000 & 0.362 $\pm$ 0.001 & 0.065 $\pm$ 0.000 \\
1524+0957  & 0.343 $\pm$ 0.003 & -1.343 $\pm$ 0.380 & 0.918 $\pm$ 0.429 & 0.366 $\pm$ 0.495 & 1.065 $\pm$ 0.000 & 0.262 $\pm$ 0.005 & 0.102 $\pm$ 0.000 \\
1357+6232  & 0.419 $\pm$ 0.004 & -1.696 $\pm$ 0.400 & 1.045 $\pm$ 1.131 & 0.385 $\pm$ 0.394 & 0.945 $\pm$ 0.015 & 0.262 $\pm$ 0.003 & 0.019 $\pm$ 0.154 \\
1354-0221  & 0.343 $\pm$ 0.003 & -1.481 $\pm$ 0.344 & 0.842 $\pm$ 0.756 & 0.477 $\pm$ 0.333 & 0.977 $\pm$ 0.009 & 0.308 $\pm$ 0.001 & 0.028 $\pm$ 0.000 \\
1120+2326  & 0.367 $\pm$ 0.002 & -1.237 $\pm$ 0.424 & 0.842 $\pm$ 0.323 & 0.439 $\pm$ 0.448 & 0.964 $\pm$ 0.000 & 0.272 $\pm$ 0.002 & 0.087 $\pm$ 0.000 \\
0956+4107  & 0.416 $\pm$ 0.002 & -1.450 $\pm$ 0.251 & 1.010 $\pm$ 0.107 & 0.388 $\pm$ 0.365 & 1.156 $\pm$ 0.000 & 0.300 $\pm$ 0.002 & 0.147 $\pm$ 0.000 \\
0328-2140  & 0.430 $\pm$ 0.004 & -1.767 $\pm$ 0.689 & 1.068 $\pm$ 1.208 & 0.480 $\pm$ 0.360 & 0.944 $\pm$ 0.000 & 0.281 $\pm$ 0.003 & 0.073 $\pm$ 0.000 \\
1120+4318  & 0.506 $\pm$ 0.006 & -1.665 $\pm$ 0.428 & 1.201 $\pm$ 0.851 & 0.506 $\pm$ 0.429 & 1.149 $\pm$ 0.000 & 0.287 $\pm$ 0.006 & 0.061 $\pm$ 0.000 \\
1334+5031  & 0.428 $\pm$ 0.005 & -1.815 $\pm$ 0.319 & 1.045 $\pm$ 0.595 & 0.517 $\pm$ 0.322 & 1.350 $\pm$ 0.000 & 0.376 $\pm$ 0.001 & 0.068 $\pm$ 0.000 \\
0542-4100  & 0.389 $\pm$ 0.005 & -1.508 $\pm$ 0.429 & 1.034 $\pm$ 0.332 & 0.464 $\pm$ 0.499 & 0.935 $\pm$ 0.000 & 0.264 $\pm$ 0.004 & 0.099 $\pm$ 0.000 \\
1202+5751  & 0.303 $\pm$ 0.002 & -1.231 $\pm$ 0.326 & 0.807 $\pm$ 0.113 & 0.409 $\pm$ 0.440 & 1.191 $\pm$ 0.000 & 0.297 $\pm$ 0.002 & 0.106 $\pm$ 0.000 \\
0405-4100  & 0.489 $\pm$ 0.004 & -1.976 $\pm$ 0.381 & 1.346 $\pm$ 0.845 & 0.448 $\pm$ 0.369 & 0.903 $\pm$ 0.000 & 0.291 $\pm$ 0.002 & 0.014 $\pm$ 0.000 \\
1221+4918  & 0.362 $\pm$ 0.002 & -1.488 $\pm$ 0.550 & 0.971 $\pm$ 1.002 & 0.282 $\pm$ 0.444 & 0.811 $\pm$ 0.000 & 0.230 $\pm$ 0.003 & 0.091 $\pm$ 0.000 \\
0230+1836  & 0.310 $\pm$ 0.002 & -1.079 $\pm$ 0.588 & 0.772 $\pm$ 0.158 & 0.442 $\pm$ 0.459 & 1.093 $\pm$ 0.000 & 0.296 $\pm$ 0.002 & 0.130 $\pm$ 0.000 \\
0152-1358  & 0.337 $\pm$ 0.003 & -1.532 $\pm$ 0.268 & 0.818 $\pm$ 0.107 & 0.510 $\pm$ 0.378 & 1.183 $\pm$ 0.000 & 0.348 $\pm$ 0.003 & 0.169 $\pm$ 0.000 \\
1226+3332  & 0.585 $\pm$ 0.006 & -1.859 $\pm$ 0.929 & 1.435 $\pm$ 0.799 & 0.277 $\pm$ 0.151 & 0.411 $\pm$ 0.018 & 0.168 $\pm$ 0.007 & 0.035 $\pm$ 0.246 \\

\hline
\end{tabular}
\label{high_par_val_vo9}
\end{small}
\end{center}
\end{table}
\end{landscape}

\begin{table*}
\caption{Three morphology parameters value of {\it radio halo} clusters, except for the V09 {\it radio halo} clusters (A754, A2256, A401, A3562, A399, and A2163). Values are listed with appropriate uncertainties of 1$\sigma$ for each parameter.}
%\begin{small}
\centering
%\hspace*{0.8cm}
\begin{center}
\begin{tabular}{cccc}
\hline
\hline
Cluster name & Gini & $M_{20}$ & Concentration  \\
\hline
A1914  & 0.621 $\pm$ 0.002 & -1.567 $\pm$ 0.247 & 1.282 $\pm$ 0.266 \\
A2218  & 0.472 $\pm$ 0.001 & -1.720 $\pm$ 0.264 & 1.122 $\pm$ 0.156 \\
A665  & 0.427 $\pm$ 0.002 & -1.320 $\pm$ 0.269 & 1.09 $\pm$ 0.229 \\
A520  & 0.357 $\pm$ 0.001 & -1.090 $\pm$ 0.154 & 0.814 $\pm$ 0.322 \\
A773  & 0.493 $\pm$ 0.003 & -1.697 $\pm$ 0.249 & 1.215 $\pm$ 0.255 \\
IE 0657-56  & 0.358 $\pm$ 0.003 & -1.073 $\pm$ 0.195 & 0.783 $\pm$ 0.156 \\
A2255  & 0.270 $\pm$ 0.000 & -1.441 $\pm$ 0.110 & 0.798 $\pm$ 0.126 \\
A2319  & 0.420 $\pm$ 0.001 & -1.491 $\pm$ 0.141 & 1.041 $\pm$ 0.088 \\
A1300  & 0.477 $\pm$ 0.004 & -1.516 $\pm$ 0.412 & 1.114 $\pm$ 0.335 \\
A1758  & 0.371 $\pm$ 0.002 & -0.883 $\pm$ 0.184 & 0.741 $\pm$ 0.414 \\
A1995  & 0.520 $\pm$ 0.002 & -1.807 $\pm$ 0.286 & 1.172 $\pm$ 0.405 \\
A2034  & 0.420 $\pm$ 0.001 & -1.645 $\pm$ 0.149 & 0.930 $\pm$ 0.101 \\
A209  & 0.449 $\pm$ 0.002 & -1.754 $\pm$ 0.248 & 1.110 $\pm$ 0.208 \\
A2219  & 0.407 $\pm$ 0.002 & -1.642 $\pm$ 0.254 & 0.986 $\pm$ 0.125 \\
A2294  & 0.463 $\pm$ 0.002 & -1.681 $\pm$ 0.305 & 1.172 $\pm$ 0.420 \\
A2744  & 0.300 $\pm$ 0.002 & -1.370 $\pm$ 0.222 & 0.821 $\pm$ 0.271 \\
A521  & 0.231 $\pm$ 0.002 & -1.055 $\pm$ 0.176 & 0.670 $\pm$ 0.417 \\
A697  & 0.382 $\pm$ 0.002 & -1.720 $\pm$ 0.265 & 1.07 $\pm$ 0.186 \\
RXCJ2003.5-2323  & 0.240 $\pm$ 0.012 & -1.023 $\pm$ 0.223 & 0.730 $\pm$ 0.267 \\

\hline
\end{tabular}
\end{center}
\label{halo_par_val_vo9}
\end{table*}

\begin{figure*}
\begin{center}
%\hspace*{-1cm}
\includegraphics[scale=0.7]{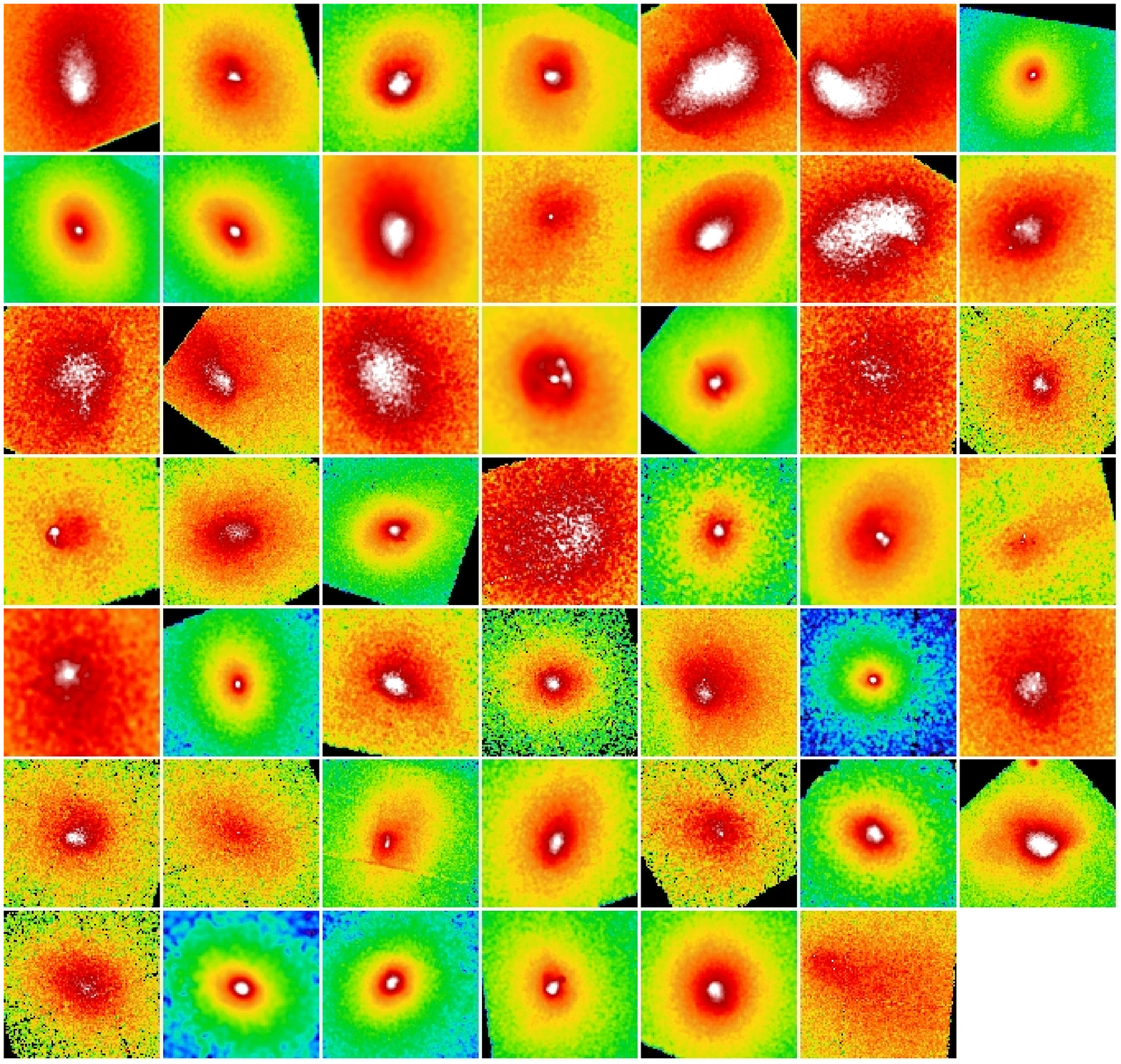}
\caption{Low-$z$ clusters of the V09. Cluster names of panels from top left to bottom right are listed as in Table \ref{tab:lowz}. {Each image has the same color, scale (log) and length}.}
\end{center}
\label{lowz vo9}
\end{figure*}

\begin{figure*}
\begin{center}
\hspace*{-1cm}
\includegraphics[scale=0.7]{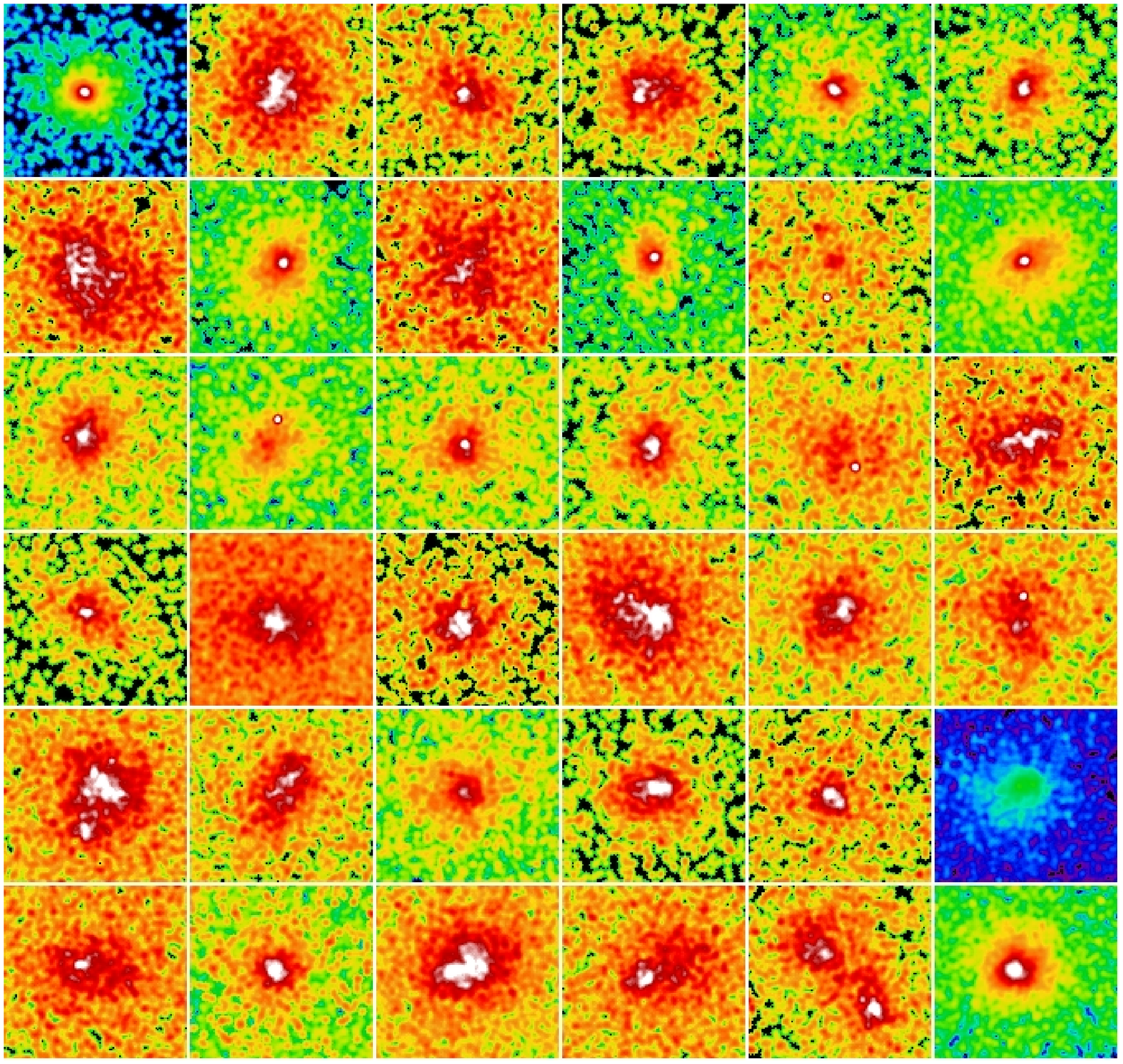}
\caption{High-$z$ clusters of the V09. Cluster names of panels from top left to bottom right are listed as in Table \ref{tab:highz}. {Each image has the same color, scale (log) and length}.}
\end{center}
\label{highz vo9}
\end{figure*}

\begin{figure*}
\begin{center}
\hspace*{-1cm}
\includegraphics[scale=0.7]{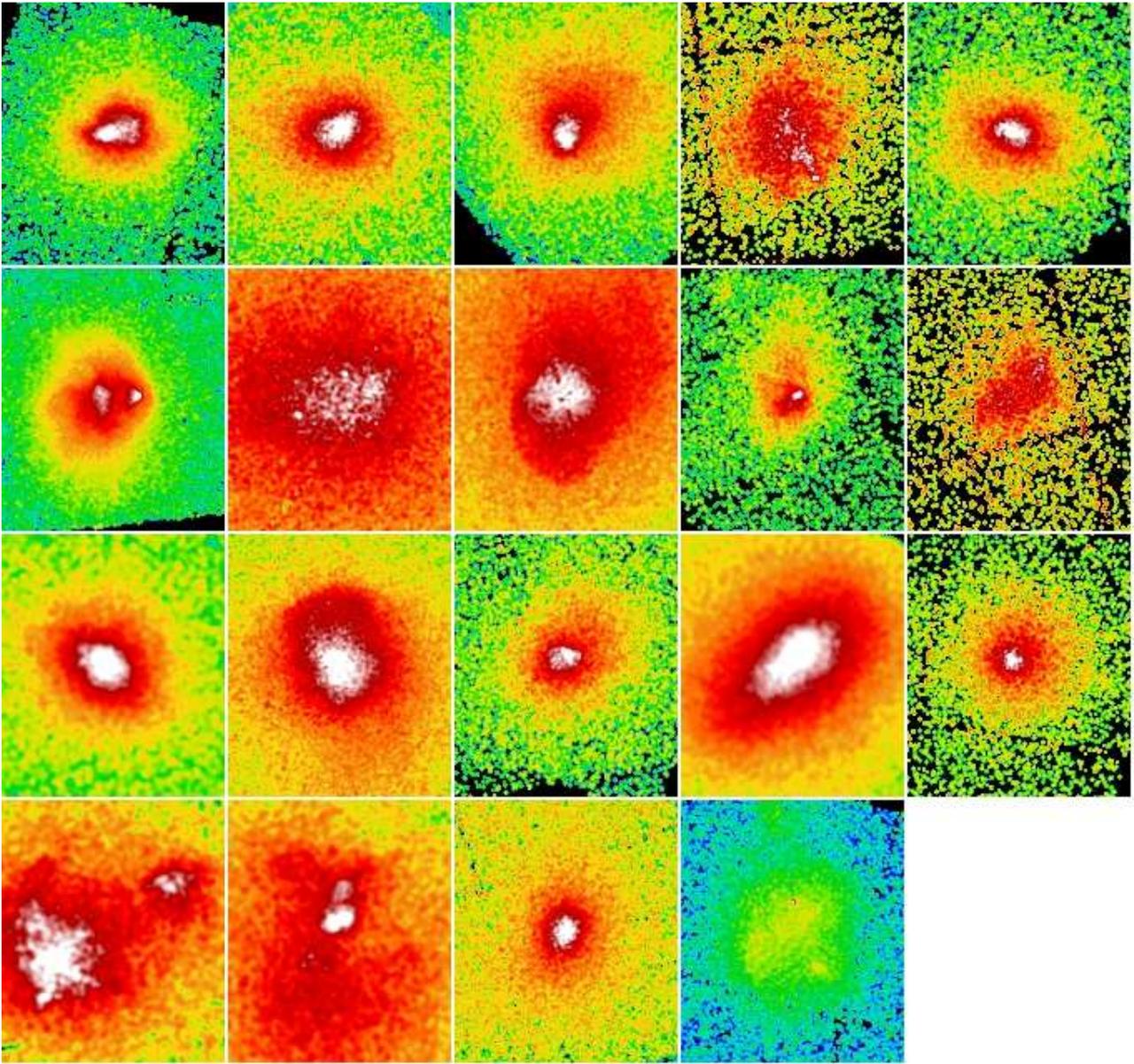}
\caption{{\it Radio halo} clusters \citep{2009A&A...507.1257G}. Cluster names of panels from top left to bottom right are listed as in Table \ref{tab:halo/relic}, except for the V09 {\it radio halo} clusters (A754, A2256, A401, A3562, A399, and A2163). {Each image has the same color, scale (log) and length}.}
\end{center}
\label{halo_img_all}
\end{figure*}
\end{appendix}

\end{document}